\documentclass[11pt,letterpaper]{article}
\usepackage[margin=1in]{geometry}
\usepackage{xcolor,enumitem}
\usepackage[T1]{fontenc}
\usepackage{lmodern}
\usepackage{amsmath,amssymb,amsthm,mathtools,aliascnt}
\usepackage{graphicx,microtype,needspace}
\usepackage{algorithm}
\usepackage[noEnd=true,indLines=true,rightComments=false]{algpseudocodex}
\usepackage[hidelinks]{hyperref}
\usepackage[capitalize,noabbrev,nameinlink]{cleveref}

\theoremstyle{plain}
\newtheorem{theorem}{Theorem}[section]
\newtheorem{lemma}{Lemma}[section]

\newtheorem{corollary}{Corollary}[section]
\theoremstyle{definition}
\newtheorem{definition}{Definition}[section]
\theoremstyle{remark}
\newtheorem{remark}{Remark}[section]
\crefname{theorem}{Theorem}{Theorems}
\Crefname{theorem}{Theorem}{Theorems}
\crefname{lemma}{Lemma}{Lemmas}
\Crefname{lemma}{Lemma}{Lemmas}
\crefname{proposition}{Proposition}{Propositions}
\Crefname{proposition}{Proposition}{Propositions}
\crefname{corollary}{Corollary}{Corollaries}
\Crefname{corollary}{Corollary}{Corollaries}
\crefname{definition}{Definition}{Definitions}
\Crefname{definition}{Definition}{Definitions}
\crefname{remark}{Remark}{Remarks}
\Crefname{remark}{Remark}{Remarks}
\newcommand{\E}{\mathbb E}
\allowdisplaybreaks

\title{Testing for a Large Induced Expander:\\[0.2em]
\Large Sublinear Algorithms and Query Lower Bounds}
\author{
  Pan Peng\thanks{University of Science and Technology of China. Email: ppeng@ustc.edu.cn} \and
  Ranran Shen\thanks{University of Science and Technology of China. Email: ranranshen@mail.ustc.edu.cn}
}
\date{}

\begin{document}

\pagenumbering{roman}   
\pagestyle{empty}       
\maketitle
\thispagestyle{empty}   

\begin{abstract}
Testing expansion asks to distinguish graphs of large conductance from graphs that are far from having large conductance. We study a relaxation in which only part of the graph is required to expand: does an $n$-vertex graph of bounded degree contain an induced expander on a constant fraction of its vertices, when the rest of the graph, including the edges leaving the expander, is arbitrary? Let $\mathcal{P}_{\alpha,\phi}$ denote the property of containing an induced subgraph on at least $\alpha n$ vertices with inner conductance at least $\phi$.

For every constant $\kappa\in(0,1)$, we give a tester that accepts graphs in $\mathcal{P}_{\alpha,\phi}$ and rejects graphs outside $\mathcal{P}_{\kappa\alpha,\mu\phi^2}$, where $\mu=\Theta_{\alpha,\kappa}(1/\log n)$, using $O_{\alpha,\kappa}(\sqrt n\,\phi^{-2}\log n)$ queries.  For fixed $\alpha\in(0,1]$, $\kappa\in(0,1)$, and $0<\phi\le9/20$, we further show that distinguishing these two classes requires $\Omega_{\alpha,\kappa,\phi}(n^{1/2+c_{\kappa,\phi}\mu})$ queries for every sufficiently large $n$ and every $0<\mu\le1$, where $c_{\kappa,\phi}>0$. The bound holds for adaptive algorithms, with its implicit constant and threshold on $n$ independent of $\mu$. Hence, for testers with $\sqrt n\,\mathrm{polylog}(n)$ queries, the logarithmic loss in conductance is necessary up to a $\log\log n$ factor. The same tester also permits distance-based rejection at a weaker conductance while retaining the original core size.

Unlike whole-graph expansion testers, our tester compares the endpoint distributions of random walks from different vertices with each other, rather than with the uniform distribution. Its analysis uses the trace chain, which observes the walk only at its visits to the unknown expanding core. We show that this chain mixes quickly, and relate it to walks in the whole graph by a stopping-time argument; the relaxation of the required size pays for the few starting vertices whose walks spend too long outside the core. Our lower bound builds on the noisy-parity construction of \cite{chiplunkar2018testing}. A single hidden sparse cut does not suffice for our property, since a large induced expander may lie on one side of it. We therefore use vector-valued parity labels in $\mathbb F_2^r$, which create many hidden classes, each too small to contain a qualifying induced expander, and show by an averaging argument that every large vertex set has a sparse cut.
\end{abstract}

\newpage
\tableofcontents
\newpage

\pagestyle{plain}       
\pagenumbering{arabic}

\section{Introduction}
Expander graphs combine sparse structure with strong connectivity: every set containing at most half the vertices has many edges to its complement, relative to its size. This combination makes bounded-degree expanders fundamental objects in theoretical computer science, with applications to derandomization, error-correcting codes, and robust network design~\cite{hoory2006expander,lubotzky2012expander,kowalski2019introduction}. Expansion also has a probabilistic interpretation: lazy random walks rapidly forget their starting vertices and approach their stationary distribution. The connection between a global connectivity requirement and the behavior of short walks raises a natural algorithmic question: how much of a graph must one explore to determine whether it expands?

Goldreich and Ron~\cite{goldreich1997property} formalized this question in the bounded-degree graph model in the framework of \emph{Property Testing}. An $n$-vertex graph $G=(V,E)$ of maximum degree at most $d$ is accessed through adjacency-list queries. In the conductance-gap formulation of expansion testing, a tester accepts graphs of conductance at least $\phi$ and rejects graphs that require more than $\varepsilon dn$ edge additions or deletions to attain conductance at least $\phi'\le\phi$, while preserving the degree bound. They showed that this requires $\Omega(\sqrt n)$ queries. In a subsequent paper~\cite{goldreich2000testing}, they proposed a random-walk collision tester whose analysis relied on a combinatorial conjecture. Its guiding intuition is that suitably regularized walks mix toward uniformity on an expander: two independent endpoints then coincide with probability close to $1/n$, whereas concentration on a smaller region produces more collisions. This viewpoint also helped motivate distribution testing~\cite{batu2000testing}. Czumaj and Sohler~\cite{czumaj2010testing} gave an unconditional analysis, obtaining $\widetilde O(\sqrt n)$ queries for fixed degree, distance, and completeness-expansion parameters, with rejection conductance $\phi'=\Theta(\phi^2/\log n)$. Kale and Seshadhri~\cite{kale2011expansion} and Nachmias and Shapira~\cite{nachmias2010testing} strengthened the conductance guarantee by allowing a query exponent above $1/2$.

The same perspective extends beyond a single expander. \emph{Clusterability} testers accept (well-clusterable) graphs whose vertex set admits a partition into a bounded number $k$ of internally expanding parts, possibly with sparse external boundaries~\cite{czumaj2015testing,chiplunkar2018testing}.  Under the bounded-degree normalization, internal expansion bounds the $(k+1)$-st smallest Laplacian eigenvalue away from zero, so a walk forgets all but at most $k$ directions quickly. Higher-order Cheeger inequalities further relate small Laplacian eigenvalues to sparse cuts~\cite{lee2014multiway}. These connections support spectral testers and, when cluster boundaries are sufficiently sparse, clustering oracles that determine cluster membership without constructing the full partition~\cite{gluch2021spectral,shen2023sublinear,shen2026sublinear,kapralov2026spectral}. Both expansion and clusterability thus impose structure on the entire graph.

A complementary line studies robustness to noise. Kale, Peres, and Seshadhri~\cite{kale2013noise} analyze graphs containing an almost-spanning induced expander with a small arbitrary remainder, and develop sublinear local reconstruction algorithms. Such an algorithm answers queries consistently with one repaired graph without first constructing all its edges. Peng~\cite{peng2020robust} studies robust clustering oracles and local reconstruction under controlled edge changes within the clusters of a well-clusterable graph. Both settings allow \emph{limited deviations} (say at most $0.1\%$ fraction of noise)  from global structure. Their analyses must control excursions into noisy regions and exclude exceptional starting vertices, since even small perturbations can trap walks and introduce slowly decaying spectral directions. The guarantees depend quantitatively on the noise level; \Cref{subsec:ov-completeness} explains the distinction relevant here.

These lines of work seek structure accounting for all, or almost all, of the input. For some purposes, however, the relevant question is whether a sufficiently large well-connected region exists, rather than whether every vertex belongs to such a description. A communication network, for example, may contain a well-connected subsystem together with a large periphery or unrelated components. Such a network is not an expander, and its periphery need not split into a few expanding clusters, yet the subsystem itself is well connected. We therefore ask whether one large expanding region exists.

Fix a constant $\alpha\in(0,1]$. We call a vertex set $U\subseteq V$ an \emph{expanding core} if $|U|\ge\alpha n$ and its induced subgraph $G[U]$ is an expander. The induced requirement is essential: the core's connectivity must be supplied by its own edges, not by paths through the exterior. Unlike an almost-spanning region in a noisy-expansion setting, the core need only contain an $\alpha$ fraction of the vertices. For example, $40\%$ of the vertices may induce an expander while the remaining $60\%$ form an arbitrary bounded-degree graph. The exterior may be connected to the core, consist of unrelated components, or contain many isolated vertices. We require neither that the exterior be small nor that the core have a sparse external boundary. We ask:

\begin{quote}
    \emph{Can one test, with a sublinear number of adjacency-list queries, whether a bounded-degree graph contains an induced expander on a given constant fraction of its vertices, when the rest of the graph and the edges leaving the expander are arbitrary?}
\end{quote}

Our goal is to detect whether an expanding core exists with a sublinear number of queries, without being given its vertices or a membership oracle. This complements full-access polynomial-time algorithms for finding induced expanders~\cite{gharan2017approximation} and constructing expander decompositions~\cite{saranurak2019expander}. Focusing on existence rather than construction is a natural starting point in the sublinear setting, since explicitly listing a linear-size core already requires linear output length.

Even an existence decision, however, becomes difficult when the size threshold must be preserved exactly. A few hidden edge changes can destroy a core that just meets the threshold, yet remain hard to detect through local queries. Indeed, we show that distinguishing graphs with an induced expander on at least $\alpha n$ vertices from graphs with no such subgraph, even of much weaker conductance, on at least $\alpha n$ vertices can require $\Omega(n)$ queries (\Cref{thrm:exact-size-lower-bound}).

This sensitivity motivates two relaxed testing formulations, both of which allow weaker expansion on the rejection side. The \emph{structural relaxation} also reduces the core size by a constant factor $\kappa$: the tester must reject graphs with no induced subgraph on at least $\kappa\alpha n$ vertices of this weaker expansion. The \emph{distance-based relaxation} instead keeps the size threshold $\alpha n$, but requires rejection only when more than $\varepsilon dn$ degree-respecting edge edits are needed to create an induced subgraph on $\alpha n$ vertices of this weaker expansion.

For the structural relaxation, we nearly determine the largest rejection threshold $\phi'$ attainable with near-square-root query complexity. For every fixed $\kappa\in(0,1)$, our tester uses $O(\phi^{-2}\sqrt n\log n)$ queries and achieves $\phi'=\Theta(\phi^2/\log n)$. Conversely, for degree bound $d=512$ and fixed $\phi\le9/20$, every tester with $\sqrt n\operatorname{polylog}(n)$ queries has $\phi'=O(\phi^2\log\log n/\log n)$. The tester walks from many starting vertices and compares their endpoint distributions with one another, which works even though the walks may spend most of their time outside the core. The same tester also solves the distance-based relaxation at the original core size.

We next define the query model and the two testing promises (\Cref{subsec:model-and-testing-promises}), state the main results (\Cref{subsec:main-results}), place them in the context of related work (\Cref{subsec:related-work}), and discuss open questions (\Cref{subsec:open-questions}).

\subsection{Query model and testing promises}
\label{subsec:model-and-testing-promises}

The input is a simple undirected graph $G=(V,E)$ on $V=[n]$, with maximum degree at most a known constant $d\ge1$. An adjacency-list query $(v,i)\in V\times[d]$ returns the $i$-th neighbor of $v$ when $i\le\deg(v)$, and the null symbol $\perp$ otherwise. Each query takes $O(1)$ time. We normalize cut sizes by the fixed degree bound $d$, rather than by the actual degrees. This convention matches the regularized lazy walk (see~\Cref{sec:preliminaries}) used by the tester and applies equally to irregular graphs.

\begin{definition}[Inner conductance]
\label{def:conductance}
    For $S\subseteq U\subseteq V$, let $E(S,U\setminus S)$ be the set of edges with one endpoint in $S$ and the other in $U\setminus S$. The \emph{inner conductance} of $G[U]$ is $\Phi_{\textup{in}}(G[U])=1$ if $|U|=1$, and otherwise
    \[
      \Phi_{\textup{in}}(G[U])
      =\min_{S\subseteq U,\ 0<|S|\le |U|/2}\frac{|E(S,U\setminus S)|}{d|S|}.
    \]
\end{definition}

We use~\Cref{def:conductance} to define our acceptance property.

\begin{definition}[Large induced expander property]
\label{def:main-property}
    For $\alpha\in(0,1]$ and $\phi\in(0,1]$, let
    \[
        \mathcal{P}_{\alpha,\phi}=\{G=(V,E):\exists U\subseteq V, |U|\ge \alpha n, \Phi_\textup{in}(G[U])\ge \phi\}.
    \]
    We call such a set $U$ an \emph{expanding core}.
\end{definition}

Both $G[V\setminus U]$ and the boundary $E(U,V\setminus U)$ are unrestricted. The property is monotone in its parameters: $\mathcal P_{\alpha,\phi}\subseteq \mathcal P_{\alpha',\phi'}$ whenever $\alpha'\le\alpha$ and $\phi'\le\phi$ (\Cref{lemma:monotonicity}).

A \emph{gap tester} accepts every graph in its YES class and rejects every graph in its NO class, each with probability at least $1-\delta$, where $\delta\in(0,1/3]$. The first requirement is called \emph{completeness}, and the second \emph{soundness}; no behavior is prescribed outside the two classes. Both testing tasks have YES class $\mathcal P_{\alpha,\phi}$. We write the rejection threshold as $\phi'=\mu\phi^2$ with a \emph{conductance multiplier} $\mu\in(0,1]$; the quadratic scale reflects the usual relation between conductance and random-walk mixing. We allow $\mu$ to depend on $n$ and ask how large it can be at a given query budget.

\vspace{0.5em}
\noindent\textbf{Structural testing.} For a size-relaxation factor $\kappa\in(0,1]$, the NO class is $\overline{\mathcal P_{\kappa\alpha,\mu\phi^2}}$. When $\kappa=1$, the size threshold is unchanged; we call this the \emph{exact-size structural promise}.

\vspace{0.5em}
\noindent\textbf{Distance testing.} For distance testing, we measure the distance between two graphs $G$ and $H$ on the same vertex set and with the same degree bound $d$ by
\[
  \textup{dist}_d(G,H)=\frac{|E(G)\triangle E(H)|}{dn},
  \qquad
  \textup{dist}_d(G,\mathcal P)=\min_{H\in\mathcal P}\textup{dist}_d(G,H).
\]
We say that $G$ is $\varepsilon$-far from $\mathcal P$ if $\textup{dist}_d(G,\mathcal P)>\varepsilon$. The NO class consists of the graphs that are $\varepsilon$-far from $\mathcal P_{\alpha,\mu\phi^2}$.

\subsection{Main results}
\label{subsec:main-results}
We give the upper and lower bounds for structural testing, and then our guarantee for distance testing. Throughout this subsection, the degree bound $d$ and the parameters $\alpha$, $\kappa$, and $\varepsilon$ are constants, whereas the conductance multiplier $\mu$ may depend on $n$. Subscripts on $O$, $\Omega$, and $\Theta$ indicate the parameters on which the hidden constants may depend.

\vspace{0.5em}
\noindent\textbf{Upper bound for structural testing.}
The first theorem gives a structural tester: local random walk samples detect one large expanding core even when most vertices lie outside it and the edges joining the two regions are unrestricted.

\begin{theorem}[Structural testing; informal]
\label{thrm:upper-bound-structural-informal}
    Let $d\ge1$ be a constant, and let $\alpha,\phi\in(0,1]$, $\kappa\in(0,1)$, and $\delta\in(0,1/3]$. There is a randomized algorithm (\Cref{algo:anchor-test}) such that, for every sufficiently large $n$ and every $n$-vertex graph $G$ of maximum degree at most $d$:
    \begin{enumerate}
        \item if $G\in\mathcal P_{\alpha,\phi}$, it accepts with probability at least $1-\delta$;
        \item if $G\notin\mathcal P_{\kappa\alpha,\mu_n\phi^2}$, it rejects with probability at least $1-\delta$, where $\mu_n=\Theta_{\alpha,\kappa}(1/\log n)$.
    \end{enumerate}
    Its worst-case adjacency-list query complexity is $O_{\alpha,\kappa}\!\left(\frac{\sqrt n\log n}{\phi^2}\log\frac1\delta\right)$.
\end{theorem}

The tester detects one induced core without a bounded cluster cover of the exterior~\cite{czumaj2015testing,chiplunkar2018testing}, and the core need not be nearly spanning as in~\cite{kale2013noise}.

The tester compares the endpoint distributions of walks from many starting vertices with one another, rather than with the uniform distribution, and its soundness follows from a sparse partition of the graph into parts smaller than the relaxed core size. For completeness, we use the trace-chain and excursion tools of~\cite{kale2013noise}, but two conclusions used there are unavailable in our setting. First, walks need not mix globally, since the graph may be disconnected or mix slowly outside the core; we show instead that walks from most core vertices approach a common, unknown distribution, which may place most of its mass outside the core. Second, the exceptional starting vertices in~\cite{kale2013noise} are bounded in terms of the size of the exterior, which here may exceed the size of the core; we instead bound the total excursion time over all core vertices and charge the exceptional starts to the size slack, which \Cref{thrm:exact-size-lower-bound} shows to be necessary. \Cref{subsec:ov-tester} sketches the argument, and \Cref{thrm:upper-bound-structural-formal} gives the full parameter dependence.

\vspace{0.5em}
\noindent\textbf{Lower bound for structural testing.} The next theorem shows that a larger multiplier requires more queries. The hard instances have maximum degree $d=512$, and conductance is normalized by this degree bound. The constant $c_{\kappa,\phi}$, the hidden constant, and the threshold $n_0$ do not depend on $\mu$, so we may take $\mu=\mu_n$ to depend on $n$, for example $\mu_n=\Theta(1/\log n)$.

\begin{theorem}[Lower bound for structural testing; informal]
\label{thrm:isotropic-lower-bound}
    Let $d=512$. Fix $\alpha\in(0,1]$, $\kappa\in(0,1)$, and $\phi\in(0,9/20]$. There are a constant $c_{\kappa,\phi}>0$ and an integer $n_0=n_0(\alpha,\kappa,\phi)$ such that, for every $n\ge n_0$ and every $0<\mu\le1$, any randomized algorithm that, on $n$-vertex graphs of maximum degree at most $d$, distinguishes $G\in\mathcal P_{\alpha,\phi}$ from $G\notin\mathcal P_{\kappa\alpha,\mu\phi^2}$ with success probability at least $2/3$ has worst-case adjacency-list query complexity $\Omega_{\alpha,\kappa,\phi}\!\left(n^{1/2+c_{\kappa,\phi}\mu}\right)$.
\end{theorem}

The construction builds upon the noisy-parity framework of~\cite{chiplunkar2018testing}, but a single hidden sparse cut does not suffice here: a qualifying core may lie entirely on one side when $\kappa$ is small. We use finite-group labels to create many hidden classes, each smaller than the relaxed core-size threshold, and control cuts in every candidate core simultaneously. This yields a lower bound for \emph{every} constant $\kappa\in(0,1)$, not just for thresholds close to $\alpha n$. 
The construction and its analysis are outlined in \Cref{subsec:ov-lower}; \Cref{thrm:isotropic-lower-bound-full} states the explicit exponent coefficient and all quantifiers.

\vspace{0.5em}
\noindent\textbf{Consequences for a near-square-root query budget.} Fix $\alpha,\kappa,\phi$ in the lower-bound range. At the multiplier $\mu_n=\Theta(1/\log n)$ of our tester, the lower bound is $\Omega(\sqrt n)$, so \Cref{thrm:upper-bound-structural-informal} is optimal up to a logarithmic factor. Since $n^{c_{\kappa,\phi}\mu_n}=\exp(c_{\kappa,\phi}\mu_n\log n)$, a tester with $\sqrt n\operatorname{polylog}(n)$ queries can only achieve $\mu_n=O(\log\log n/\log n)$. Closing the remaining $\log\log n$ gap is an open question (\Cref{subsec:open-questions}).

\vspace{0.5em}
\noindent\textbf{Distance testing at the original size.} Without size relaxation, structural testing requires $\Theta(n)$ queries, even when the rejection threshold tends to zero arbitrarily fast (\Cref{thrm:exact-size-lower-bound}). The hard instances, however, differ by only a constant number of edge edits, so they need not be rejected under a distance promise. Indeed, the same tester keeps the size threshold $\alpha n$ on both sides in the distance formulation.

\begin{theorem}[Distance testing; informal]
\label{thrm:upper-bound-distance-soundness-informal}
    Let $d\ge3$ be a constant, and let $\alpha,\phi\in(0,1]$, $\varepsilon\in(0,1)$, and $\delta\in(0,1/3]$. There is a randomized algorithm (\Cref{algo:anchor-test}) such that, for every sufficiently large $n$ and every $n$-vertex graph $G$ of maximum degree at most $d$:
    \begin{enumerate}
        \item if $G\in\mathcal P_{\alpha,\phi}$, it accepts with probability at least $1-\delta$;
        \item if $G$ is $\varepsilon$-far from $\mathcal P_{\alpha,\mu_n^{\mathrm{dist}}\phi^2}$, it rejects with probability at least $1-\delta$, where $\mu_n^{\mathrm{dist}}=\Theta_{\alpha,\varepsilon,d}(1/\log n)$.
    \end{enumerate}
    Its worst-case adjacency-list query complexity is $O_{\alpha,\varepsilon}\!\left(\frac{\sqrt n\log n}{\phi^2}\log\frac1\delta\right)$.
\end{theorem}

For $\alpha=1$, the YES class $\mathcal P_{1,\phi}$ consists of the graphs of conductance at least $\phi$, and \Cref{thrm:upper-bound-distance-soundness-informal} becomes expansion testing in the conductance-gap formulation, with a gap of the same order $\Theta(\phi^2/\log n)$ as~\cite{czumaj2010testing}. The theorem thus generalizes the expansion tester of~\cite{czumaj2010testing} from the whole graph to an induced core on any constant fraction of the vertices. 

The proof reduces to \Cref{thrm:upper-bound-structural-informal}. If a graph contains a core that is slightly smaller than $\alpha n$, we can enlarge it by inserting vertices from outside into distinct internal edges. This costs few edits, preserves the degree bound, and loses only a constant factor in conductance for fixed $d$. Hence every graph that is far from $\mathcal P_{\alpha,\mu_n^{\mathrm{dist}}\phi^2}$ also lacks the smaller core excluded by the structural tester. \Cref{subsec:ov-distance} explains the reduction, and \Cref{thrm:upper-bound-distance-soundness-formal} gives its quantitative form.

\subsection{Related work}
\label{subsec:related-work}
We discuss further related work.

\vspace{0.5em}
\noindent\textbf{Cluster covers versus an arbitrary exterior.} Our setting differs from spectral clusterability testing in its completeness promise. \cite{chiplunkar2018testing} obtained a logarithm-free conductance gap with query exponent $1/2+O(\mu)$ at weaker conductance $\mu\phi^2$, for fixed cluster and distance parameters. Sparse boundaries between the parts are not required, but a bounded number of internally expanding parts must cover the graph. Their resulting bound on the number of slowly decaying spectral directions is unavailable for one core with an arbitrary exterior.

\vspace{0.5em}
\noindent\textbf{Componentwise expansion.} Dinur and Goldreich~\cite{dinur2025expansion} allow arbitrarily many components, provided that \emph{every} component expands. They prove an $\Omega(\sqrt n)$ lower bound and an $n^{2/3+\eta}\operatorname{poly}(1/\varepsilon)$-query pseudotester for every fixed $\eta>0$, with a quadratic expansion gap. Their method combines expansion testing with generalized uniformity testing---testing whether a distribution is uniform on some unknown support. Our core need not be a component, and its ambient endpoint distributions need not be nearly uniform on any set. Their analysis therefore does not apply to our completeness promise.

\vspace{0.5em}
\noindent\textbf{Mixing certificates and local outputs.} \cite{chakraborti2022well} showed that sufficiently many vertices mixing to global uniformity in a regular graph imply an almost-spanning induced expander. Our analysis instead uses an unknown common distribution, potentially nonuniform and supported outside the core, and proves soundness through a sparse partition. The reconstruction and membership oracles discussed in the introduction provide stronger outputs under global structure or controlled noise; our tester supplies only an existence decision. Trace-chain techniques also occur in tolerant Unique Games testing~\cite{yoshida2026tolerant}, with a different testing objective. For the relation between edit noise and an almost-spanning core, \Cref{sec:graphs-close-to-expanders} records the complementary pruning implication to our size-repair argument.

\vspace{0.5em}
\noindent\textbf{Full-access induced expanders and decompositions.} Krivelevich gives sufficient conditions for expanding subgraphs based on local sparsity, with applications to paths, cycles, and minors~\cite{krivelevich2018finding,krivelevich2019expanders}; Louf and Skerman~\cite{louf2020finding} relate large expanding topological minors to large induced expanders. Oveis Gharan and Rezaei~\cite{gharan2017approximation} give polynomial-time bicriteria algorithms for induced spectral expansion, approximating the size and expansion of an explicitly returned set. These works use full graph access, whereas we seek only a decision from sublinear queries. Expander decompositions organize the entire graph into expanding parts while controlling interpart edges~\cite{saranurak2019expander}. Our analytical partition instead stops when every part is smaller than the required core: the parts need not expand, and the tester never constructs them. 

\vspace{0.5em}
\noindent\textbf{Other walk-based and existential tests.} Random walks also support forbidden-minor finding, minor-closed property testing, and partition oracles~\cite{kumar2018finding,kumar2019random,kumar2022random}, as well as sublinear Max-Cut algorithms on expanders and clusterable graphs~\cite{peng2023sublinear,jha2024sublinear}. For bipartiteness, following the walk-based tester of Goldreich and Ron~\cite{goldreich1998sublinear}, Fei and Rubinfeld~\cite{fei2026testing} show that a near-square-root number of logarithmic-length walks suffice for bipartiteness testing using an SDP-based analysis; our acceptance criterion instead compares shared endpoint mass. Testing for a clique on a linear fraction of the vertices is another existential induced-subgraph problem, studied in the dense model~\cite{goldreich1998property,feige2004graphs,blais2025testing}. There, a small vertex sample exposes many relevant edges directly. In a bounded-degree graph, even a large expanding core may leave almost no edges among the sampled vertices; walks probe connectivity beyond that induced sample.

\subsection{Open questions}
\label{subsec:open-questions}
For fixed $\alpha$, $\kappa$, and $\phi$, our tester achieves the multiplier $\mu_n=\Theta(1/\log n)$ with $O(\sqrt n\log n)$ queries, whereas the lower bound allows at most $\mu_n=O(\log\log n/\log n)$ with $\sqrt n\operatorname{polylog}(n)$ queries. Can this $\log\log n$ gap be closed?

A second question is whether a fixed positive multiplier $\mu$ can be achieved with $\widetilde O(n^{1/2+O_{\kappa,\phi}(\mu)})$ queries. The lower bound does not rule this out, and such tradeoffs are known for whole-graph expansion and clusterability~\cite{kale2011expansion,nachmias2010testing,chiplunkar2018testing}. Those testers compare walk distributions with a known reference at a finer scale. In our setting the reference distribution is unknown, and excursions into the exterior can change individual endpoint probabilities even on YES inputs, so a stronger test must remain complete for such walks.

\subsection{Organization}
\label{subsec:organization}

\Cref{sec:overview} presents the main ideas behind the tester, the distance reduction, and the lower bounds. \Cref{sec:preliminaries} introduces the notation and standard random-walk estimates. \Cref{sec:tester} proves the structural testing guarantee, and \Cref{sec:distance} proves the distance testing guarantee at the original core size. \Cref{sec:many-label-lower} proves the query lower bound for every size relaxation, and \Cref{sec:exact-size-obstruction} proves the linear lower bound without size relaxation.

\section{Technical overview}
\label{sec:overview}

This section explains the ideas behind our results. \Cref{subsec:ov-tester} describes the tester of \Cref{thrm:upper-bound-structural-informal} and why it works. \Cref{subsec:ov-distance} explains why the size relaxation is needed under the structural promise (\Cref{thrm:exact-size-lower-bound-full}), and why it is not needed under the distance promise (\Cref{thrm:upper-bound-distance-soundness-informal}). \Cref{subsec:ov-lower} describes the lower bound of \Cref{thrm:isotropic-lower-bound}, which shows that, for testers with $\widetilde O(\sqrt n)$ queries, the logarithmic loss in our tester is nearly necessary.

All walks are lazy random walks on $G$: from a vertex $x$, the walk moves to each neighbor of $x$ with probability $1/(2d)$ and stays at $x$ with the remaining probability. We write $P$ for its transition matrix. Our tester runs this walk for a random number of steps, chosen uniformly from $\{0,1,\dots,\ell-1\}$. The transition matrix of this averaged walk is $Q_\ell=\frac1\ell\sum_{j=0}^{\ell-1}P^j$, and we write $q_x=e_xQ_\ell$ for the distribution of its endpoint when it starts at $x$. Both $P$ and $Q_\ell$ are symmetric, so the uniform distribution on $V$ is stationary for both (see \Cref{sec:preliminaries}). In a YES instance, $U$ denotes a fixed expanding core, which is used only in the analysis.

In the discussion of the tester, we write
\[
  \beta=(1-\kappa)\alpha
\]
for the \emph{size slack}, so that the YES class requires a core on at least $\alpha n$ vertices and the NO class excludes cores on at least $\kappa\alpha n=(\alpha-\beta)n$ vertices. The analysis divides this gap of $\beta n$ vertices among several thresholds, which are easier to compare in terms of $\beta$. We assume $\kappa\ge3/4$, that is, $\beta\le\alpha/4$; smaller values of $\kappa$ reduce to this case by monotonicity (see \Cref{sec:tester}). In the discussion of the lower bound, we use $\kappa$ directly.

\subsection{The tester for structural testing}
\label{subsec:ov-tester}

Expansion testers~\cite{goldreich2000testing,czumaj2010testing,kale2011expansion,nachmias2010testing} sample a few vertices $x$, estimate the collision probability of walk endpoints, and compare it with $1/n$, its value for the uniform distribution on $V$. Sufficiently long walks on an expander approach uniformity closely enough for this comparison. Under our promise there is no such reference value. A walk from a vertex $x\in U$ may leave $U$, stay for a long time in $V\setminus U$, and return to $U$ at a distant vertex, so $q_x$ depends on the part of the graph outside the core, which is arbitrary. For example, if $V\setminus U$ consists of isolated vertices, then the endpoint distribution of walks from $x$ approaches the uniform distribution on $U$, whose collision probability is $1/|U|$ rather than $1/n$, and $|U|$ is unknown. If instead $U$ has many edges to a large exterior, then most of the mass of $q_x$ may lie outside $U$, and $q_x$ need not be close to any distribution we know in advance.

Moreover, the walk from a single vertex cannot tell a YES instance from a NO instance. Let $G$ be a disjoint union of expanders, each on slightly fewer than $\kappa\alpha n$ vertices. Every set of at least $\kappa\alpha n$ vertices meets two components and induces a disconnected graph, so $G$ is a NO instance. Yet the walk from every vertex mixes inside its component, exactly as it would inside a core. What separates this graph from a YES instance is the number of vertices whose walks have similar endpoint distributions. In a YES instance, most vertices of $U$, about $\alpha n$ in total, have nearly the same endpoint distribution. In the disjoint union $G$ above, walks from different components never meet, so the endpoint distribution of a vertex is close only to those of the fewer than $\kappa\alpha n$ vertices in its own component.

Our tester estimates this number (\Cref{algo:anchor-test}). It samples $R$ uniformly random vertices, called \emph{anchors}. For each anchor $x$, it samples $k$ uniformly random vertices $y$, called \emph{partners}, and checks for each partner whether a list of walk endpoints from $x$ and a list of walk endpoints from $y$ share a vertex. If at least an $(\alpha-\beta/2)$ fraction of the partners pass this check, the anchor $x$ is \emph{flagged}. The tester accepts if at least an $\alpha/2$ fraction of the anchors are flagged, and rejects otherwise.

The analysis has two parts. For completeness, we show that some $(\alpha-\beta/8)n$ vertices of $U$ have pairwise close endpoint distributions (\Cref{subsec:ov-completeness}). An anchor among these vertices collides with almost all of them, so it is flagged with good probability, and roughly an $\alpha$ fraction of the anchors are flagged. For soundness, we partition a NO instance into parts of fewer than $(\alpha-\beta)n$ vertices with few edges between them (\Cref{subsec:ov-soundness}). Every anchor then has fewer than $(\alpha-\beta)n$ potential partners in its own part, and on average collides with only a small fraction of the vertices in other parts, so only a small fraction of the anchors are flagged. The flag threshold $\alpha-\beta/2$ lies between $\alpha-\beta/8$, the fraction of good vertices in a YES instance, and $\alpha-\beta$, the bound on the size of a part in a NO instance. The gap between them comes from the size relaxation, which cannot be removed under the structural promise (\Cref{subsec:ov-distance}). \Cref{subsec:ov-soundness} also explains where the logarithmic loss in the conductance comes from, and \Cref{subsec:ov-collision} explains how the tester checks for collisions with few samples.

\subsubsection{Completeness: a common endpoint distribution on most of the core}
\label{subsec:ov-completeness}

Let $\pi_U$ be the uniform distribution on $U$. The goal is to show that, for all but $\beta n/8$ vertices $x\in U$ (\Cref{lemma:transfer}),
\begin{equation}\label{eq:ov-common-law}
 \lVert q_x-\pi_UQ_\ell\rVert_\textup{TV}\le 1/8, \qquad \ell=\Theta\biggl(\Bigl(1+\frac{1-\alpha}{\beta}\Bigr)\frac{\log n}{\phi^2}\biggr).
\end{equation}

The expansion of $G[U]$ says nothing directly about walks in $G$, which may leave $U$ at any step. Let $X_0,X_1,\dots$ be the walk with transition matrix $P$. Following~\cite{kale2013noise}, we look at the walk only at its visits to $U$. For $x,y\in U$, let $P^U(x,y)$ be the probability that the walk started at $x$ is at $y$ at its next visit to $U$, after at least one step; we call $P^U$ the \emph{trace chain} on $U$. It is symmetric, since every trip outside $U$ can be reversed, and it is lazy. Every edge of $G[U]$ is also a transition of the trace chain, so the trace chain has large conductance, no matter how many edges leave $U$. Hence it mixes to $\pi_U$ after $t=O(\phi^{-2}\log n)$ steps from every start (\Cref{lemma:inherited-conductance-and-mixing}).

The trace chain counts visits to $U$, whereas the tester samples the length of a walk in $G$ uniformly from $\{0,\ldots,\ell-1\}$, independently of the walk; the time $\tau_t$ of the $t$-th visit to $U$ is random and unbounded. Averaging over the walk length relates the two. By \Cref{lemma:stopping-time}, for every stopping time $\tau$ with $\E_x[\tau]<\infty$, if $\nu$ is the distribution of $X_\tau$, then

\[
    \lVert e_xQ_\ell-\nu Q_\ell\rVert_\textup{TV}\le\E_x[\tau]/\ell.
\]
Intuitively, a walk of uniformly random length from $x$ and one from $X_\tau$ differ only in a time shift of $\tau$ steps, which matters with probability about $\tau/\ell$. Bounds of this kind go back to the stopping rules of Lov\'asz and Winkler~\cite{lovasz1998mixing} and were used in~\cite{kale2013noise}. Here $\nu$ is arbitrary, so $e_xQ_\ell$ is compared with $\nu Q_\ell$ rather than with a stationary distribution. With $\tau=\tau_t$ we get $\nu=e_x(P^U)^t\approx\pi_U$, and therefore $q_x\approx\pi_UQ_\ell$ as long as $\E_x[\tau_t]\le\ell/16$.

The condition $\E_x[\tau_t]\le\ell/16$ can fail, and this is where the analysis of noisy expanders~\cite{kale2013noise} \emph{does not apply} to our setting. The walk from some $x$ may enter a large region outside $U$ and stay there for a long time, so that $\E_x[\tau_t]$ is much larger than $t$. A noisy expander consists of an expander part $D$ and a small noise part $B$; here $D$ plays the role of $U$, and $B$ the role of $V\setminus U$. They show that there is a set $\tilde B\subseteq D$ of at most $|B|$ vertices such that, from every start in $D\setminus\tilde B$, $t$ trace steps take at most $2t$ steps in expectation (see also~\cite{peng2020robust}). So the vertices that may behave badly, those in $B\cup\tilde B$, number at most $2|B|$. This is enough when $B$ is small. In our setting, $|V\setminus U|$ may be as large as $(1-\alpha)n$. For example, if $\alpha=0.4$, then $|V\setminus U|$ may be $0.6n$, and the bound $2|B|$ allows $1.2n$ bad vertices, more than the whole graph; in particular, $\tilde B$ may contain all of $U$. Even when $\alpha$ is larger, an exceptional set of $(1-\alpha)n$ vertices of $U$ can be much larger than the $\beta n/8$ vertices of $U$ that the size slack allows us to lose.

We therefore choose the threshold according to $\beta$. By a Kac-type formula, $\sum_{x\in U}\E_x[\tau_1]=|C_U|$, where $C_U$ is the connected component of $G$ containing $U$ (\Cref{lemma:Kac}). Since $\pi_U$ is stationary for $P^U$, it follows that
\[
    \sum_{x\in U}\bigl(\E_x[\tau_t]-t\bigr)=t\,(|C_U|-|U|)\le t(1-\alpha)n.
\]
By Markov's inequality, for every $c>1$, at most $(1-\alpha)n/(c-1)$ vertices $x\in U$ have $\E_x[\tau_t]>ct$. The threshold $c=2$ gives an exceptional set of at most $(1-\alpha)n$ vertices of $U$, as in the noisy-expander setting. We take $c=c_\textup{exc}=1+8(1-\alpha)/\beta$ instead, which reduces the exceptional set to at most $\beta n/8$ vertices of $U$. In other words, we make the walks longer by the factor $c_\textup{exc}$ and pay for the exceptional set with the size slack $\beta$, not with a bound on $|V\setminus U|$. Taking $\ell=\lceil16c_\textup{exc}t\rceil$ gives~\eqref{eq:ov-common-law}.

The distribution $\pi_UQ_\ell$ is not known to the tester. In general it is far from uniform, and most of its mass may lie outside $U$. This does not matter for the tester, which compares walks only with each other: by~\eqref{eq:ov-common-law}, $\lVert q_x-q_y\rVert_\textup{TV}\le1/4$ for all $(\alpha-\beta/8)n$ good vertices $x,y$. This is also the second difference from~\cite{kale2013noise}, where walks from most vertices of the expander part mix in the whole graph. In our setting the walk eventually converges to the uniform distribution on its connected component $C_U$, which is a stationary distribution of $P$, but its convergence may be slow. If the exterior consists of isolated vertices, this limiting distribution is uniform on $U$, rather than on $V$. The tester needs only the common distribution at the chosen walk length.

\subsubsection{Soundness: a sparse partition and cross-part overlap}
\label{subsec:ov-soundness}

Suppose $G\notin\mathcal{P}_{\kappa\alpha,\psi}$, that is, $G\notin\mathcal{P}_{\alpha-\beta,\psi}$. Then every induced subgraph on at least $(\alpha-\beta)n$ vertices has a cut of conductance below $\psi$. Different subgraphs may be cut in different places, and a candidate core may lie on one side of any single cut, so one cut is not enough. We therefore use a partition. We repeatedly split every part of size at least $(\alpha-\beta)n$ along a cut of conductance below $\psi$, until all parts are smaller (\Cref{lemma:sparse-cut-partition}). Since every cut is sparse, the set $F$ of edges between the final parts is small: $|F|<\psi dnL$ with $L=O(\log\frac{2}{\alpha-\beta})$. Because the splitting stops at parts of linear size, $L$ does not depend on $n$.

We then relate this partition to the walks used by the tester (\Cref{lemma:overlap-bounded-by-cross-flow}). Since $Q_\ell$ is symmetric, a collision at $z$ between walks from $x$ and $y$ corresponds to two walks from $z$, one ending at $x$ and one ending at $y$. If $x$ and $y$ lie in different parts, one of these walks must cross an edge of $F$, which is unlikely because $F$ is small. This gives (\Cref{coro:overlap-bounded-by-cross-flow})
\begin{equation}\label{eq:ov-cross}
    \E_{x,y\sim\textup{Unif}(V)}\bigl[\mathbf 1_{C(x)\ne C(y)}\langle q_x,q_y\rangle\bigr]<\frac{\ell\psi L}{n},
\end{equation}
where $C(x)$ is the part containing $x$. The bound holds on average rather than for every pair, which suffices since the tester rejects whenever only a small fraction of the anchors are flagged.

The logarithmic loss in the conductance comes from this bound. An anchor's own part contains fewer than $(\alpha-\beta)n$ vertices, so for the anchor to be flagged, it must collide with a $\Omega(\beta)$ fraction of partners in other parts. By~\eqref{eq:ov-cross}, this is rare as long as $\psi\ell$ is small, which forces $\psi=\Theta_{\alpha,\kappa}(1/\ell)$ (see~\eqref{eq:psi}). Since completeness requires $\ell=\Theta(\phi^{-2}\log n)$, this gives $\psi=\Theta_{\alpha,\kappa}(\phi^2/\log n)$. In short, completeness needs walks long enough for the trace chain to mix, while in a NO instance the probability of crossing a sparse cut grows linearly with the walk length. By \Cref{thrm:isotropic-lower-bound}, a tester with $\sqrt n\,\mathrm{polylog}(n)$ queries cannot reduce this loss by more than a $\log\log n$ factor.

\subsubsection{The collision test and reuse of anchor samples}
\label{subsec:ov-collision}

For each anchor $x$, the tester stores a list $A$ of $m_A$ endpoints of independent walks from $x$ and reuses it for all $k$ partners of $x$. For each partner $y$, it draws a new list $D$ of $m_D$ endpoints, and the check succeeds if $A\cap D\ne\emptyset$. A good partner whose check fails is not counted, so for a good pair the failure probability must be smaller than $\beta$. Moreover, the two endpoint distributions are compared with each other, not with the uniform distribution, and they may put large probability on a few vertices. With $\Lambda=\log(128/\beta)$, \Cref{lemma:collision-experiment} shows that lists of lengths $m_A\approx32\sqrt{n\Lambda\omega}$ and $m_D\approx32\sqrt{n\Lambda/\omega}$, for any $1\le\omega\le n/\Lambda$, miss each other with probability at most $e^{-\Lambda}=\beta/128$ when $\lVert p-q\rVert_\textup{TV}\le1/4$, and meet with probability at most $1089\,n\Lambda\langle p,q\rangle$ for arbitrary $p$ and $q$. For the first bound, the vertices that carry the common mass of $p$ and $q$ are split into light ones, which the longer list is unlikely to contain, and heavy ones. On light vertices, a birthday-type bound depends only on the product $m_Am_D=\Theta(n\Lambda)$. Each heavy vertex appears in the longer list with constant probability, so the shorter list has to hit the heavy vertices itself, which requires $m_D=\Omega(\Lambda)$; this is why $\omega\le n/\Lambda$. The second bound is a union bound over pairs of samples.

With $\omega=k$ (recall that $k$ is the number of partners per anchor), each anchor uses $m_A+km_D=O(\sqrt{n\Lambda k})$ samples instead of $k\sqrt{n\Lambda}$, which saves a factor $\sqrt k$. The $k$ checks of one anchor share the list $A$ and are not independent, and even a good anchor may draw an atypical list. We therefore condition on the \emph{anchor state} $(x,A)$ and consider its \emph{score}
\[
    f(x,A)=\Pr_{y,D}\bigl[A\cap D\ne\emptyset\bigm| x,A\bigr],
\]
the probability that the check with a fresh partner succeeds (\Cref{lem:reusable-scores}). In a YES instance, the expected fraction of good partners missed by the list of a good anchor is at most $\beta/128$, so by Markov's inequality, $f(x,A)\ge\alpha-\beta/4$ with probability at least $15/16$ over $A$. In a NO instance, $f(x,A)$ is at most the fraction of partners in the part of $x$, which is less than $\alpha-\beta$, plus the fraction of colliding partners in other parts, whose average is at most $\alpha\beta/64$ by~\eqref{eq:ov-cross}; so by Markov's inequality, $f(x,A)>\alpha-3\beta/4$ for at most an $\alpha/16$ fraction of anchor states.

Once the anchor state $(x,A)$ is fixed, the $k$ checks with fresh partners are independent Bernoulli trials with mean $f(x,A)$, so their average estimates $f(x,A)$. The thresholds $\alpha-\beta/4$ and $\alpha-3\beta/4$ lie $\beta/4$ above and below the flag threshold $\alpha-\beta/2$, and $k=\Theta(\alpha\beta^{-2}\log(2/\alpha))$ partners suffice to decide on which side the score lies, except with probability at most $\alpha/64$ (\Cref{lemma:score-estimation}). Finally, $R=\Theta(\alpha^{-1}\log(1/\delta))$ independent anchors suffice to decide whether at least an $\alpha/2$ fraction of the anchors are flagged. The tester makes $O(R\ell\sqrt{n\Lambda k})=O_{\alpha,\kappa}(\phi^{-2}\sqrt n\log n\log(1/\delta))$ queries (\Cref{thrm:upper-bound-structural-formal}).

\subsection{The size relaxation and distance soundness}
\label{subsec:ov-distance}
\label{subsec:ov-exact}

Under the structural promise, the size relaxation cannot be removed (\Cref{thrm:exact-size-lower-bound}). Let $G_{\mathrm Y}$ consist of a $d$-regular expander on exactly $\lceil\alpha n\rceil$ vertices together with isolated vertices. To get a NO graph, we pick a random vertex $v$ of the expander and delete its $d$ edges. Now every component has fewer than $\alpha n$ vertices, so every set of at least $\alpha n$ vertices induces a disconnected graph, and the graph is a NO instance of the exact-size structural promise for every positive conductance threshold. The two graphs differ only at $v$ and its $d$ neighbors. If we run an algorithm on both graphs with the same random bits, the two runs are the same until the algorithm queries one of these $d+1$ vertices. Since $v$ is random and independent of the run on $G_{\mathrm Y}$, each query of that run hits one of them with probability at most $(d+1)/\lceil\alpha n\rceil$, and a union bound over these queries shows that the algorithm needs $\Omega(\alpha n/d)$ queries in expectation on $G_{\mathrm Y}$ (\Cref{sec:exact-size-obstruction}).

These NO graphs are very close to $G_{\mathrm Y}$: their edit distance is only $1/n$. Under the distance promise, a NO instance must be $\varepsilon$-far from the YES class, so such graphs need not be rejected, and the size relaxation is no longer needed (\Cref{thrm:upper-bound-distance-soundness-informal}). The same tester works. We run it with $\kappa=1-\min\{1/4,\varepsilon/(8\alpha)\}$, i.e., $\beta=\min\{\alpha/4,\varepsilon/8\}$, so that it rejects every graph outside $\mathcal P_{\alpha-\beta,\psi}$. It then suffices to show that every graph in $\mathcal P_{\alpha-\beta,\psi}$ is close to $\mathcal P_{\alpha,\psi/(2d+2)}$; taking the contrapositive, every graph that is $\varepsilon$-far from $\mathcal P_{\alpha,\psi/(2d+2)}$ is rejected.

To show this, we enlarge a core of size at least $(\alpha-\beta)n$ to one of size $\alpha n$ (\Cref{lemma:adding-vertices-by-subdivision}). We take $b\le\beta n+1$ vertices outside the core, delete all their edges, and insert each of them into a different edge of the core, so that the edge $uw$ becomes the path $u$--$v$--$w$. This takes at most $(d+3)b$ edits and keeps the maximum degree at most $d$. An edge of the old core that crosses a cut becomes a path that crosses the corresponding cut of the new core, and a counting argument shows that the inner conductance decreases by at most a factor $2d+2$. So every graph in $\mathcal P_{\alpha-\beta,\psi}$ is $(5\beta/2)$-close to $\mathcal P_{\alpha,\psi/(2d+2)}$ (\Cref{coro:size-slack-implies-small-edit-distance}), and $5\beta/2\le5\varepsilon/16<\varepsilon$.

\subsection{The lower bound for structural testing}
\label{subsec:ov-lower}

We construct two input distributions. The first is concentrated on $\mathcal P_{\alpha,\phi}$, the second on graphs outside $\mathcal P_{\kappa\alpha,\mu\phi^2}$, and no algorithm with $o(n^{1/2+c_{\kappa,\phi}\mu})$ queries can tell them apart. We build on the NoisyParities problem of~\cite{chiplunkar2018testing}. There, the edges of a random $d$-regular graph on $N$ vertices carry bits; in our construction, $N$ will be a constant fraction of $n$. The bits are either independent and uniform, or \emph{planted}: $Y_{uv}=X_u+X_v+Z_{uv}$, where $X$ is a hidden uniform labeling of the vertices and the noise $Z_{uv}$ is $1$ with probability $p$, independently over the edges. The noise rate $p$ will be proportional to $\mu$. On a forest, the planted labels are also uniform, since the hidden labeling $X$ can absorb any values; only the sum of the labels around a cycle carries information, and under the planted law this sum equals the sum of the noise on the cycle. An algorithm with few queries sees few cycles, all of them long, and the noise on a long cycle hides its parity.

To turn the labels into a graph, take the $2$-lift they define: replace each vertex $v$ by $(v,0)$ and $(v,1)$, and each edge $uv$ by the two edges $(u,a)$--$(v,a+Y_{uv})$. Uniform labels give an expander, while planted labels split the lift into the two hidden halves $\{(v,a):a+X_v=0\}$ and $\{(v,a):a+X_v=1\}$, with about a $p$ fraction of the edges between them. For our property, one sparse cut is not enough, because a qualifying induced core may lie entirely on one side. For example, two disjoint expanders form a sparse bipartition, yet either one can meet a size threshold below half of the graph. In the planted lift, each half has about half of the vertices, so when $\kappa<1/2$ it is larger than the relaxed threshold $\kappa\alpha n$, and the sparse cut between the halves says nothing about the edges inside it. The NO argument must instead exclude every set of at least $\kappa\alpha n$ vertices, including sets that meet several hidden parts.

We therefore use labels in the group $K=\mathbb F_2^r$ of $r$-bit vectors, with $q=2^r\ge2/\kappa$. The lift replaces each base vertex $v$ by the $q$ vertices $(v,a)$, $a\in K$, and each base edge $uv$ by the perfect matching $(u,a)$--$(v,a+Y_{uv})$; we delete fewer than $q$ vertices from one fiber to leave $\lceil\alpha n\rceil$ lift vertices, and isolated vertices pad it to $n$ vertices. Under the planted law, color the vertex $(v,a)$ by $a+X_v$. Then there are $q$ color classes of at most $N$ vertices each, where $N$ is about $\alpha n/q\le\kappa\alpha n/2$, and a lifted edge joins two different classes only if its noise is nonzero (\Cref{lem:lower-colors}).

All candidate cores are then excluded at once by an averaging bound (\Cref{lem:lower-class-cut}). Let $G$ be any graph whose vertex set is partitioned into classes of size at most $s$, and let $F$ be the set of edges between classes. For every set $U$ with $|U|\ge2s$, the intersection of $U$ with each class has at most $|U|/2$ vertices and is one side of a cut of $G[U]$. Summing the conductance inequality over these cuts counts every edge of $F$ at most twice, so $\Phi_\textup{in}(G[U])\le2|F|/(d|U|)$ for all such $U$ simultaneously, without a union bound over sets. We apply this bound to the color classes of the planted lift, with $s=N$ and $U$ any set of at least $\kappa\alpha n$ vertices; the choice $q\ge2/\kappa$ makes $\kappa\alpha n\ge2N$ up to rounding (\Cref{subsec:lower-no}). For $\mu\ge N^{-1/2}$, the noise rate $p=\kappa\mu\phi^2/2$ ensures with high probability that every such $U$ has inner conductance below $\mu\phi^2$. For smaller $\mu$ we take zero noise, so every such $U$ induces a disconnected graph.

Under uniform labels, the lift is an expander. Its adjacency matrix decomposes along the characters $\chi_\sigma(a)=(-1)^{\sigma\cdot a}$ of $K$: the part for $\sigma=0$ is the base graph, and each part for $\sigma\ne0$ is a signed copy of the base graph with independent uniform signs. The base graph has small nontrivial eigenvalues by Friedman's theorem~\cite{friedman2008proof,bordenave2015new}. The signed copies have small eigenvalues by a theorem on random signings of graphs in which every ball of radius $\Theta(\log N)$ contains at most one cycle~\cite{mohanty2020explicit}, a property that random regular graphs have with high probability~\cite{bordenave2015new}. This gives inner conductance above $9/20$ for $d=512$ (\Cref{lem:lower-yes-simple}).

To show that the two label laws are indistinguishable, we compute the likelihood ratio of the labels $y$ on an exposed multigraph $H$. A \emph{circulation} on $H$ is an assignment $\gamma$ of elements of $K$ to the edges such that the values at every vertex sum to zero. We use \emph{isotropic} noise: $Z_e=0$ with probability $1-p$, and otherwise $Z_e$ is uniform on the nonzero elements of $K$. Then every nontrivial character of $Z_e$ has the same bias $\rho=1-qp/(q-1)$, and Fourier analysis over $K$ gives
\[
    D_H(y)=\frac{\Pr_{\mathrm{planted}}(y)}{\Pr_{\mathrm{uniform}}(y)}=\sum_{\gamma:\,\partial\gamma=0}\chi_\gamma(y)\,\rho^{|\operatorname{supp}\gamma|}.
\]
 The term $\gamma=0$ equals $1$, so the two laws are close if there are few nonzero circulations and each has large support. For $K=\mathbb F_2$, circulations are unions of edge-disjoint cycles, and it suffices that few cycles are exposed and each is long. For $r>1$, the support of a circulation need not be a union of cycles, since three nonzero vectors can sum to zero at a vertex. We therefore bound the whole sum at once (\Cref{lem:isotropic-closure}). Fix a spanning forest of $H$, and let $R$ be the number of edges of $H$ outside it; $R$ depends on $H$ and we bound $R$ in the next paragraph. A circulation is determined by its values on these $R$ edges. If their endpoints are pairwise at distance at least $D_{\mathrm{sep}}$ in the forest, then a circulation that is nonzero on $j$ of them has support of size at least $jD_{\mathrm{sep}}$, as one sees by walking around each tree of its support in the forest. Hence
\[
    |D_H(y)-1|\le\exp\bigl((q-1)R\rho^{D_{\mathrm{sep}}}\bigr)-1.
\]

It remains to bound $R$ and $D_{\mathrm{sep}}$ for the graph exposed by an adaptive algorithm. The algorithm makes \emph{source queries}: a query to a base vertex returns its incident edges and their labels. By \Cref{lem:lower-simple}, each adjacency-list query to the lift costs at most one source query. Consider an algorithm with $T=sN^{1/2+\xi}$ source queries (\Cref{lem:isotropic-source}). Although the exposed graph depends on the labels the algorithm has seen, the probability of a complete record of queries and answers factors into a part that depends only on the exposed graph and the likelihood of its labels, so its likelihood ratio is still $D_H(y)$. We generate the base graph in the configuration model, in which each new partner half-edge is uniform among the unmatched ones, and condition on simplicity at the end. Then the expected number of edges outside the forest is $O(T^2/N)$, and by Markov's inequality $R\le aN^{2\xi}$ except with probability $O(s^2/a)$. To keep the endpoints of these edges apart, we put guards on the balls of radius $D_{\mathrm{sep}}=\lfloor\ln N/(4\ln d_0)\rfloor$ around existing endpoints; each ball has $O(N^{1/4})$ vertices. Exposing an edge into a guarded ball is unlikely, also when the algorithm queries guarded vertices on purpose, because each vertex has only $d_0$ half-edges. Since $\rho^{D_{\mathrm{sep}}}\le2N^{-b\lambda}$ with $b=1/(4\ln d_0)$ and $\lambda=-\ln\rho$, every record with $R\le aN^{2\xi}$ and separated endpoints satisfies $(q-1)R\rho^{D_{\mathrm{sep}}}=O(a)$ whenever $2\xi\le b\lambda$. When $\mu\ge N^{-1/2}$, the noise rate is $p=\kappa\mu\phi^2/2$, so $\lambda=\Theta(\kappa\mu\phi^2)$ and we can take $\xi=c_{\kappa,\phi}\mu$, which gives the bound $n^{1/2+c_{\kappa,\phi}\mu}$.

Compared with~\cite{chiplunkar2018testing}, the essential new ingredients are the many hidden classes and the averaging bound above, because the NO condition must exclude every large induced expander rather than a single sparse cut. With isotropic noise, the likelihood bound must also handle branching circulations. Independent binary noise on the $r$ coordinates would instead reduce to the binary case by a hybrid argument, but with a coefficient smaller by a factor $\Theta(\log(2/\kappa))$ (\Cref{rem:coordinate-hybrid}). Finally, the bounds on bad records do not involve the noise rate, and the condition $2\xi\le b\lambda$ gives $(q-1)R\rho^{D_{\mathrm{sep}}}=O(a)$ for every admissible noise rate, so the constants in \Cref{lem:isotropic-source} do not depend on the noise. As a result, \Cref{thrm:isotropic-lower-bound} holds for every $0<\mu\le1$ with the same constants, including $\mu$ that tends to zero arbitrarily fast with $n$.   When $\mu<N^{-1/2}$, we take $p=0$: every set of at least $\kappa\alpha n$ vertices then induces a disconnected graph, and since $n^{c_{\kappa,\phi}\mu}=O(1)$ in this range, the bound $\Omega(\sqrt n)$ suffices.

\section{Preliminaries}
\label{sec:preliminaries}

This section collects the notation and random-walk estimates used in the proofs. We retain the graph model and conductance normalization of \Cref{subsec:model-and-testing-promises}; in particular, a \emph{$d$-bounded} graph has maximum degree at most $d$. We first record how the expanding-core property changes when its parameters are relaxed, then define the walk and the two analytic estimates needed in \Cref{sec:tester}.

\begin{lemma}[Monotonicity]
\label{lemma:monotonicity}
    If $\alpha^\prime \le \alpha$ and $\phi^\prime \le \phi$, then $\mathcal{P}_{\alpha,\phi}\subseteq \mathcal{P}_{\alpha^\prime,\phi^\prime}$.
\end{lemma}
\begin{proof}
    Any core of size at least $\alpha n$ and inner conductance at least $\phi$ also has size at least $\alpha^\prime n$ and inner conductance at least $\phi^\prime$.
\end{proof}

\paragraph{Vectors and distributions.}
Endpoint distributions are row vectors and evolve by right multiplication by a transition matrix. We regard functions in spectral arguments as column vectors; the intended convention will be stated when needed. For $x\in V$, let $e_x$ be the point mass at $x$, and for a nonempty set $S\subseteq V$, let $\mathrm{Unif}(S)$ be the uniform distribution on $S$.  For a real vector $a$, write
\[
    \|a\|_1=\sum_x|a(x)|, \qquad \|a\|_{\mathrm{TV}}=\tfrac12\|a\|_1.
\]
For probability distributions $p,q$, their total variation distance is the largest difference between the probabilities they assign to the same event. The inner product $\langle p,q\rangle=\sum_zp(z)q(z)$ is the probability that independent samples from $p$ and $q$ coincide.

A matrix $K$ is \emph{stochastic} if its entries are nonnegative and each row sums to one. When $K$ is symmetric, its columns also sum to one, so the uniform distribution is stationary. For a distribution $\pi$, we use the weighted inner product and norm
\[
 \langle f,g\rangle_\pi=\sum_x\pi(x)f(x)g(x), \qquad \|f\|_{2,\pi}^2=\langle f,f\rangle_\pi.
\]
Without a subscript, the inner product and Euclidean norm are unweighted.

\paragraph{The walk and its random-length version.}
As in \Cref{subsec:ov-tester}, we use the \emph{$d$-regularized lazy walk} on the input graph. Its transition matrix is
\[
    P(x,y)=
    \begin{cases}
        \dfrac{1}{2d} & \{x,y\}\in E,\ x\neq y\\[6pt]
        1-\dfrac{\deg(x)}{2d} & x=y\\[6pt]
        0 & \text{otherwise}
    \end{cases}.
\]

From each vertex, the walk moves along each incident edge with probability $1/(2d)$ and stays put with the remaining probability. The walk is therefore symmetric and \emph{lazy}, meaning that it stays put with probability at least $1/2$. One step uses at most one adjacency-list query: with probability $1/2$ stay; otherwise it chooses $i\in[d]$ uniformly at random, makes the query $(v,i)$, and moves to the returned neighbor unless the answer is $\perp$.

The \emph{uniform averaging walk} uses a random number of steps. For an integer $\ell\ge1$, its transition matrix is
\[
    Q_\ell=\frac{1}{\ell}\sum_{j=0}^{\ell-1}P^j.
\]

To sample its endpoint from $x$, choose $j$ uniformly from $\{0,\ldots,\ell-1\}$ and take $j$ steps of $P$. We write $q_x=e_xQ_\ell$ for the resulting distribution. Each sample uses fewer than $\ell$ neighbor queries. Thus $\ell$ is the averaging-window length, not a fixed number of steps taken by every sampled walk. 

Both $P$ and $Q_\ell$ are symmetric and stochastic. Uniform probability on $V$ is stationary even when $G$ is disconnected; this does not mean that a walk from an individual vertex mixes to that distribution.

For a walk $(X_s)_{s\ge0}$, write $\Pr_x$ and $\E_x$ when $X_0=x$. A \emph{stopping time} $\tau$ is a rule for stopping that can be decided from the trajectory observed so far: the event $\{\tau=s\}$ depends only on $X_0,\ldots,X_s$. We call it \emph{integrable} when $\E_x[\tau]<\infty$.

\paragraph{Contraction and mixing.}
We use two standard estimates. First, applying a stochastic matrix to a vector cannot increase
its $\ell_1$ norm. In particular, applying the same stochastic matrix to two distributions
cannot increase their total variation distance. 

\begin{lemma}[Contraction of total variation]
\label{lemma:contraction-of-total-variation}
    Let $K\in\mathbb{R}^{n\times n}$ be a stochastic matrix and let $a\in\mathbb{R}^n$ be a vector. Then
    \[
        \lVert aK\rVert_\textup{TV}\le \lVert a\rVert_\textup{TV}.
    \]
\end{lemma}
\begin{proof}
    $\lVert aK\rVert_1\le \sum_{x\in[n]}{\sum_{y\in[n]}{|a(y)|}\cdot K(y,x)} =\sum_{y\in[n]}{|a(y)|\cdot \sum_{x\in[n]}{K(y,x)}}=\lVert a\rVert_1.$
    Dividing by $2$ gives the claim.
\end{proof}

Second, conductance controls the mixing of a lazy symmetric chain. To state the relevant inequality, recall that a chain is \emph{reversible} with stationary distribution $\pi$ if $\pi(x)K(x,y)=\pi(y)K(y,x)$ for every pair of states. This equality says that the stationary probability flow is the same in both directions.

\begin{definition}[Conductance of Markov chains]
\label{defn:conductance-of-chain}
    For a reversible Markov chain $K$ with stationary distribution $\pi$, define its conductance by
    \[
        \Phi(K)=\min_{0<\pi(S)\le 1/2}{\frac{\sum_{x\in S,y\notin S}{\pi(x)\cdot K(x,y)}}{\pi(S)}},
    \]
    where $\pi(S)=\sum_{x\in S}{\pi(x)}$. The ratio is the probability of leaving $S$ in one step when the start has law $\pi$ conditioned on $S$.
\end{definition}

The \emph{Dirichlet form} $\mathcal E_K(f)$ below measures half the expected squared change of $f$ in one stationary step. Cheeger's inequality says that a function with substantial variation must have substantial energy when every cut expands. For a lazy chain, this energy bound implies decay of the nonconstant directions and hence the stated mixing estimate.

\begin{lemma}[Cheeger bound for symmetric chains] 
\label{lemma:mixing-time}
    Let $K$ be a symmetric stochastic matrix on a finite state space $\Omega$, and let $\pi$ be the uniform distribution on $\Omega$. Let $h\in(0,1]$ and assume that $\Phi(K)\ge h$. Then, for every real function $f$ of $\pi$-mean zero, i.e., $\sum_{x}{\pi(x)f(x)}=0$,
    \[
        \mathcal{E}_K(f)=\frac{1}{2}\sum_{x,y}{\pi(x)K(x,y)}\left(f(x)-f(y)\right)^2\ge \frac{h^2}{2}\lVert f\rVert_{2,\pi}^2.
    \]
    
    If $K$ is also lazy, then for every state $x$ and every integer $t\ge 0$,
    \[
        \lVert e_xK^t-\pi\rVert_\textup{TV}\le \frac{1}{2}\sqrt{|\Omega|}e^{-h^2t/2}.
    \]
\end{lemma}

The proof of~\Cref{lemma:mixing-time} is deferred to~\Cref{subsec:proof-of-mixing-time}. We will apply this lemma to the trace chain rather than infer global mixing of $G$.

\section{The tester for structural testing}
\label{sec:tester}

This section proves \Cref{thrm:upper-bound-structural-informal}, following the outline of \Cref{subsec:ov-tester}. We present the tester in \Cref{algo:anchor-test} and prove completeness, soundness, and the query bound in \Cref{subsec:tester-analysis}.

\begin{theorem}[Structural testing; formal version of~\Cref{thrm:upper-bound-structural-informal}]
\label{thrm:upper-bound-structural-formal}
    Let $d\ge1$ be a constant. Let $\alpha,\phi\in(0,1]$ and $\delta\in(0,1/3]$. Let $\kappa\in[3/4,1)$ and $n\ge4/((1-\kappa)\alpha)$. There is a randomized algorithm (\Cref{algo:anchor-test}) such that, for every $n$-vertex graph $G$ of maximum degree at most $d$:
    \begin{enumerate}
        \item if $G\in\mathcal{P}_{\alpha,\phi}$, it accepts with probability at least $1-\delta$;
        \item if $G\notin\mathcal{P}_{\kappa\alpha,\mu_n\phi^2}$, it rejects with probability at least $1-\delta$, where $\mu_n=\psi/\phi^2$ and $\psi$ is defined in~\eqref{eq:psi}. In particular,
        \[
           \mu_n=\Theta\!\left(
           \frac{(1-\kappa)^2\alpha^3}
           {(1-\kappa\alpha)\log n\,\log\frac{1}{(1-\kappa)\alpha}\,\log\frac{2}{\alpha}}
           \right).
        \]
    \end{enumerate}
    The algorithm makes
    \[
        \min\Bigl\{dn,\ O\Bigl(\frac{\sqrt n\log n}{\phi^2}\cdot
        \frac{1-\kappa\alpha}{(1-\kappa)^2\alpha^{5/2}}
        \sqrt{\log\tfrac{2}{\alpha}\,\log\tfrac{1}{(1-\kappa)\alpha}}\cdot
        \log\tfrac{1}{\delta}\Bigr)\Bigr\}
    \]
    adjacency-list queries in the worst case. For fixed $\alpha,\kappa,\phi,\delta$ and all sufficiently large $n$, its running time is bounded by the same order as the second term.
\end{theorem}

The restriction $\kappa\ge3/4$ is only a convenience for the analysis. For $\kappa<3/4$, run the tester with $\kappa^\prime=\max\{\kappa,3/4\}$ and evaluate all parameters at $\kappa^\prime$. By \Cref{lemma:monotonicity}, nonmembership in $\mathcal P_{\kappa\alpha,\psi}$ implies nonmembership in $\mathcal P_{\kappa^\prime\alpha,\psi}$. This recovers the full range of \Cref{thrm:upper-bound-structural-informal}.

Throughout the rest of this section, we write
\[
    \beta=(1-\kappa)\alpha
\]
for the size slack, so that $\kappa\alpha=\alpha-\beta$. The analysis divides the gap of $\beta n$ vertices between the YES and NO core sizes among several thresholds, which are easier to compare in terms of $\beta$.

\subsection{The tester}
\label{subsec:tester-algorithm}
We specify the random-walk comparisons and their deterministic query budget. The parameters have three roles: $t,c_{\mathrm{exc}},\ell$ ensure the common-distribution guarantee; $\Lambda,k,\omega,m_A,m_D$ control the accuracy and cost of each anchor's comparisons; and $R$ controls the final failure probability. The trace time $t$ is used only in the analysis: every sampled endpoint is obtained from a walk in $G$ of random length less than $\ell$.

For each anchor, its list $A$ is stored once and reused with $k$ fresh partner lists. All randomness for different anchors is independent. Reading the full adjacency table is an alternative implementation when it is cheaper than the sampling bound; the subsequent experiment is unchanged. The parameter restrictions ensure $1\le\omega\le n/\Lambda$, as checked in \Cref{lem:reusable-scores}.

\begin{algorithm}[H]
\caption{\textsc{AnchorTest}$(G,\alpha,\kappa,\phi,\delta)$}
\label{algo:anchor-test}
    \begin{algorithmic}[1]
        \State $\beta\coloneqq (1-\kappa)\alpha$\quad\Comment{size slack parameter}
        
        \State $t\coloneqq \left\lceil \frac{8\log {8\sqrt{n}}}{\phi^2} \right\rceil$, $c_\textup{exc}\coloneqq1+\frac{8(1-\alpha)}{\beta}$, $\ell \coloneqq\left\lceil 16 c_\textup{exc}t \right\rceil$\quad\Comment{$t$ trace steps suffice for mixing in a core}
        
        \State $\Lambda \coloneqq \log{\frac{128}{\beta}}$, $k\coloneqq \left\lceil \frac{64\alpha}{\beta^2}\log {\frac{128}{\alpha}} \right\rceil$, $\omega \coloneqq\min\{k,\frac{n}{\Lambda}\}$
        
        \State $m_A\coloneqq \left\lceil 32\sqrt{n\Lambda \omega} \right\rceil$, $m_D\coloneqq \left\lceil 32\sqrt{n\Lambda/ \omega} \right\rceil$\quad\Comment{lengths of the anchor and partner lists}
        
        \State $R\coloneqq \left\lceil \frac{48}{\alpha}\log{\frac{2}{\delta}} \right\rceil$

        \If{$k>n/\Lambda$ or $66R\ell\sqrt{n\Lambda k}>dn$}\label{line:full-table}
            \State read the adjacency table and answer all subsequent graph queries from this table
        \EndIf

        \State $N_\textup{flag}\coloneqq 0$
        \For{$i=1$ to $R$}
            \State draw a vertex $x$, called \emph{anchor}, uniformly at random from $V$, and a list $A$ of $m_A$ independent samples from $q_x=e_xQ_\ell$

            \State $N_\textup{collision}\coloneqq  0$
            
            \For{$j=1$ to $k$}
                \State draw a vertex $y$, called \emph{partner}, uniformly at random from $V$, and a list $D$ of $m_D$ independent samples from $q_y=e_yQ_\ell$

                \State $N_\textup{collision}\coloneqq N_\textup{collision} + \mathbf{1}_{A\cap D\ne \emptyset}$\quad\Comment{number of partners whose lists intersect $A$}
            \EndFor

            \If{$N_\textup{collision}/k\ge \alpha-\beta/2$}
                \State $N_\textup{flag}\coloneqq N_\textup{flag} + 1$
            \EndIf
        \EndFor

        \State \Return accept if $N_\textup{flag}/R\ge\alpha /2$, and reject otherwise
    \end{algorithmic}
\end{algorithm}

\subsection{Analysis}
\label{subsec:tester-analysis}

The proof follows the three steps outlined in \Cref{subsec:ov-tester}: establish a common endpoint distribution on most of a YES core, bound average cross-part overlap in a NO graph, and turn these facts into a reliable collision test. \Cref{subsec:trace-chain,subsec:trace-transfer} prove completeness of the walk comparison, \Cref{subsec:partition-flow} supplies its soundness input, and \Cref{subsec:collision-bound,subsec:structural-proof} analyze sampling and concentration.

\subsubsection{The trace chain on a core}
\label{subsec:trace-chain}

Suppose $G[U]$ has positive inner conductance and $|U|\ge 2$. Then $G[U]$ is connected, so $U$ is contained in a single connected component of $G$, denoted by $C_U$. For a walk $(X_s)_{s\ge0}$ started in $U$, define successive visit times to $U$ by
\[
    \tau_0=0, \qquad \tau_{j+1}=\min\{s>\tau_j: X_s\in U\}.
\]

The chain $(X_{\tau_j})_{j\ge 0}$ is the \textit{trace chain} on $U$, with transition matrix $P^U$. Each trace step goes from one visit to $U$ to the next. The walk may leave $U$ before returning, but the whole trip counts as one trace step. A holding step in $U$ also counts as one trace step and costs one step of $P$. Thus, $t$ trace steps cost $\tau_t$ steps of $P$, and $\tau_t-t$ is the extra cost of trips outside $U$. We write $\pi_U=\operatorname{Unif}(U)$.

The next lemma shows that the trace chain is symmetric and lazy, so that $\pi_U$ is stationary.

\begin{lemma}[Symmetry and laziness]
\label{lemma:trace-chain-symmetry-laziness}
    The matrix $P^U$ is symmetric and stochastic. Hence $\pi_U$ is stationary. Moreover, $P^U(x,x)\ge 1/2$ for every $x\in U$.
\end{lemma}
\begin{proof}
    For $x,y\in U$, we can write $P^U(x,y)$ as
    \[
    P^U(x,y) = P(x,y) + \sum_{k\ge 2}{\sum_{z_1,\dots,z_{k-1}\in C_U\setminus U}{P(x,z_1)P(z_1,z_2)\dots P(z_{k-1},y)}}.
    \]

    Reversing each path leaves its probability unchanged, so $P^U(x,y)=P^U(y,x)$ for all $x,y\in U$.
    
    For each $x\in U$, the row sum is $\Pr_x(\tau_1<\infty)=1$, because the finite connected component $C_U$ is irreducible. Thus $P^U$ is stochastic. Symmetry also gives unit column sums, so $\pi_U$ is stationary.

    Finally, holding at $x$ for one step of $P$ is a one-step return to $U$, so $P^U(x,x)\ge P(x,x)=1-\textup{deg}(x)/(2d)\ge 1/2$.
\end{proof}

The next lemma ensures that the expectations used below are finite. From every vertex of $C_U$, the walk hits $U$ within a fixed number of steps with probability bounded away from zero, which gives a geometric tail. \Cref{lemma:Kac} later computes the average excursion cost exactly.

\begin{lemma}[Integrability of the trace times]
\label{lemma:integrability-of-trace-times}
    For every $z\in C_U$, the hitting time $T_z^U=\min\{s\ge 0:X_s\in U\}$ (with $X_0=z$) has finite expectation, i.e., $\E[T_z^U]<\infty$. Consequently, for every $x\in U$ and every integer $i\ge 1$, $\E[\tau_i]<\infty$.
\end{lemma}
\begin{proof}
    Every vertex of the finite connected graph $G[C_U]$ has a path to $U$ of length at most $|C_U|-1$. A prescribed edge step has probability $1/(2d)$, and a holding step has at least that probability. Thus, with $p=(2d)^{-|C_U|}$, uniformly over the starting vertex $z\in C_U$,
    \[
        \Pr[T_z^U\le |C_U|]\ge p.
    \]

    Applying the Markov property at the deterministic times $|C_U|,2|C_U|,\dots$, we obtain
    \[
        \Pr(T^U_z>k|C_U|)\le (1-p)^k
        \qquad
        \text{for every }z\in C_U\text{ and }k\ge0.
    \]

    Sum the tail probabilities in successive blocks of length $|C_U|$:
    \begin{align*}
        \mathbb{E}[T_z^U]
        &=\sum_{s\ge0}\Pr(T_z^U>s)\\
        &\le
        |C_U|\sum_{k\ge0}\Pr(T_z^U>k|C_U|)\\
        &\le
        |C_U|\sum_{k\ge0}(1-p)^k\\
        &
        \le\frac{|C_U|}{p}<\infty.
    \end{align*}

    It remains to consider the successive visit times. By the strong Markov property at time $\tau_i$, after one step from $X_{\tau_i}\in U$, the walk is at some state of $C_U$, and the additional time needed to hit $U$ is a hitting time of $U$ from that state. Thus, for every $i\ge0$,
    \[
        \E[\tau_{i+1}-\tau_i]\le 1 + \max_{z\in C_U}\E[T_z^U]\le 1+\frac{|C_U|}{p}<\infty.
    \]
    Summing over $0,\dots,i-1$ and using $\tau_0=0$ gives
    \[
        \mathbb{E}[\tau_i]
        \le
        i\left(1+\frac{|C_U|}{p}\right)
        <\infty,
    \]
    as desired.
\end{proof}

The next lemma shows that the trace chain inherits the expansion of $G[U]$, regardless of the edges leaving $U$, and hence mixes quickly on $U$.

\begin{lemma}[Inherited conductance and mixing]
\label{lemma:inherited-conductance-and-mixing}
    If $\Phi_\textup{in}(G[U])\ge \phi$, then $\Phi(P^U)\ge \frac{\phi}{2}$. Moreover, for every integer $t\ge 0$ and every vertex $x\in U$,
    \[
        \lVert e_x (P^U)^t-\pi_U\rVert_\textup{TV}\le \frac{1}{2}\sqrt{|U|}\cdot e^{-\phi^2t/8}.
    \]  
\end{lemma}
\begin{proof}
    Fix a nonempty set $S\subseteq U$ with $|S|\le |U|/2$. The trace transition $P^U(x,y)$ includes the direct one-step transition $P(x,y)$. Therefore,
    \[
        \sum_{x\in S,y\in U\setminus S}{\pi_U(x)\cdot P^U(x,y)}\ge \frac{1}{|U|}\sum_{x\in S,y\in U\setminus S}{P(x,y)}=\frac{|E(S,U\setminus S)|}{2d|U|}\ge \frac{d\phi|S|}{2d|U|}=\frac{\phi}{2}\pi_U(S).
    \]
    Dividing by $\pi_U(S)$ proves $\Phi(P^U)\ge\phi/2$.

    The trace chain is lazy by \Cref{lemma:trace-chain-symmetry-laziness}, so \Cref{lemma:mixing-time} yields, for every $x\in U$ and $t\ge0$,
    \[
        \lVert e_x (P^U)^t-\pi_U\rVert_\textup{TV}\le \frac{1}{2}\sqrt{|U|}\cdot e^{-(\phi/2)^2t/2}=\frac{1}{2}\sqrt{|U|}\cdot e^{-\phi^2t/8}.
    \]
\end{proof}

Fast trace mixing does not by itself bound the number of steps of $P$: a single excursion may last a long time. The next identity bounds the average number of steps of $P$ used by $t$ trace steps.

\begin{lemma}[Kac formula and trace cost]
\label{lemma:Kac}
    Let $c(x)=\E_x[\tau_1]$ be the expected number of steps of $P$ used by one trace transition. Then, $\sum_{x\in U}{c(x)}=|C_U|$. Consequently, 
    for every integer $t\ge 1$, 
    \[
        \sum_{x\in U}{(\E_x[\tau_t]-t)}=t(|C_U|-|U|).
    \]    
\end{lemma}
\begin{proof}
    For $y\in C_U\setminus U$, let $h(y)=\E[T_y^U]$, which is finite by \Cref{lemma:integrability-of-trace-times}. We first express the total time spent outside $U$ through first-step equations and then sum over the starting vertices.
        
    First-step conditioning gives, for $y\in C_U\setminus U$, 
    \begin{align}
    \label{eq:h(y)}
        h(y)=\E[T_y^U]=\sum_{z\in U}{P(y,z)\cdot 1} + \sum_{z\in C_U\setminus U}{P(y,z)\cdot (1+h(z))}=1+\sum_{z\in C_U\setminus U}{P(y,z)h(z)}.
    \end{align}
    Summing \eqref{eq:h(y)} over the exterior of $U$ within $C_U$ gives
    \[
        \sum_{y\in C_U\setminus U}{h(y)}=|C_U\setminus U| + \sum_{y\in C_U\setminus U}{\sum_{z\in C_U\setminus U}{P(y,z)h(z)}}.
    \]

    Thus,
    \begin{align}
    \label{eq:C-U}
        |C_U\setminus U| &= \sum_{y\in C_U\setminus U}{h(y)} - \sum_{y\in C_U\setminus U}{\sum_{z\in C_U\setminus U}{P(y,z)h(z)}}\notag\\
        &=\sum_{z\in C_U\setminus U}{\left(h(z)-\sum_{y\in C_U\setminus U}{P(y,z)h(z)}\right)}\notag\\
        &=\sum_{z\in C_U\setminus U}{h(z)\left(1-\sum_{y\in C_U\setminus U}{P(y,z)}\right)}\notag\\
        &=\sum_{z\in C_U\setminus U}{h(z)\sum_{x\in U}{P(z,x)}}.
    \end{align}

    For $x\in U$, we have
    \begin{align}
    \label{eq:c(x)}
        c(x)=\E_x[\tau_1]=\sum_{z\in U}{P(x,z)\cdot 1} + \sum_{z\in C_U\setminus U}{P(x,z) \cdot (1+h(z))}=1+\sum_{z\in C_U\setminus U}{P(x,z) h(z)}.
    \end{align}
    Sum \eqref{eq:c(x)} over $x\in U$ and use symmetry together with \eqref{eq:C-U}:
    \[
        \sum_{x\in U}{c(x)}=|U|+\sum_{x\in U}{\sum_{z\in C_U\setminus U}{P(x,z)h(z)}}=|U|+|C_U\setminus U|=|C_U|.
    \]

    This proves the one-step identity.

    For the $t$-step identity, start with $X_0\sim\pi_U$ and write $Y_i=X_{\tau_i}$. Stationarity of the trace chain makes every $Y_i$ uniform on $U$. The strong Markov property at each visit time gives
    \[
        \E[\tau_{i+1}-\tau_i|Y_i]=c(Y_i).
    \]

    Thus, 
    \[
        \E[\tau_{i+1}-\tau_i]=\E[\E[\tau_{i+1}-\tau_i|Y_i]]=\E[c(Y_i)]=\frac{1}{|U|}\sum_{x\in U}{c(x)}=\frac{|C_U|}{|U|}.
    \]

    Since $\tau_0=0$, the total elapsed time is the sum of the successive return times:
    \[
        \tau_t = (\tau_1-\tau_0) + (\tau_2-\tau_1) + \dots +(\tau_t-\tau_{t-1})=\sum_{i=0}^{t-1}{(\tau_{i+1}-\tau_i)}.
    \]

    Finally,
    \[
        \E[\tau_t] = \sum_{i=0}^{t-1}{\E[\tau_{i+1}-\tau_i]}=t\cdot \frac{|C_U|}{|U|}.
    \]

    Thus, 
    \[
    \sum_{x\in U}(\E_x[\tau_t]-t)
    =|U|\E[\tau_t]-t|U|
    =t(|C_U|-|U|).
    \]
\end{proof}

\subsubsection{From trace mixing to walks in the input graph}
\label{subsec:trace-transfer}

We convert mixing after a fixed number of visits to $U$ into a common-distribution guarantee for the random-length walks actually sampled by the tester. The difficulty is that $t$ trace steps use $\tau_t$ steps of $P$, and $\tau_t$ can depend strongly on the start.

Averaging over walk lengths resolves this. If a walk from $x$ is stopped at an integrable time $\tau$, we compare the averaging walk started at $x$ with the one started at $X_\tau$. The expected occupation vector records the visits at times $0,\ldots,\tau-1$, and applying $P$ shifts these times to $1,\ldots,\tau$. Their difference is the initial distribution minus the stopping distribution. Multiplication by $Q_\ell$ telescopes this identity, leaving an error at most $\E_x[\tau]/\ell$.

\begin{lemma}[Stopping time potential identity]
\label{lemma:stopping-time}

    Let $P$ be any finite Markov transition matrix, and let $(X_s)_{s\ge0}$ be the chain with transition matrix $P$ started at $X_0=x$. Let $\tau$ be a stopping time with $\E_x[\tau]<\infty$, and let $\nu$ be the distribution of $X_\tau$, that is, $\nu(y)=\Pr_x[X_\tau=y]$. For $Q_\ell=\frac{1}{\ell}\sum_{j=0}^{\ell-1}P^j$,
    \[
        \lVert e_xQ_\ell-\nu Q_\ell\rVert_\textup{TV}\le \frac{\E_x[\tau]}{\ell}.
    \]
\end{lemma}
\begin{proof}
    We consider the stopped path $(X_0,X_1,\dots,X_\tau)$. For $z\in V$, define $h(z)$ to be the expected number of times the chain visits $z$ strictly before $\tau$, i.e.,
    \[
        h(z)=\E_x\left[\sum_{s=0}^{\tau-1}{\mathbf{1}_{\{X_s=z\}}}\right].
    \]

    Every time before $\tau$ contributes one visit, so $\sum_{z}{\sum_{s=0}^{\tau-1}{\mathbf{1}_{\{X_s=z\}}}}=\tau$. Note that $h$ is nonnegative, thus we get that $\|h\|_1=\sum_{z}{h(z)}=\sum_{z}{\E_x[\sum_{s=0}^{\tau-1}{\mathbf{1}_{\{X_s=z\}}}]}=\E_x[\sum_{z}{\sum_{s=0}^{\tau-1}{\mathbf{1}_{\{X_s=z\}}}}]=\E_x[\tau]<\infty$. This integrability justifies the sums of expected visits below.

    Because $\{s<\tau\}$ is determined by the trajectory through time $s$, the Markov property applies inside each summand. Consequently,
    \begin{align*}
        (hP)(y)&=\sum_{z}{h(z)P(z,y)}\\
        &=\sum_{z}{\E_x\left[\sum_{s=0}^{\tau-1}{\mathbf{1}_{\{X_s=z\}}}\right]\cdot P(z,y)}\\
        &=\sum_{z}{\E_x\left[\sum_{s=0}^{\tau-1}{\mathbf{1}_{\{X_s=z\}}}\cdot P(z,y)\right]}\\
        &=\E_x\left[\sum_{s=0}^{\tau-1}{\sum_{z}{\mathbf{1}_{\{X_s=z\}}\cdot P(z,y)}}\right]\\
        &=\E_x\left[\sum_{s=0}^{\tau-1}{P(X_s,y)}\right]\\
        &=\E_x\left[\sum_{s=0}^{\tau-1}{\mathbf{1}_{\{X_{s+1}=y\}}}\right]\\
        &=\E_x\left[\sum_{s=1}^{\tau}{\mathbf{1}_{\{X_{s}=y\}}}\right].\\
    \end{align*}

    Therefore,
    \[
        h(y)-(hP)(y) = \E_x\left[\sum_{s=0}^{\tau-1}{\mathbf{1}_{\{X_s=y\}}}\right] - \E_x\left[\sum_{s=1}^{\tau}{\mathbf{1}_{\{X_{s}=y\}}}\right]=\mathbf{1}_{\{x=y\}}-\Pr_x[X_\tau=y].
    \]

    Thus, we have that
    \[
        h(I-P) = h-hP = e_x - \nu.
    \]

    Multiplying on the right by $Q_\ell$ telescopes the intermediate powers of $P$:
    \begin{align*}
        (e_x-\nu)Q_\ell = h(I-P)Q_\ell = h(I-P)\frac{1}{\ell}\sum_{j=0}^{\ell-1}{P^j}=\frac{1}{\ell} h\left(\sum_{j=0}^{\ell-1}{P^j}-\sum_{j=0}^{\ell-1}{P^{j+1}}\right)=\frac{1}{\ell}h(I-P^\ell).
    \end{align*}
    
    Recall that $\lVert h\rVert_1=\E_x[\tau]$. Since $h$ is nonnegative and $P^\ell$ is stochastic, we have that $\lVert hP^\ell\rVert_1=\lVert h\rVert_1$. The triangle inequality therefore gives
    \[
        \lVert h(I-P^\ell)\rVert_1\le \lVert h\rVert_1 + \lVert h\rVert_1=2\lVert h\rVert_1\le 2\E_x[\tau].
    \]

    Therefore, 
    \[
        \lVert (e_x-\nu)Q_\ell \rVert_1 = \frac{1}{\ell}\lVert h(I-P^\ell)\rVert_1\le \frac{2\E_x[\tau]}{\ell}.
    \]

    Since $\lVert p-q\rVert_\textup{TV}=\frac{1}{2}\lVert p-q\rVert_1$, we get that
    \[
        \lVert e_xQ_\ell - \nu Q_\ell\rVert_\textup{TV}\le \frac{\E_x[\tau]}{\ell}.
    \]
\end{proof}

\begin{remark}[Relation to stopping rules]
Stopping-time comparisons for averaging walks are developed in Lov\'asz and Winkler~\cite{lovasz1998mixing} and used in noisy-expander and robust-clustering analyses~\cite{kale2013noise,peng2020robust}. \Cref{lemma:stopping-time} is an elementary algebraic form of this existing technique. Its useful feature here is that the stopping distribution $\nu$ may be arbitrary: both $e_x$ and $\nu$ are propagated by the same averaging kernel. We do not replace $\nu Q_\ell$ by global uniformity.
\end{remark}

Apply the stopping-time estimate at the $t$-th trace visit. The trace endpoint is close to $\pi_U$, but we must first discard starts whose expected elapsed time is too large. Kac's identity bounds the sum of their excess costs by $t(1-\alpha)n$. Choosing the cutoff $c_{\mathrm{exc}}t$ spends only $\beta n/8$ of the size slack.

For every retained start, the stopping-time estimate and contraction then compare $q_x$ with the same distribution $\pi_UQ_\ell$. This distribution can put mass outside $U$ and need not be stationary. The tester uses only its existence, not its explicit form.

\begin{lemma}[Relative transfer with a slack-controlled exceptional set]
\label{lemma:transfer}
    Let $\phi\in(0,1]$. Let $\alpha\in(0,1]$, $\kappa\in[3/4,1)$, $n\ge4/((1-\kappa)\alpha)$, $\beta=(1-\kappa)\alpha$ and $c_\textup{exc}=1+\frac{8(1-\alpha)}{\beta}$. Assume that $U\subseteq V$ satisfies $|U|\ge \alpha n$ and $\Phi_\textup{in}(G[U])\ge \phi$. Let $t=\lceil \frac{8\log (8\sqrt{n})}{\phi^2} \rceil$ and $\ell=\lceil 16c_\textup{exc} t\rceil$. For all $x\in V$, let $q_x=e_xQ_{\ell}$. Then, there is a set $\mathcal{G}\subseteq U$ such that $|\mathcal{G}|\ge (\alpha-\frac{\beta}{8})n$ and for all $x\in \mathcal{G}$, 
    \[
        \lVert q_x-\pi_UQ_\ell\rVert_\textup{TV}\le 1/8.
    \]
    In particular, for all $x,y\in\mathcal{G}$,
    \[
        \lVert q_x-q_y\rVert_\textup{TV}\le \frac{1}{4}.
    \]
\end{lemma}
\begin{proof}
   If $\alpha <1$, by~\Cref{lemma:Kac}, we get that $\sum_{x\in U}{(\E_x[\tau_t]-t)}=t(|C_U|-|U|)\le t(1-\alpha)n$. The number of vertices in $U$ with $\E_x[\tau_t]>c_\textup{exc}t$ is at most $\beta n/8$. Otherwise, more than $\beta n/8$ vertices $x\in U$ satisfy $\E_x[\tau_t]-t>(c_\textup{exc}-1)t$. Since $\tau_t\ge t$, every term in the sum is nonnegative, and thus
    \[
        \sum_{x\in U}\bigl(\E_x[\tau_t]-t\bigr)
        > \frac{\beta n}{8}\cdot (c_\textup{exc}-1)\,t
        = \frac{\beta n}{8}\cdot \frac{8(1-\alpha)}{\beta}\,t
        = t(1-\alpha)n,
    \]
    which contradicts the bound from~\Cref{lemma:Kac}. Thus, if $\alpha<1$, we take $\mathcal{G}=\{x\in U: \E_x[\tau_t]\le c_\textup{exc}t\}$ and $\mathcal{G}$ satisfies that $|\mathcal{G}|\ge (\alpha-\frac{\beta}{8})n$.

    If $\alpha=1$, then $U=V$ and $\tau_t=t$ deterministically. In this case, we take $\mathcal{G}= U$ and $|\mathcal{G}|\ge (\alpha-\frac{\beta}{8})n$.

    For each $x\in\mathcal{G}$, let $\nu_x$ be the distribution of $X_{\tau_t}$ when the walk starts at $x$. Since $X_{\tau_t}$ is the position of the trace chain after $t$ steps, $\nu_x=e_x(P^U)^t$. By~\Cref{lemma:inherited-conductance-and-mixing} and our value of $t$, we get that 
    
    \[
        \lVert \nu_x-\pi_U\rVert_\textup{TV}=\lVert e_x(P^U)^t-\pi_U\rVert_\textup{TV}\le\frac{1}{2}\sqrt{|U|}\cdot e^{-\phi^2t/8} \le\frac{1}{16}.
    \]
    
    The stopping-time bound (\Cref{lemma:stopping-time}) and $\ell\ge16c_{\mathrm{exc}}t$ give
    \[
        \lVert e_xQ_\ell -\nu_xQ_\ell\rVert_\textup{TV}\le \frac{\E_x[\tau_t]}{\ell} \le\frac{c_\textup{exc}t}{\ell}\le \frac{1}{16}.
    \]

    Combining these estimates with contraction gives
    \begin{align*}
        \lVert q_x -\pi_UQ_\ell\rVert_\textup{TV}&=\lVert e_xQ_\ell -\pi_UQ_\ell\rVert_\textup{TV}\\
        &\le \lVert e_xQ_\ell -\nu_xQ_\ell \rVert_\textup{TV}+\lVert \nu_xQ_\ell -\pi_U Q_\ell \rVert_\textup{TV}\\
        &\le \frac{1}{16} + \lVert \nu_x -\pi_U  \rVert_\textup{TV}&&\text{(\Cref{lemma:contraction-of-total-variation})}\\
        &\le \frac{1}{16}+\frac{1}{16}=\frac{1}{8}.
    \end{align*}

    Finally, by triangle inequality, for all $x,y\in\mathcal{G}$,
    \[
        \lVert q_x-q_y\rVert_\textup{TV}\le \lVert q_x -\pi_UQ_\ell\rVert_\textup{TV} + \lVert \pi_UQ_\ell-q_y \rVert_\textup{TV}\le \frac{1}{4}.
    \]
\end{proof}

\subsubsection{A sparse partition and cross-part overlap}
\label{subsec:partition-flow}

We now turn to the soundness input. If $G\notin\mathcal{P}_{\alpha_0,\psi}$, then every induced subgraph on at least $\alpha_0n$ vertices has a cut of inner conductance below $\psi$. Splitting along such cuts recursively, until every part has fewer than $\alpha_0n$ vertices, gives a partition of $V$ with few edges between parts (\Cref{lemma:sparse-cut-partition}). The parts need not be expanders. We only use that they are small and that few edges join them.

We then relate this partition to the walks used by the tester (\Cref{lemma:overlap-bounded-by-cross-flow}). Recall that $\langle q_x,q_y\rangle$ is the probability that independent walks from $x$ and $y$ end at the same vertex. We show that this probability is small on average over pairs $x,y$ in different parts. The proof uses the symmetry of $Q_\ell$: a collision between walks from $x$ and $y$ corresponds to probability flow across the partition, which is small because few edges cross it. The bound holds on average rather than for every pair, and this is all that the soundness analysis in \Cref{lem:reusable-scores} requires.

\begin{lemma}[Sparse-cut partition]
\label{lemma:sparse-cut-partition}
    Suppose $G\notin \mathcal{P}_{\alpha_0,\psi}$, where $\alpha_0\in(0,1)$ and $\psi\in(0,1)$. Then there is a partition $V=C_1\cup C_2\cup\dots\cup C_s$ with
    \[
        |C_i|< \alpha_0 n \textup{ for every } i\in[s],\quad |F|<\psi dnL,\quad L=\left\lfloor\log_2(1/\alpha_0)\right\rfloor+1,
    \]
    where $F$ is the set of edges between distinct parts.
\end{lemma}
\begin{proof}
    Start with the single part $V$. Whenever a current part $C$ has size $|C|\ge \alpha_0 n$, the hypothesis implies $\Phi_\textup{in}(G[C])<\psi$. Therefore, there is a nonempty set $S\subseteq C$ with $|S|\le |C|/2$ and 
    \begin{align}
    \label{eq:no-case-E}
        |E(S,C\setminus S)|< \psi d|S|.
    \end{align}

    Replace $C$ by $S$ and $C\setminus S$. Continue until all parts have size below $\alpha_0 n$. We use $C_1,\dots,C_s$ to denote all these parts. Note that every edge in $F$ is cut at exactly one split. Therefore, summing~\cref{eq:no-case-E} over all splits gives
    \begin{align}
    \label{eq:no-case-F}
        |F|< \psi d \sum_{\textup{splits}}{|S_\textup{small}|},
    \end{align}
    where $S_{\mathrm{small}}$ denotes the smaller side of the split.

    We now bound the sum on the right-hand side of~\cref{eq:no-case-F}. At each split, assign one unit of charge to every vertex in $S_{\mathrm{small}}$. Thus, $\sum_{\text{splits}} |S_{\mathrm{small}}|$ is exactly the total number of charges assigned to all vertices. Consider a fixed vertex $x$. Whenever $x$ receives a charge, it lies in the smaller side of the current split. Consequently, the size of the part containing $x$ decreases by at least a factor of two. After $k$ charges, the size of this part is therefore at most $n/(2^k)$. Since the recursive procedure stops once every part has size less than $\alpha_0 n$, the vertex can receive at most
    \[
        L=\left\lfloor\log_2(1/\alpha_0)\right\rfloor+1
    \]
    charges. Since there are $n$ vertices and each receives at most $L$ charges,
    we have $\sum_{\text{splits}} {|S_{\mathrm{small}}|}
        \le nL$. Combining this with~\cref{eq:no-case-F} yields
    \[
        |F|<\psi d\,nL.
    \]
\end{proof}

The next lemma translates the edge bound into an overlap bound. Symmetry lets us reverse the two walks contributing to a common endpoint $z$. If their other endpoints lie in different parts, at least one reversed walk leaves the part of $z$. A change of part requires crossing an edge of $F$, whose stationary traversal probability is easy to count.

\begin{lemma}[Overlap is bounded by cross flow]
\label{lemma:overlap-bounded-by-cross-flow}
    For a symmetric stochastic matrix $Q$, write $Q(S,T)=\sum_{x\in S,y\in T}{Q(x,y)}$ and $q_x=e_xQ$. Then for every $A\subseteq V$,
    \[
        \sum_{x\in A,y\notin A}{\langle q_x,q_y\rangle}\le 2Q(A,V\setminus A).
    \]
     Let $C_1,\dots,C_s$ be a partition, and let $F$ be the set of edges between different parts. For the uniform averaging walk $Q_\ell=\frac{1}{\ell}\sum_{j=0}^{\ell-1}{P^j}$,
     \[
        \frac{1}{n^2}\sum_{\substack{x,y\in V\\ C(x)\ne C(y)}}{\langle q_x,q_y\rangle}\le \frac{\ell |F|}{dn^2},
     \]
     where $C(x)$ denotes the part containing $x$.
\end{lemma}
\begin{proof}
    For the first claim, symmetry and stochasticity give
     \begin{align*}
        \sum_{x\in A,y\notin A}{\langle q_x,q_y\rangle} &= \sum_{x\in A,y\notin A}{\sum_{z}{q_x(z)q_y(z)}}\\
        &=\sum_{x\in A,y\notin A}{\sum_{z}{Q(x,z)Q(y,z)}}\\
        &=\sum_{z}{\sum_{x\in A}{\sum_{y\notin A}{Q(z,x)Q(z,y)}}}\\
        &=\sum_{z}{\left(\sum_{x\in A}{Q(z,x)}\right)\left(\sum_{y\notin A}{Q(z,y)}\right)}\\
        &=\sum_z{Q(z,A)Q(z,V\setminus A)}\\
        &=\sum_{z\in A}{Q(z,A)Q(z,V\setminus A)} +\sum_{z\notin A}{Q(z,A)Q(z,V\setminus A)}\\
        &\le \sum_{z\in A}{Q(z,V\setminus A)} + \sum_{z\notin A}{Q(z,A)}\\
        &=Q(A,V\setminus A) + Q(V\setminus A,A)\\
        &=2Q(A, V\setminus A).
    \end{align*}

    Now we prove the second claim. Recall that $Q_\ell$ is symmetric and stochastic. For each $i\in[s]$, we apply $\sum_{x\in A,y\notin A}{\langle q_x,q_y\rangle}\le 2Q(A,V\setminus A)$ with $A=C_i$. Every ordered pair $(x,y)$ with $C(x)\ne C(y)$ appears in exactly one of these $s$ inequalities, namely the one for the unique $i$ with $x\in C_i$. Summing over $i\in[s]$, we get
    
    \begin{align}
    \label{eq:overlap-cross-flow}
        \sum_{\substack{x,y\in V\\ C(x)\ne C(y)}}\langle q_x,q_y\rangle\le 2\sum_{i=1}^{s}Q_\ell(C_i,V\setminus C_i).
    \end{align}

    We now bound the right-hand side of~\cref{eq:overlap-cross-flow}. Fix an integer $j\ge 0$. Let $X_0,X_1,\dots$ be the random walk with transition matrix $P$, where $X_0$ is uniform on $V$. Then
    \begin{align}
    \label{eq:Pj-cross}
        \sum_{i=1}^{s}P^j(C_i,V\setminus C_i)
        =\sum_{x\in V}P^j\bigl(x,V\setminus C(x)\bigr)
        =n\Pr\bigl[C(X_j)\ne C(X_0)\bigr].
    \end{align}
    If $C(X_j)\ne C(X_0)$, then the walk traverses an edge in $F$ at some step $k\in[j]$, that is, $\{X_{k-1},X_k\}\in F$. Since the uniform distribution is stationary for $P$, the vertex $X_{k-1}$ is uniform on $V$ for every $k$. Moreover, $P(u,v)=P(v,u)=\frac{1}{2d}$ for every edge $\{u,v\}$. Hence, the walk traverses a fixed edge $\{u,v\}\in F$ at step $k$ with probability $\frac{1}{n}P(u,v)+\frac{1}{n}P(v,u)=\frac{1}{dn}$, and it traverses some edge in $F$ at step $k$ with probability $\frac{|F|}{dn}$. By a union bound over $k\in[j]$,
    
    \[
        \Pr\bigl[C(X_j)\ne C(X_0)\bigr]\le\frac{j|F|}{dn}.
    \]
    Together with~\cref{eq:Pj-cross}, this gives $\sum_{i=1}^{s}P^j(C_i,V\setminus C_i)\le j|F|/d$. Therefore,
    \[
        \sum_{i=1}^{s}Q_\ell(C_i,V\setminus C_i)
        =\frac{1}{\ell}\sum_{j=0}^{\ell-1}\sum_{i=1}^{s}P^j(C_i,V\setminus C_i)
        \le\frac{1}{\ell}\sum_{j=0}^{\ell-1}\frac{j|F|}{d}
        =\frac{(\ell-1)|F|}{2d}.
    \]
    Plugging this into~\cref{eq:overlap-cross-flow}, we get
    \[
        \sum_{\substack{x,y\in V\\ C(x)\ne C(y)}}\langle q_x,q_y\rangle\le\frac{(\ell-1)|F|}{d}\le\frac{\ell|F|}{d}.
    \]
    Dividing both sides by $n^2$ completes the proof.
 \end{proof}

\begin{corollary}
\label{coro:overlap-bounded-by-cross-flow}
    Assume the hypotheses of \Cref{lemma:sparse-cut-partition}, and let $C(x)$ be the part of $x$ in the partition supplied by that lemma. With $L=\lfloor\log_2(1/\alpha_0)\rfloor+1$, we have
    \[
        \E_{x,y\sim \textup{Unif}(V)}[\mathbf{1}_{C(x)\ne C(y)} \langle q_x,q_y\rangle]<\frac{\ell\psi L}{n}.
    \]
\end{corollary}
\begin{proof}
    Substitute the cut-edge bound from \Cref{lemma:sparse-cut-partition} into the second inequality of \Cref{lemma:overlap-bounded-by-cross-flow}, and interpret the normalized sum as an expectation over independent uniform vertices.
\end{proof}

\subsubsection{The collision primitive}
\label{subsec:collision-bound}

We now analyze the collision test used by the tester.

\begin{definition}[Collision experiment]
\label{def:collision-experiment}
    Let $p,q$ be distributions on $V$, and let $m_A,m_D\ge 1$ be integers. The \emph{$(m_A,m_D)$-collision experiment} for $(p,q)$ draws a list $A$ of $m_A$ independent samples from $p$ and, independently, a list $D$ of $m_D$ independent samples from $q$. The experiment returns $1$ if $A\cap D\ne\emptyset$ and $0$ otherwise.
\end{definition}

In \Cref{algo:anchor-test}, the experiment is run with $p=q_x$ for an anchor $x$ and $q=q_y$ for a partner $y$. The anchor list $A$ is reused for all $k$ partners, so it is longer than each partner list $D$; up to rounding, $m_A/m_D=\omega$. The next lemma gives two guarantees for these list lengths. We use the first for completeness, when the anchor and the partner both lie in the good set $\mathcal{G}$ from \Cref{lemma:transfer}, and the second for soundness, when they lie in different parts of the sparse partition.

\begin{lemma}[High-confidence intersection from shared mass]
\label{lemma:collision-experiment}
    Let $p,q$ be distributions on $V$, and let $\Lambda\ge 1$ and $1\le\omega\le n/\Lambda$. Let $m_A=\lceil 32\sqrt{n\Lambda\omega}\,\rceil$ and $m_D=\lceil 32\sqrt{n\Lambda/\omega}\,\rceil$. Then the $(m_A,m_D)$-collision experiment for $(p,q)$ satisfies the following.
    \begin{enumerate}
        \item If $\lVert p-q\rVert_\textup{TV}\le 1/4$, then it returns $0$ with probability at most $e^{-\Lambda}$.
        \item For arbitrary $p$ and $q$, it returns $1$ with probability at most $1089\,n\Lambda\langle p,q\rangle$.
    \end{enumerate}
\end{lemma}

\paragraph{Why both list sizes matter.}
For the analysis only, replace the fixed list lengths by independent Poisson counts with means $\lambda_p=16\sqrt{n\Lambda\omega}$ and $\lambda_q=16\sqrt{n\Lambda/\omega}$. Counts at different vertices then become independent. Split the common mass $s_z=\min\{p(z),q(z)\}$ according to whether $\lambda_ps_z\ge1$. If the light vertices carry substantial common mass, the product $\lambda_p\lambda_q=256n\Lambda$ gives a birthday-paradox bound. If the heavy vertices carry substantial common mass, the shorter list must itself have $\Omega(\Lambda)$ samples; this is ensured by $\omega\le n/\Lambda$. Finally, couple the Poisson lists to prefixes of the fixed-size lists. This last step restores a deterministic sample budget. The second claim follows separately from a union bound over pairs of samples.

\begin{proof}[Proof of~\Cref{lemma:collision-experiment}]
    Write $A=(a_1,\dots,a_{m_A})$ and $D=(d_1,\dots,d_{m_D})$ for the two lists drawn in the experiment of~\Cref{def:collision-experiment}. We view $A$ and $D$ as sets when we write $A\cap D$. The experiment returns $1$ if and only if $A\cap D\ne\emptyset$.

    Since $\omega\le n/\Lambda$, we have $n\Lambda/\omega\ge\Lambda^2\ge 1$. Since $\omega\ge 1$, we also have $n\Lambda\omega\ge n\Lambda/\omega$. Hence
    \begin{align}
    \label{eq:coll-sqrt}
        \sqrt{n\Lambda\omega}\ \ge\ \sqrt{n\Lambda/\omega}\ \ge\ \Lambda\ \ge\ 1.
    \end{align}
    For every real number $r\ge 1$, we have $\lceil 32r\rceil\le 32r+1\le 33r$. Together with~\cref{eq:coll-sqrt}, this gives
    \begin{align}
    \label{eq:coll-m-upper}
        m_A\le 33\sqrt{n\Lambda\omega}
        \quad\text{and}\quad
        m_D\le 33\sqrt{n\Lambda/\omega}\le 33\sqrt{n\Lambda\omega}.
    \end{align}
    In particular, the experiment draws $m_A+m_D\le 66\sqrt{n\Lambda\omega}$ samples.

    We first prove item 1.
    
    Assume that $\lVert p-q\rVert_\textup{TV}\le 1/4$. For every $z\in V$, let $s_z=\min\{p(z),q(z)\}$. Since $p(z)-s_z=\max\{p(z)-q(z),0\}$, we have
    \[
        \sum_{z\in V}s_z=1-\sum_{z\in V}\max\{p(z)-q(z),0\}=1-\lVert p-q\rVert_\textup{TV}\ge\frac34.
    \]
    We now show that $\Pr[A\cap D=\emptyset]\le e^{-\Lambda}$.

    Let $\lambda_p=16\sqrt{n\Lambda\omega}$ and $\lambda_q=16\sqrt{n\Lambda/\omega}$. 
    
    By~\cref{eq:coll-sqrt}, we have
    \begin{align}
    \label{eq:coll-lambda}
        \lambda_p\ge\lambda_q\ge 16\Lambda
        \quad\text{and}\quad
        \lambda_p\lambda_q=256\,n\Lambda.
    \end{align}
    Moreover, $m_A\ge 32\sqrt{n\omega\Lambda}=2\lambda_p$ and $m_D\ge 32\sqrt{n\Lambda/\omega}=2\lambda_q$.

    Let $a_1,a_2,\dots$ be an infinite sequence of independent samples from $p$, and let $d_1,d_2,\dots$ be an infinite sequence of independent samples from $q$. Let $N_A\sim\mathrm{Poi}(\lambda_p)$ and $N_D\sim\mathrm{Poi}(\lambda_q)$. All these random variables are independent. The first $m_A$ elements of the first sequence and the first $m_D$ elements of the second sequence have the same joint distribution as the lists $A$ and $D$ in the experiment. So we may take $A=(a_1,\dots,a_{m_A})$ and $D=(d_1,\dots,d_{m_D})$. We also define the Poisson lists
    \[
        A'=(a_1,\dots,a_{N_A})
        \quad\text{and}\quad
        D'=(d_1,\dots,d_{N_D}).
    \]
    If $N_A\le m_A$ and $N_D\le m_D$, then $A'$ is a prefix of $A$ and $D'$ is a prefix of $D$. In this case, $A\cap D=\emptyset$ implies $A'\cap D'=\emptyset$. Hence
    \begin{align}
    \label{eq:coll-coupling}
        \Pr[A\cap D=\emptyset]\le\Pr[A'\cap D'=\emptyset]+\Pr[N_A>m_A]+\Pr[N_D>m_D].
    \end{align}

    For $z\in V$, let $N_A(z)$ be the number of indices $i\le N_A$ with $a_i=z$, and define $N_D(z)$ in the same way. By the standard Poisson splitting property \cite{mitzenmacher2005probability}, the variables $N_A(z)$, $z\in V$, are independent and $N_A(z)\sim\mathrm{Poi}(\lambda_p p(z))$. Similarly, the variables $N_D(z)$, $z\in V$, are independent and $N_D(z)\sim\mathrm{Poi}(\lambda_q q(z))$, and the two families are independent of each other.

    The lists $A'$ and $D'$ are disjoint if and only if, for every $z\in V$, we have $N_A(z)=0$ or $N_D(z)=0$. By the independence above,
    \[
        \Pr[A'\cap D'=\emptyset]
        =\prod_{z\in V}\Bigl(1-\bigl(1-e^{-\lambda_p p(z)}\bigr)\bigl(1-e^{-\lambda_q q(z)}\bigr)\Bigr).
    \]
    Each factor is decreasing in $p(z)$ and in $q(z)$. Since $p(z)\ge s_z$ and $q(z)\ge s_z$, we can replace both $p(z)$ and $q(z)$ by $s_z$ and get an upper bound. Let $u_z=\lambda_p s_z$ and $v_z=\lambda_q s_z$. Then $u_z\ge v_z\ge 0$ by~\cref{eq:coll-lambda}, and
    \begin{align}
    \label{eq:coll-product}
        \Pr[A'\cap D'=\emptyset]\le\prod_{z\in V}h(u_z,v_z),
        \quad\text{where } h(u,v)=1-(1-e^{-u})(1-e^{-v}).
    \end{align}
    For all $u,v\ge 0$, we have $e^{-v}\le h(u,v)\le 1$. So $-\log h(u,v)$ is well defined and nonnegative.
    
    We claim that for all $u\ge v\ge 0$,
    \begin{align}
    \label{eq:coll-key}
        -\log h(u,v)\ge\frac14\,v\min\{u,1\}.
    \end{align}
    We first note that for every $x\ge 0$,
    \begin{align}
    \label{eq:coll-exp}
        1-e^{-x}\ge\frac12\min\{x,1\}.
    \end{align}
    Indeed, the function $x\mapsto 1-e^{-x}$ is concave and equals $0$ at $x=0$. So for $0\le x\le 1$, we have $1-e^{-x}\ge(1-e^{-1})x\ge x/2$. For $x\ge 1$, we have $1-e^{-x}\ge 1-e^{-1}\ge 1/2$.
    
    Now we prove~\cref{eq:coll-key}. We consider two cases.
    \begin{itemize}
        \item Case $0\le v\le 1$. Let $x=(1-e^{-u})(1-e^{-v})$. Then $0\le x<1$. Using $-\log(1-x)\ge x$ and applying~\cref{eq:coll-exp} twice, we get
        \[
            -\log h(u,v)=-\log(1-x)\ge x\ge\frac12\min\{u,1\}\cdot\frac12\min\{v,1\}=\frac14\,v\min\{u,1\},
        \]
        where the last step uses $v\le 1$.
        \item Case $v>1$. Then $u\ge v>1$, so $\min\{u,1\}=1$. Since $h(u,v)=e^{-u}+e^{-v}-e^{-u-v}\le e^{-u}+e^{-v}\le 2e^{-v}$, we get
        \[
            -\log h(u,v)\ge v-\log 2\ge\frac{v}{4}=\frac14\,v\min\{u,1\},
        \]
        where the second inequality holds because $\frac34 v>\frac34>\log 2$.
    \end{itemize}
    
    By~\cref{eq:coll-product} and~\cref{eq:coll-key},
    \begin{align}
    \label{eq:coll-sum}
        -\log\Pr[A'\cap D'=\emptyset]
        \ge\sum_{z\in V}-\log h(u_z,v_z)
        \ge\frac14\sum_{z\in V}v_z\min\{u_z,1\}.
    \end{align}
    Let $\mathcal{Z}=\{z\in V:u_z<1\}$ be the set of light points; the other vertices are heavy. Since $\sum_{z\in V}s_z\ge 3/4$, we have $\sum_{z\notin\mathcal{Z}}s_z\ge 3/8$ or $\sum_{z\in\mathcal{Z}}s_z\ge 3/8$.
    \begin{itemize}
        \item Suppose that the heavy vertices carry common mass at least $3/8$, that is, $\sum_{z\notin\mathcal{Z}}s_z\ge 3/8$. For every $z\notin\mathcal{Z}$, we have $\min\{u_z,1\}=1$. Since every term in~\cref{eq:coll-sum} is nonnegative, we get
        \[
            -\log\Pr[A'\cap D'=\emptyset]
            \ge\frac14\sum_{z\notin\mathcal{Z}}v_z
            =\frac{\lambda_q}{4}\sum_{z\notin\mathcal{Z}}s_z
            \ge\frac{3\lambda_q}{32}
            \ge \frac{3}{2}\Lambda,
        \]
        where the last step uses $\lambda_q\ge 16\Lambda$.
        
        \item Suppose that the light vertices carry common mass at least $3/8$, that is, $\sum_{z\in\mathcal{Z}}s_z\ge 3/8$. In particular, $\mathcal{Z}\ne\emptyset$. For every $z\in\mathcal{Z}$, we have $\min\{u_z,1\}=u_z$. By the Cauchy--Schwarz inequality and $|\mathcal{Z}|\le n$,
        \[
            \sum_{z\in\mathcal{Z}}s_z^2\ge\frac{1}{|\mathcal{Z}|}\Bigl(\sum_{z\in\mathcal{Z}}s_z\Bigr)^2\ge\frac{9}{64n}.
        \]
        Hence, using $\lambda_p\lambda_q=256\,n\Lambda$, we get
        \[
            -\log\Pr[A'\cap D'=\emptyset]
            \ge\frac14\sum_{z\in\mathcal{Z}}u_zv_z
            =\frac{\lambda_p\lambda_q}{4}\sum_{z\in\mathcal{Z}}s_z^2
            \ge\frac{9\lambda_p\lambda_q}{256\,n}
            =9\Lambda
            \ge \frac{3}{2}\Lambda.
        \]
    \end{itemize}
    In both cases,
    \begin{align}
    \label{eq:coll-poisson-disjoint}
        \Pr[A'\cap D'=\emptyset]\le e^{-3\Lambda/2}.
    \end{align}
    
    Let $Y\sim\mathrm{Poi}(\lambda)$ with $\lambda>0$. Then
    \[
        \E\bigl[2^Y\bigr]=\sum_{k\ge 0}e^{-\lambda}\frac{\lambda^k 2^k}{k!}=e^{\lambda}.
    \]
    By Markov's inequality,
    \[
        \Pr[Y\ge 2\lambda]=\Pr\bigl[2^Y\ge 4^\lambda\bigr]\le\frac{e^\lambda}{4^\lambda}=e^{-(\log 4-1)\lambda}\le e^{-\lambda/4},
    \]
    where the last step uses $\log 4-1\ge 1/4$. Since $m_A\ge 2\lambda_p$ and $\lambda_p\ge 16\Lambda$, we get
    
    \[
        \Pr[N_A>m_A]\le\Pr[N_A\ge 2\lambda_p]\le e^{-\lambda_p/4}\le e^{-4\Lambda}.
    \]
    In the same way, $\Pr[N_D>m_D]\le e^{-4\Lambda}$.
    
    By~\cref{eq:coll-coupling},~\cref{eq:coll-poisson-disjoint}, $\Pr[N_A>m_A]\le e^{-4\Lambda}$ and $\Pr[N_D>m_D]\le e^{-4\Lambda}$,
    
    \[
        \Pr[A\cap D=\emptyset]\le e^{-3\Lambda/2}+2e^{-4\Lambda}
    =e^{-\Lambda}\bigl(e^{-\Lambda/2}+2e^{-3\Lambda}\bigr)
    \le e^{-\Lambda}\bigl(e^{-1/2}+2e^{-3}\bigr)\le e^{-\Lambda},
    \]
    
    where the last step uses $e^{-1/2}+2e^{-3}<0.71$. This proves item $1$.
            
    We now prove item 2. For arbitrary $p$ and $q$, by a union bound over all pairs $(i,j)\in[m_A]\times[m_D]$,
    \[
        \Pr[A\cap D\ne\emptyset]\le\sum_{i=1}^{m_A}\sum_{j=1}^{m_D}\Pr[a_i=d_j].
    \]
    Since $a_i\sim p$ and $d_j\sim q$ are independent, we have $\Pr[a_i=d_j]=\sum_{z\in V}p(z)q(z)=\langle p,q\rangle$. Therefore, by~\cref{eq:coll-m-upper},
    \[
        \Pr[A\cap D\ne\emptyset]
        \le m_Am_D\langle p,q\rangle
        \le 33^2\sqrt{n\omega\Lambda}\cdot\sqrt{n\Lambda/\omega}\cdot\langle p,q\rangle
        =1089\,n\Lambda\langle p,q\rangle.
    \]
\end{proof}

\subsubsection{Proof of \Cref{thrm:upper-bound-structural-formal}}
\label{subsec:structural-proof}

In each iteration of the outer loop of~\Cref{algo:anchor-test}, the tester compares the list $A$ of an anchor $x$ with the lists of $k$ partners. Fix a partner $y$. Then $A$ and $D$ are independent lists of $m_A$ samples from $q_x$ and $m_D$ samples from $q_y$, respectively. So each comparison is exactly one run of the $(m_A,m_D)$-collision experiment for $(q_x,q_y)$ (\Cref{def:collision-experiment}), and \Cref{lemma:collision-experiment} applies to it. The comparisons with different partners, however, are not independent, since they share the list $A$. To simplify the analysis, we condition on $A$ and introduce the following notions.

We call the pair $(x,A)$ an \emph{anchor state}. For a partner $y$ with list $D$, let
\[
    Z(x,y;A)=\mathbf{1}_{\{A\cap D\ne\emptyset\}}.
\]
The \emph{conditional score} of $(x,A)$ is
\begin{equation}
\label{eq:score}
    f(x,A)=\Pr_{y,D}\bigl[Z(x,y;A)=1\bigm| x,A\bigr],\qquad y\sim\textup{Unif}(V).
\end{equation}
In other words, $f(x,A)$ is the probability that a comparison with a fresh partner succeeds, once the anchor state is fixed. Conditioned on $(x,A)$, the comparisons with the $k$ partners are independent Bernoulli trials with mean $f(x,A)$. For an anchor, let $N_\textup{collision}$ be the number of its partners whose lists intersect $A$, and let $\widehat f=N_\textup{collision}/k$. We say that the anchor is \emph{flagged} if $\widehat f\ge\alpha-\beta/2$.

The proof has three steps. First, we show that conditional scores separate the two cases: in the YES case many anchor states have a high score, and in the NO case few do (\Cref{lem:reusable-scores}). Second, since the comparisons with the $k$ partners are independent given the anchor state, $\widehat f$ estimates $f(x,A)$, so an anchor is flagged with high probability if its score is high and with low probability if it is low (\Cref{lemma:score-estimation}). Third, the $R$ anchors are independent, and the tester estimates the fraction of flagged anchors.

We begin with the first step. We bound the scores of anchor states rather than of anchors, since even a good anchor may draw an atypical list $A$. In both the YES and the NO case, we first bound the average of a suitable quantity over $A$ and then apply Markov's inequality.

\begin{lemma}[Separation of conditional scores]
\label{lem:reusable-scores}
    Let $\alpha\in(0,1]$, $\phi\in(0,1]$, $\kappa\in[3/4,1)$, $n\ge 4/((1-\kappa)\alpha)$ and $\beta=(1-\kappa)\alpha$. Let $\Lambda,\ell,\omega,m_A$ and $m_D$ be the parameters set in~\Cref{algo:anchor-test}, let $L=\lfloor\log_2\frac{1}{\alpha-\beta}\rfloor+1$, and let $\psi=\frac{\alpha\beta}{64\times 1089 \Lambda \ell L}$. Let $x$ be a uniform vertex, and, given $x$, let $A$ be a list of $m_A$ independent samples from $q_x$.
    \begin{enumerate}
        \item If $G\in\mathcal{P}_{\alpha,\phi}$, then $\Pr\bigl[f(x,A)\ge\alpha-\frac{\beta}{4}\bigr]\ge\frac{15}{16}\bigl(\alpha-\frac{\beta}{8}\bigr)$.
        \item If $G\notin\mathcal{P}_{\alpha-\beta,\psi}$, then $\Pr\bigl[f(x,A)>\alpha-\frac{3\beta}{4}\bigr]\le\alpha/16$.
    \end{enumerate}
\end{lemma}
\begin{proof}  
    We first check the conditions of the lemmas we use. Since $\kappa\ge3/4$ and $\alpha\le1$, we have $(1-\kappa)\alpha\le1/4$. Together with $n\ge4/((1-\kappa)\alpha)$, this gives $n\ge16$ and $1<\log512\le\Lambda\le\log(32n)\le n$. Hence $\Lambda\ge1$ and $1\le\omega\le n/\Lambda$, so~\Cref{lemma:collision-experiment} applies. Moreover, $\alpha n\ge4(1-\kappa)\alpha n\ge16>2$, so~\Cref{lemma:transfer} applies.
    
    \textit{Proof of item 1.} Let $G\in\mathcal P_{\alpha,\phi}$. Let $U$ be a core for $G\in\mathcal{P}_{\alpha,\phi}$, that is, $|U|\ge\alpha n$ and $\Phi_\textup{in}(G[U])\ge\phi$. Let $\mathcal{G}\subseteq U$ be the set given by \Cref{lemma:transfer}. Then $|\mathcal{G}|\ge(\alpha-\frac{\beta}{8})n$, and for all $x,y\in\mathcal{G}$, $\lVert q_x-q_y\rVert_\textup{TV}\le 1/4$.

    Fix $x\in\mathcal{G}$. For every $y\in\mathcal{G}$, item $1$ of \Cref{lemma:collision-experiment} gives
    \[
        \Pr_{A,D}[A\cap D=\emptyset]\le e^{-\Lambda}=\frac{\beta}{128}.
    \]
    
    We define the \emph{lost good-partner mass} of the anchor state $(x,A)$ as
    \[
        \textup{loss}(x,A)=\Pr_{y,D}\bigl[y\in\mathcal{G}\text{ and }A\cap D=\emptyset\bigm| x,A\bigr].
    \]
    
    Averaging over $A$ and using the bound above for each $y\in\mathcal{G}$, we get
    \[
        \E_A\bigl[\textup{loss}(x,A)\bigr]
        =\frac{1}{n}\sum_{y\in\mathcal{G}}\Pr_{A,D}[A\cap D=\emptyset]
        \le\frac{\beta}{128}.
    \]
    
    By Markov's inequality, $\textup{loss}(x,A)\le\beta/8$ with probability at least $15/16$ over $A$. For every such $A$,
    \[
        f(x,A)
        \ge\Pr_{y,D}\bigl[y\in\mathcal{G}\text{ and }A\cap D\ne\emptyset\bigm| x,A\bigr]
        =\frac{|\mathcal{G}|}{n}-\textup{loss}(x,A)
        \ge\Bigl(\alpha-\frac{\beta}{8}\Bigr)-\frac{\beta}{8}
        =\alpha-\frac{\beta}{4}.
    \]
    Since $x$ is uniform, we have $x\in\mathcal{G}$ with probability at least $\alpha-\frac{\beta}{8}$. Together with the bound above, this proves item $1$.

    \textit{Proof of item 2.} Let $G\notin \mathcal{P}_{\alpha-\beta,\psi}$. Let $\alpha_0=\alpha-\beta$ and $L=\lfloor\log_2(1/\alpha_0)\rfloor+1$. Since $\kappa\ge3/4$, we have $\alpha_0=\alpha-\beta=\kappa\alpha\ge3(1-\kappa)\alpha>0$, and thus $\alpha_0n\ge3(1-\kappa)\alpha n\ge12>1$. By \Cref{lemma:sparse-cut-partition}, there is a partition $V=C_1\cup\dots\cup C_s$ such that $|C_i|<\alpha_0 n$ for every $i\in[s]$, and the set $F$ of edges between different parts satisfies $|F|<\psi dnL$. Let $C(x)$ denote the part containing $x$.

    For an anchor state $(x,A)$, we define its \emph{cross-part score} as
    \[
        b_\times(x,A)=\Pr_{y,D}\bigl[C(y)\ne C(x)\text{ and }A\cap D\ne\emptyset\bigm| x,A\bigr].
    \]
    
    A partner in $C(x)$ is chosen with probability $|C(x)|/n<\alpha-\beta$. Hence
    \begin{equation}
    \label{eq:score-cross}
        f(x,A)\le\frac{|C(x)|}{n}+b_\times(x,A)<\alpha-\beta+b_\times(x,A).
    \end{equation}
    We now bound the average of $b_\times$. By item $2$ of \Cref{lemma:collision-experiment}, for all $x,y\in V$,
    \[
        \Pr_{A,D}[A\cap D\ne\emptyset]\le 1089\,n\Lambda\langle q_x,q_y\rangle.
    \]
    Averaging over $x$ and $A$, we get
    \[
        \E_{x,A}\bigl[b_\times(x,A)\bigr]
        =\frac{1}{n^2}\sum_{\substack{x,y\in V\\ C(x)\ne C(y)}}\Pr_{A,D}[A\cap D\ne\emptyset]
        \le 1089\,n\Lambda\cdot\frac{1}{n^2}\sum_{\substack{x,y\in V\\ C(x)\ne C(y)}}\langle q_x,q_y\rangle.
    \]
    By \Cref{lemma:overlap-bounded-by-cross-flow} and $|F|<\psi dnL$, the last average is at most $\ell|F|/(dn^2)<\psi\ell L/n$. By the choice of $\psi$:
    \begin{align}
        \label{eq:psi}
        \psi=\frac{\alpha\beta}{64\times 1089 \Lambda \ell L},
    \end{align}
    
    we have $1089\,\Lambda\ell L\psi\le\alpha\beta/64$. Therefore,
    \[
        \E_{x,A}\bigl[b_\times(x,A)\bigr]<1089\,\Lambda\ell L\psi\le\frac{\alpha\beta}{64}.
    \]
    By Markov's inequality, $b_\times(x,A)>\beta/4$ with probability at most $\alpha/16$. For every other anchor state, \cref{eq:score-cross} gives $f(x,A)<\alpha-\beta+\beta/4=\alpha-\frac{3\beta}{4}$. This ends the proof.
\end{proof}

Conditional on an anchor state, fresh partners estimate its score. The YES and NO thresholds in \Cref{lem:reusable-scores} lie $\beta/4$ on either side of the flag threshold $\alpha-\beta/2$. The chosen $k$ makes the estimation error smaller than this margin with the probability required for the outer loop.

\begin{lemma}[Estimating a conditional score]
\label{lemma:score-estimation}
    Let $\alpha\in(0,1]$, $\kappa\in[3/4,1)$ and $\beta=(1-\kappa)\alpha$, and let $k$ be as in~\Cref{algo:anchor-test}. Fix an anchor state $(x,A)$, and let $\widehat f=N_\textup{collision}/k$ be the fraction of its $k$ partners whose lists intersect $A$.
    \begin{enumerate}
        \item If $f(x,A)\ge\alpha-\frac{\beta}{4}$, then the anchor is flagged with probability at least $1-\alpha/64$.
        \item If $f(x,A)\le\alpha-\frac{3\beta}{4}$, then the anchor is flagged with probability at most $\alpha/64$.
    \end{enumerate}
\end{lemma}

To prove~\Cref{lemma:score-estimation}, we need the following Chernoff bound for Bernoulli variables, whose proof is deferred to~\Cref{subsec:proof-of-Bernoulli-tails}.

\begin{lemma}[Bernoulli tails]
\label{lemma:Bernoulli-tails}
    Let $S$ be a sum of $k$ independent Bernoulli trials of common mean $p$. For $0<a\le1$,
    \[
        \Pr[S\le(1-a)kp]\le e^{-a^2kp/2},\qquad \Pr[S\ge(1+a)kp]\le e^{-a^2kp/3}.
    \]
\end{lemma}

\begin{proof}[Proof of~\Cref{lemma:score-estimation}]
    Conditioned on $(x,A)$, the count $N_\textup{collision}$ is a sum of $k$ independent Bernoulli variables with mean $f(x,A)$, that is, $N_\textup{collision}\sim\mathrm{Bin}(k,f(x,A))$. For every $h$, the probability $\Pr[\mathrm{Bin}(k,p)\ge h]$ is nondecreasing in $p$. To see this, write each Bernoulli variable with mean $p$ as $\mathbf{1}_{\{Y\le p\}}$, where $Y$ is uniform on $[0,1]$. So it suffices to consider the boundary means
    \[
        p_+=\alpha-\frac{\beta}{4}
        \quad\text{and}\quad
        p_-=\alpha-\frac{3\beta}{4}.
    \]
    Let $u=\beta/4$. The flag threshold is $\alpha-\frac{\beta}{2}=p_+-u=p_-+u$. Since $\kappa\ge3/4$ and $\beta=(1-\kappa)\alpha\le \alpha/4$, we have $0<u\le p_-\le p_+\le\alpha$. Recall that $k\ge\frac{64\alpha}{\beta^2}\log\frac{128}{\alpha}$.

    For item $1$, let $X_+\sim\mathrm{Bin}(k,p_+)$. By the Chernoff bound (\Cref{lemma:Bernoulli-tails}),
    \[
        \Pr\bigl[X_+<(p_+-u)k\bigr]
        \le\exp\Bigl(-\frac{ku^2}{2p_+}\Bigr)
        \le\exp\Bigl(-\frac{k\beta^2}{32\alpha}\Bigr)
        \le\Bigl(\frac{\alpha}{128}\Bigr)^{2}
        \le\frac{\alpha}{64}.
    \]
    For item $2$, let $X_-\sim\mathrm{Bin}(k,p_-)$. Since $u\le p_-$, the Chernoff bound gives
    \[
        \Pr\bigl[X_-\ge(p_-+u)k\bigr]
        \le\exp\Bigl(-\frac{ku^2}{3p_-}\Bigr)
        \le\exp\Bigl(-\frac{k\beta^2}{48\alpha}\Bigr)
        \le\Bigl(\frac{\alpha}{128}\Bigr)^{4/3}
        \le\frac{\alpha}{64}.
        \qedhere
    \]
\end{proof}

It remains to analyze the outer loop and the cost. The following proof establishes \Cref{thrm:upper-bound-structural-formal}; the monotonicity reduction at the start of the section then gives \Cref{thrm:upper-bound-structural-informal}.

\begin{proof}[Proof of \Cref{thrm:upper-bound-structural-formal}]\leavevmode
    \textbf{Correctness.}
    The outer-loop iterations use independent anchors, endpoint lists, and partner randomness. Their flags are therefore independent Bernoulli variables with a common probability $\pi_{\mathrm{flag}}$, so $N_{\mathrm{flag}}\sim\mathrm{Bin}(R,\pi_{\mathrm{flag}})$. Monotonicity of binomial tails allows us to use a lower bound on $\pi_{\mathrm{flag}}$ for completeness and an upper bound for soundness.

    In the YES case, item $1$ of \Cref{lem:reusable-scores} and item $1$ of \Cref{lemma:score-estimation} give
    \[
        \pi_\textup{flag}
        \ge\frac{15}{16}\Bigl(\alpha-\frac{\beta}{8}\Bigr)\Bigl(1-\frac{\alpha}{64}\Bigr)
        \ge\frac{15}{16}\cdot\frac{31\alpha}{32}\cdot\frac{63}{64}
        \ge\frac{3\alpha}{4},
    \]
    where the second inequality uses $\beta\le\alpha/4$, which follows from $\kappa\ge3/4$, and $\alpha\le1$. The tester rejects only if $N_\textup{flag}<\alpha R/2$, and $\alpha R/2$ is two thirds of $3\alpha R/4$. By the lower Chernoff bound with mean $3\alpha R/4$,
    \[
        \Pr[\text{reject}]
        \le\Pr\Bigl[\mathrm{Bin}\Bigl(R,\frac{3\alpha}{4}\Bigr)<\frac23\cdot\frac{3\alpha R}{4}\Bigr]
        \le\exp\Bigl(-\frac{\alpha R}{24}\Bigr)
        \le\delta,
    \]
    where the last step uses $R\ge\frac{48}{\alpha}\log\frac{2}{\delta}$.
    
    In the NO case, item $2$ of \Cref{lem:reusable-scores} and item $2$ of \Cref{lemma:score-estimation} give
    \[
        \pi_\textup{flag}
        \le\Pr\Bigl[f(x,A)>\alpha-\frac{3\beta}{4}\Bigr]+\frac{\alpha}{64}
        \le\frac{\alpha}{16}+\frac{\alpha}{64}
        \le\frac{\alpha}{8}
        \le\frac{\alpha}{4}.
    \]
    The tester accepts only if $N_\textup{flag}\ge\alpha R/2$, which is twice $\alpha R/4$. By the upper Chernoff bound with mean $\alpha R/4$,
    \[
        \Pr[\text{accept}]
        \le\Pr\Bigl[\mathrm{Bin}\Bigl(R,\frac{\alpha}{4}\Bigr)\ge 2\cdot\frac{\alpha R}{4}\Bigr]
        \le\exp\Bigl(-\frac{\alpha R}{12}\Bigr)
        \le\delta.
    \]
    
    The proofs of \Cref{lem:reusable-scores,lemma:score-estimation} use only $1\le\omega\le n/\Lambda$, so they hold for both values $\omega=k$ and $\omega=n/\Lambda$. On the full-table branch, the tester simulates every random walk exactly from the adjacency table. This does not change the distribution of any sample, so the argument above also holds on this branch.
    
    \paragraph{Query complexity and $\psi$.} If the full-table branch is taken, the tester makes $dn$ queries. This branch is taken either because $66R\ell\sqrt{n\Lambda k}>dn$, or because $k>n/\Lambda$, in which case $R\ell\sqrt{n\Lambda k}\ge\sqrt{n\Lambda k}>n$. In both cases, $dn=O(R\ell\sqrt{n\Lambda k})$, since $d$ is a constant.
    
    Otherwise, $k\le n/\Lambda$ and $66R\ell\sqrt{n\Lambda k}\le dn$. Then $\omega=k$, and by \Cref{lemma:collision-experiment} each anchor draws
    \[
        m_A+km_D\le 33\sqrt{n\Lambda k}+33k\sqrt{n\Lambda/k}=66\sqrt{n\Lambda k}
    \]
    samples, each costing at most $\ell$ queries. So the tester makes at most $66R\ell\sqrt{n\Lambda k}\le dn$ queries.

    Combining the branches gives the minimum of the full-table cost and $O(R\ell\sqrt{n\Lambda k})$. Substituting the parameter values proves the stated query bound. Substituting $\Lambda,\ell,L$ into \eqref{eq:psi} gives the claimed conductance multiplier $\mu_n=\psi/\phi^2$.

    \paragraph{Running time.} Suppose that the full-table branch is not taken. Then $\omega=k$, and the tester draws $R(m_A+km_D)$ samples. Each sample is a walk of fewer than $\ell$ steps, and each step takes $O(1)$ time, so sampling takes $O(R\ell(m_A+km_D))$ time. Since $m_A+km_D\le66\sqrt{n\Lambda k}$, this is $O(R\ell\sqrt{n\Lambda k})$. To detect intersections, we sort $A$ once for each anchor and look up each element of each partner list by binary search in $A$; this takes $O(R(m_A+km_D)\log n)=O(R\sqrt{n\Lambda k}\log n)$ time. Since $\ell\ge 8\log n$, the total running time is $O(R\ell\sqrt{n\Lambda k})$, the same order as the bound on the number of queries.

    For fixed $\alpha,\kappa,\phi,\delta$ and all sufficiently large $n$, the full-table branch is not taken: $k$ and $\Lambda$ depend only on $\alpha,\kappa$, so $k\le n/\Lambda$, and $66R\ell\sqrt{n\Lambda k}=O(\sqrt n\log n)$, so $66R\ell\sqrt{n\Lambda k}\le dn$. Hence the running time is $O(R\ell\sqrt{n\Lambda k})$, as claimed in \Cref{thrm:upper-bound-structural-formal}.
\end{proof}

\section{Distance testing at the original core size}
\label{sec:distance}

This section proves \Cref{thrm:upper-bound-distance-soundness-informal} by converting a relaxed-size core into a nearby core of the original required size (see \Cref{subsec:ov-distance}).
    
\begin{theorem}[Same-size distance testing; formal version of~\Cref{thrm:upper-bound-distance-soundness-informal}]
    \label{thrm:upper-bound-distance-soundness-formal}
    Let $d\ge3$ be a constant. Let $\alpha,\phi\in(0,1]$, $\varepsilon\in(0,1)$ and $\delta\in(0,1/3]$. Let $\kappa=1-\min\{1/4,\varepsilon/(8\alpha)\}$ and $n\ge4/((1-\kappa)\alpha)$. Run \Cref{algo:anchor-test} with this $\kappa$. Then, for every $n$-vertex graph $G$ of maximum degree at most $d$:
    \begin{enumerate}
        \item if $G\in\mathcal{P}_{\alpha,\phi}$, it accepts with probability at least $1-\delta$;
        \item if $G$ is $\varepsilon$-far from $\mathcal{P}_{\alpha,\mu_n^{\mathrm{dist}}\phi^2}$, it rejects with probability at least $1-\delta$, where
        \[
            \mu_n^{\mathrm{stru}}:=\psi/\phi^2,\qquad \mu_n^{\mathrm{dist}}:=\frac{\mu_n^{\mathrm{stru}}}{2d+2},
        \]
        and $\mu_n^{\mathrm{stru}}$ is the multiplier of \Cref{thrm:upper-bound-structural-formal} at the chosen $\kappa$.
    \end{enumerate}
    Its query complexity and running time are as in \Cref{thrm:upper-bound-structural-formal}.
\end{theorem}

Throughout the rest of this section, we also write $\beta=(1-\kappa)\alpha$ for the size slack as in~\Cref{sec:tester}. With $\kappa$ as in \Cref{thrm:upper-bound-distance-soundness-formal}, we have $\beta=\min\{\alpha/4,\varepsilon/8\}$. The repair below adds about $\beta n$ vertices to a core, so both its cost and the resulting edit distance are easier to state in terms of $\beta$.

Completeness follows directly from the structural theorem. For soundness, we prove the contrapositive: if $G$ contains a core of size at least $\kappa\alpha n=(\alpha-\beta)n$ and conductance at least $\psi$, then a small number of degree-bound-preserving edits creates a core of size at least $\alpha n$ and conductance at least $\psi/(2d+2)$.

We enlarge the core by subdividing edges: place each added vertex in the middle of a distinct internal edge. Old degrees do not increase, and an original edge crossing a cut becomes a path that must still cross the corresponding enlarged cut. The proof below accounts separately for original and added vertices.

\begin{lemma}[Adding vertices by subdivision]
\label{lemma:adding-vertices-by-subdivision}
    Let $G$ be a simple graph of maximum degree at most $d\ge 2$. Suppose $H=G[U]$ has $m=|U|\ge 2$ vertices and $\Phi_\textup{in}(H)\ge \theta>0$. Let $W\subseteq V\setminus U$ have $b\le m-1$ vertices. There is a simple graph $G^\prime$ on the same vertex set $V$, of maximum degree at most $d$, with 
    \[
        |E(G)\triangle E(G^\prime)|\le (d+3)b
    \]
    and
    \[
        \Phi_\textup{in}(G^\prime[U\cup W])\ge \frac{\theta}{2d+2}.
    \]
\end{lemma}
\begin{proof}
    First, we delete every edge incident to $W$, using at most $db$ deletions. Since $H$ is connected, it has at least $m-1$ edges. Choose $b$ distinct edges of $H$ and match them bijectively to the vertices of $W$. For the edge $(u,v)$ assigned to vertex $w\in W$, delete $(u,v)$ and add $(u,w), (w,v)$. There are at most three further edits per new vertex, proving that $|E(G)\triangle E(G^\prime)|\le (d+3)b$. Original endpoints keep their internal degree; each new vertex has degree $2\le d$ and no other edges. All other degrees can only decrease. The resulting graph is simple because the new middle vertices are distinct and all their old edges were deleted.

    We now bound the inner conductance. Let $T \subset U \cup W$ be nonempty with $|T| \le (m+b)/2$, and let $F_T$ be the number of edges of $G^\prime[U\cup W]$ crossing this cut. Of the two sides of the cut, choose a side $S$ containing at most $m/2$ original vertices. Write $S_U = S \cap U$ and $s = |S_U|$. Each original edge of $H$ crossing $(S_U,U\setminus S_U)$ becomes an edge-disjoint path of length one or two crossing the new cut. Hence
    \[
        F_T\ge \theta ds.
    \]
    This inequality is also valid when $s=0$.

    There are at most $ds$ new vertices in $S$ with an original neighbor in $S_U$, since they use distinct incidences at vertices of $S_U$. Every other new vertex in $S$ has both original neighbors on the other side and contributes two distinct cut edges. It follows that
    \[
        |T|\le |S|\le (d+1)s+F_T/2\le \left(\frac{d+1}{\theta d}+\frac{1}{2}\right)F_T.
    \]
    Here $|T|\le |S|$ holds because $T$ is a smaller side in total size, whether or not it is the chosen side $S$. Thus
    \[
        \frac{F_T}{d|T|}\ge \frac{\theta}{d+1+\theta d/2}\ge \frac{\theta}{2d+2},
    \]
    where we used $\theta\le 1$. Taking the minimum over $T$ ends the proof.
\end{proof}

The next corollary translates the number of added vertices into normalized edit distance. Its rounding terms ensure that the repaired core has at least $\lceil\alpha n\rceil$ vertices while respecting the degree bound.

\begin{corollary}[Size slack implies small edit distance]
\label{coro:size-slack-implies-small-edit-distance}

Let $d\ge3$, $\alpha\in(0,1]$, $\kappa\in[3/4,1)$, $n\ge4/((1-\kappa)\alpha)$ and $\beta=(1-\kappa)\alpha$.  For every $\theta\in(0,1]$,
\[
    G\in\mathcal{P}_{\alpha-\beta,\theta} \quad\Rightarrow \quad \textup{dist}_d(G,\mathcal{P}_{\alpha,\theta/(2d+2)})\le\frac{5\beta}{2}.
\]
\end{corollary}
\begin{proof}
    Suppose that $G\in\mathcal{P}_{\alpha-\beta,\theta}$. Then $G$ has a core of size $m\ge(\alpha-\beta)n$. If $m\ge \alpha n$, no edits are needed. Otherwise let $b=\lceil \alpha n\rceil-m\le \beta n+1$ and choose any $b$ vertices outside the core. Note that there are enough such vertices because $\lceil \alpha n\rceil \le n$. Also $m\ge 3\beta n$ and $\beta n\ge 4$, so $b\le m-1$ and $m\ge 2$. Applying~\Cref{lemma:adding-vertices-by-subdivision}, since $(d+3)/d\le 2$, we get
    \[
        \textup{dist}_d(G,\mathcal{P}_{\alpha, \theta/(2d+2)})\le \frac{(d+3)b}{dn}\le 2\frac{b}{n}\le 2\beta +\frac{2}{n}\le \frac{5\beta}{2}.
    \]
\end{proof}

\begin{proof}[Proof of~\Cref{thrm:upper-bound-distance-soundness-formal}]
    \textbf{Correctness.} Our choice gives $\kappa\in[3/4,1)$, so \Cref{thrm:upper-bound-structural-formal} applies with this $\kappa$. Completeness follows from~\Cref{thrm:upper-bound-structural-formal}.

    For soundness, let $\psi=\mu_n^{\mathrm{stru}}\phi^2$, and suppose that $G\in\mathcal{P}_{\kappa\alpha,\psi}$, that is  $G\in\mathcal{P}_{\alpha-\beta,\psi}$. By \Cref{coro:size-slack-implies-small-edit-distance},
    \[
        \textup{dist}_d\bigl(G,\mathcal{P}_{\alpha,\psi/(2d+2)}\bigr)
        \le\frac{5\beta}{2}\le\frac{5\varepsilon}{16}<\varepsilon.
    \]
    Hence every graph that is $\varepsilon$-far from $\mathcal{P}_{\alpha,\psi/(2d+2)}$ lies outside $\mathcal{P}_{\alpha-\beta,\psi}$, and it is rejected with probability at least $1-\delta$ by item~2 of \Cref{thrm:upper-bound-structural-formal}.

    \paragraph{Query complexity and the two multipliers.} The tester is \Cref{algo:anchor-test} with $\kappa=1-\min\{1/4,\varepsilon/(8\alpha)\}$, so its query complexity and running time are those of \Cref{thrm:upper-bound-structural-formal}. Substituting, it makes at most
    \[
        \min\Bigl\{dn,\ O\Bigl(\frac{\sqrt n\log n}{\phi^2}\cdot
        \frac{1-\kappa\alpha}{(1-\kappa)^2\alpha^{5/2}}
        \sqrt{\log\tfrac{2}{\alpha}\,\log\tfrac{1}{(1-\kappa)\alpha}}\cdot
        \log\tfrac{1}{\delta}\Bigr)\Bigr\}
    \]
    adjacency-list queries, and
    \[
        \mu_n^{\mathrm{stru}}=\Theta\!\left(\frac{(1-\kappa)^2\alpha^3}{(1-\kappa\alpha)\,\log n\,\log\frac{1}{(1-\kappa)\alpha}\,\log\frac{2}{\alpha}}\right).
    \]
    The rejection threshold is therefore
    \[
        \frac{\psi}{2d+2} =\mu_n^{\mathrm{dist}}\phi^2 =\frac{\mu_n^{\mathrm{stru}}\phi^2}{2d+2} =\Theta_{\alpha,\varepsilon}\!\left(\frac{\phi^2}{d\log n}\right).
    \]
    In particular, $\mu_n^{\mathrm{dist}}=\Theta_{\alpha,\varepsilon,d}(1/\log n)$, as stated in \Cref{thrm:upper-bound-distance-soundness-informal}.
\end{proof}

\section{The lower bound for structural testing}
\label{sec:many-label-lower}

This section proves \Cref{thrm:isotropic-lower-bound}. Its formal version is the following.

\begin{theorem}[Formal version of \cref{thrm:isotropic-lower-bound}]
\label{thrm:isotropic-lower-bound-full}
    Let $d_0=512$. Fix $\alpha\in(0,1]$, $\kappa\in(0,1)$, and $\phi\in(0,9/20]$, and put
    \begin{equation}
    \label{eq:isotropic-explicit-coefficient}
        q=2^{\lceil\log_2(2/\kappa)\rceil}, \qquad c_{\kappa,\phi} =\frac{q\kappa\phi^2}{16(q-1)\ln d_0}.
    \end{equation}
    There exist a constant $A=A(\alpha,\kappa,\phi)>0$ and an integer $n_0=n_0(\alpha,\kappa,\phi)$ such that the following holds for every $n\ge n_0$ and every $0<\mu\le1$. Suppose that a randomized adaptive algorithm $\mathcal A$, on simple graphs of order $n$ and maximum degree at most $d_0$, accepts every graph in $\mathcal P_{\alpha,\phi}$ and rejects every graph outside $\mathcal P_{\kappa\alpha,\mu\phi^2}$, each with probability at least $2/3$. Then $\mathcal A$ has worst-case adjacency-list query complexity greater than
    \[
        A\,n^{1/2+c_{\kappa,\phi}\mu},
    \]
    where the worst case is over graphs satisfying either promise and over the algorithm's internal randomness. Computation is unrestricted, and the constants $A,n_0$ are independent of $\mu$.
\end{theorem}

We construct two distributions over graphs obtained from a random $d_0$-regular base graph on $N=\Theta_{\alpha,\kappa}(n)$ vertices whose edges carry labels in $K=\mathbb F_2^r$, where $q=2^r$. Each label determines a perfect matching between two groups of $q$ vertices, and the input graph is the resulting lift. In the YES distribution the labels are independent and uniform, and the lift is an expander. In the NO distribution the labels are noisy parities of hidden vertex labels. The lift then splits into $q$ hidden classes, each with about $\alpha n/q\le\kappa\alpha n/2$ vertices and joined to the others by few edges, so that no set of $\kappa\alpha n$ vertices induces a subgraph of conductance $\mu\phi^2$. An algorithm with few queries sees only a few cycles of the base graph, all of them long, and the noise hides the sum of the labels around a long cycle. The noise rate is proportional to $\mu$. When $\mu<N^{-1/2}$, we use no noise, and every set of at least $\kappa\alpha n$ vertices induces a disconnected graph.

\Cref{subsec:lower-inputs} defines the labeled base graph and the two label distributions. \Cref{subsec:lower-lift} defines the lift, reduces adjacency-list queries to queries on the base graph, and proves the YES property. \Cref{subsec:lower-no} proves the NO property. \Cref{subsec:isotropic-lower} shows that the two label distributions are hard to distinguish with few adaptive queries, and \Cref{subsubsec:lower-completion} puts the parts together.

\paragraph{Relation to \cite{chiplunkar2018testing}.}
\cite{chiplunkar2018testing} proved an $n^{1/2+\Omega(\epsilon)}$ query lower bound for their NoisyParities problem. There, the edges of a random $d$-regular graph carry bits that are either independent and uniform, or noisy parities $X_u+X_v+Z_e$ with noise rate $\epsilon$. Their analysis~\cite[Section~4.3]{chiplunkar2018testing} shows that the edges seen by a typical query sequence form a forest together with a few edges whose endpoints are far apart in the forest, and bounds the resulting bias by Fourier analysis. We follow this approach and depart from it in three places. First, our labels lie in $\mathbb F_2^r$ rather than in $\mathbb F_2$, so that the planted lift has $q$ hidden classes, each too small to contain a qualifying core (\Cref{lem:lower-class-cut}). Second, for $r>1$ the circulations that determine the likelihood ratio may branch, and \Cref{lem:isotropic-closure} bounds their total contribution directly. Third, the constants in \Cref{lem:isotropic-source} do not depend on the noise rate, which gives the uniformity in $\mu$. Using $r$ independent binary coordinates instead of our noise also gives a bound of the form $n^{1/2+\Omega_{\kappa,\phi}(\mu)}$, with a coefficient smaller by a factor $\Theta(\log(2/\kappa))$ (\Cref{rem:coordinate-hybrid}).

Throughout this section, $d_0=512$ and $0<\phi\le9/20$. The base graph has $N$ vertices and degree $d_s$. The final construction uses $d_s=d_0$, but \Cref{lem:isotropic-source} holds for every $d_s\ge3$. Some letters from \Cref{sec:tester}, such as $p$, $q$, $m$, $s$, $D$, $F$ and $R$, have different meanings here. This slight abuse of notation causes no ambiguity, since the two parts of the paper do not interact.

\subsection{The labeled base graph}
\label{subsec:lower-inputs}

We first prove the lower bound for a labeled graph, called the \emph{source}, and then transfer it to graph inputs through the lift of \Cref{subsec:lower-lift}.

We generate the base graph in the \emph{configuration model} with $N$ vertices and degree $d_s\ge3$, where $d_sN$ is even. Each vertex $v\in[N]$ has $d_s$ half-edges $(v,1),\ldots,(v,d_s)$, and a uniformly random perfect matching of all half-edges determines the edges. We call each matched pair an \emph{edge occurrence}. Loops and parallel edges are allowed, and a loop has two incidences at its vertex. Conditional on the event that the resulting multigraph is simple, it is a uniformly random simple $d_s$-regular graph on $[N]$~\cite{bollobas1980probabilistic}. The matching can be exposed one pair at a time, which we use in \Cref{subsubsec:lower-adaptive-record} to analyze adaptive queries.

Let $r\ge1$, and let $K=\mathbb F_2^r$ be the group of $r$-bit vectors under coordinatewise addition modulo two. Its order is $q=2^r$, and since $-a=a$ in $K$, edges need no orientation. Every edge occurrence $e=uv$ receives a label $Y_e\in K$ in one of two ways. Under the \emph{uniform law} $\mathsf U$, the labels are independent and uniform on $K$. Under the \emph{planted law} $\mathsf P$, every vertex $v$ receives an independent uniform \emph{potential} $X_v\in K$, every edge occurrence $e$ receives an independent \emph{noise} $Z_e\in K$, and
\[
    Y_{uv}=X_u+X_v+Z_{uv}.
\]
Under both laws, the randomness that generates the labels (the uniform labels under $\mathsf U$, and the potentials and noises under $\mathsf P$) is sampled independently of the matching. Conditional on the matching, the joint law of the labels on a specified set of edge occurrences depends only on those occurrences and their endpoints, not on how the remaining half-edges are matched. Under $\mathsf P$, every potential appears twice in the sum of the labels around a cycle, so this sum equals the sum of the noises on the cycle. Under $\mathsf U$, the sum is uniform. Cycles are therefore what distinguishes the two laws, and the noise weakens the information carried by long cycles.

\paragraph{Noise and characters.}
We use \emph{isotropic} noise, which is $0$ with probability $1-p$ and otherwise uniform on the nonzero elements of $K$:
\begin{equation}\label{eq:isotropic-noise}
    \Pr(Z=0)=1-p,\qquad
    \Pr(Z=z)=\frac{p}{q-1}\quad(z\ne0),\qquad
    \rho=1-\frac{qp}{q-1},\qquad \lambda=-\ln\rho.
\end{equation}
Throughout, $0\le p\le1/4$. Since $q/(q-1)\le2$, we have $1/2\le\rho\le1$ and $0\le\lambda\le\ln2$. For $p=0$, we have $\rho=1$ and $\lambda=0$.

For $\sigma,a\in K$, let $\chi_\sigma(a)=(-1)^{\sigma\cdot a}$, where $\sigma\cdot a$ is the inner product modulo two. Thus $\chi_\sigma(a)$ is $1$ or $-1$ according to whether the coordinates of $a$ selected by $\sigma$ have even or odd sum. These \emph{characters} satisfy
\begin{equation}
\label{eq:lower-character-facts}
    \chi_\sigma(a+b)=\chi_\sigma(a)\chi_\sigma(b),\qquad
    \chi_\sigma\chi_\tau=\chi_{\sigma+\tau},\qquad
    \sum_{a\in K}\chi_\sigma(a)=q\,\mathbf{1}_{\{\sigma=0\}},
\end{equation}
and every $f:K\to\mathbb R$ satisfies the Fourier inversion formula
\begin{equation}
\label{eq:lower-fourier-inversion}
    f(a)=\frac1q\sum_{\sigma\in K}\widehat f(\sigma)\chi_\sigma(a),
    \qquad
    \widehat f(\sigma)=\sum_{b\in K}f(b)\chi_\sigma(b).
\end{equation}
For $\sigma\ne0$, the last identity in \eqref{eq:lower-character-facts} gives $\sum_{z\ne0}\chi_\sigma(z)=-1$, and hence
\begin{equation}
\label{eq:isotropic-bias}
    \E\left[\chi_\sigma(Z)\right] =(1-p)+\frac{p}{q-1}\sum_{z\ne0}\chi_\sigma(z) =1-p-\frac{p}{q-1} =\rho.
\end{equation}
So every nontrivial character of the noise has the same bias $\rho$. If $Z_1,\ldots,Z_\ell$ are independent noises and $\sigma\ne0$, then multiplicativity gives
\[
    \E\left[\chi_\sigma(Z_1+\cdots+Z_\ell)\right]=\rho^\ell=e^{-\lambda\ell}.
\]
Under $\mathsf P$, this is the bias that remains in a nontrivial character of the label sum around an $\ell$-edge cycle. We relate $\lambda$ to $p$ by
\begin{equation}
\label{eq:isotropic-lambda-lower}
    \lambda=-\ln\bigl(1-(1-\rho)\bigr)\ge 1-\rho=\frac{qp}{q-1}.
\end{equation}

\begin{remark}[Choice of noise]
For any noise with $\Pr(Z\ne0)=p$, the identity $\sum_{\sigma\in K}\chi_\sigma(z)=q\,\mathbf 1_{\{z=0\}}$, which follows from \eqref{eq:lower-character-facts} since $\chi_\sigma(z)=\chi_z(\sigma)$, gives
\[
    \sum_{\sigma\ne0}\E\left[\chi_\sigma(Z)\right]=q(1-p)-1=(q-1)\rho.
\]
Hence some nontrivial character of $Z$ has bias at least $\rho$. Isotropic noise is the noise of rate $p$ for which every nontrivial character has exactly this bias.
\end{remark}

\paragraph{Locality and source queries.}
The labels on a set of edge occurrences do not depend on how the other half-edges are matched. Formally, let $H$ be a multigraph whose edge occurrences are specified pairs of half-edges. For $y\in K^{E(H)}$, let
\[
    \Pr\nolimits^H_{\mathsf U}(y)=q^{-|E(H)|},
    \qquad
    \Pr\nolimits^H_{\mathsf P}(y)=\Pr\bigl(X_u+X_v+Z_e=y_e \text{ for all }e=uv\in E(H)\bigr),
\]
where the potentials of $V(H)$ and the noises of $E(H)$ are independent and distributed as above. Since the label randomness is sampled independently of the matching, for $\mathsf L\in\{\mathsf U,\mathsf P\}$ and every perfect matching $\mathcal M$ that contains all pairs of $H$,
\begin{equation}
\label{eq:lower-label-locality}
    \Pr\nolimits_{\mathsf L}\bigl(Y_H=y\bigm|\mathcal M\bigr) =\Pr\nolimits^H_{\mathsf L}(y).
\end{equation}
In \Cref{subsubsec:lower-adaptive-record}, we apply this identity to the edges exposed by an adaptive algorithm.

An algorithm accesses the source through \emph{source vertex queries}. A query specifies a vertex $v\in[N]$ and returns, for every $i\in[d_s]$, the half-edge matched to $(v,i)$ and the label of the corresponding edge occurrence. Queries may be adaptive. By \Cref{lem:lower-simple}, one adjacency-list query to the input graph costs at most one source vertex query, so it suffices to prove the lower bound for source vertex queries.

\subsection{The lift and the YES instances}
\label{subsec:lower-lift}

Let $B$ be a simple $d_s$-regular graph on $[N]$ with labels $(Y_e)$. Its \emph{lift} $H$ has vertex set $[N]\times K$, and we call $\{v\}\times K$ the \emph{fiber} of $v$. Every base edge $uv$ is replaced by the $q$ edges
\begin{equation}
\label{eq:lower-lift}
    (u,a)\,(v,a+Y_{uv})\qquad(a\in K),
\end{equation}
which form a perfect matching between the fibers of $u$ and $v$. For example, if $r=2$ and $Y_{uv}=(1,0)$, the copies $00,01,10,11$ of $u$ are joined to the copies $10,11,00,01$ of $v$, respectively. Since $B$ is simple, $H$ is a simple $d_s$-regular graph on $qN$ vertices, and every vertex of $H$ has at most one neighbor in each fiber. The adjacency lists of $H$ follow those of $B$: if the $i$-th half-edge of $v$ belongs to an edge $e=vw$, then the $i$-th neighbor of $(v,a)$ is $(w,a+Y_e)$.

\paragraph{Padded lifts.}
The input graph must have exactly $n$ vertices and contain a core on at least $\alpha n$ of them. Given $n$, set
\begin{equation}
\label{eq:lower-padding}
    m=\lceil\alpha n\rceil,
    \qquad N=\left\lceil\frac{m}{q}\right\rceil,
    \qquad r_{\rm del}=qN-m\in\{0,\ldots,q-1\}.
\end{equation}
Fix an ordering of $K$, and let $D$ be the set of vertices $(1,a)$ for the first $r_{\rm del}$ elements $a$ in this ordering. The \emph{padded lift} $G$ consists of $H'=H-D$ together with $n-m$ isolated vertices, identified with $[n]$ by a fixed bijection. Thus $G$ has $n$ vertices, $H'$ has $m\ge\alpha n$ vertices, and neither $D$ nor the bijection depends on the base graph or the labels.

\begin{lemma}[Local simulation of the padded lift]
\label{lem:lower-simple}
    Any algorithm making $Q$ adjacency-list queries to the padded lift can be simulated using at most $Q$ source vertex queries. The simulation requires neither the potentials nor the noises.
\end{lemma}

\begin{proof}
    The simulation stores the answers to all source queries made so far. A query to an isolated vertex returns $\perp$ without accessing the source. For a query $((v,a),i)$ with $(v,a)\in V(H')$, query the source at $v$ unless this was done before. The answer gives, for every half-edge of $v$, the base neighbor $w$ and the label $Y_e$ of the corresponding edge, and hence the adjacency list of $(v,a)$ in $H$. Remove the vertices of $D$ from this list, keep the order of the remaining entries, and return the $i$-th entry, or $\perp$ if fewer than $i$ entries remain. These answers are the adjacency lists of $H'$, and each query uses at most one new source query. Vertices that the algorithm chooses at random require no queries.
\end{proof}

\paragraph{Expansion under uniform labels.}
We analyze the spectrum of $H$ through the characters of $K$. For $\sigma\in K$ and $g:[N]\to\mathbb R$, let $(g\otimes\chi_\sigma)(v,a)=g(v)\chi_\sigma(a)$, and let $W_\sigma$ be the space of these functions. By \eqref{eq:lower-character-facts}, the spaces $W_\sigma$ are pairwise orthogonal and together span all real functions on $[N]\times K$. By \eqref{eq:lower-lift} and multiplicativity, the adjacency matrix $A_H$ of $H$ satisfies
\[
    \bigl(A_H(g\otimes\chi_\sigma)\bigr)(u,a)=\sum_{v:\,uv\in E(B)}g(v)\chi_\sigma(a+Y_{uv})=\chi_\sigma(a)\,(A_\sigma g)(u),
\]
where
\begin{equation}
\label{eq:lower-character-blocks}
    (A_\sigma g)(u)=\sum_{v:\,uv\in E(B)}\chi_\sigma(Y_{uv})\,g(v).
\end{equation}
Hence each $W_\sigma$ is invariant under $A_H$, and the spectrum of $A_H$ is the union, with multiplicity, of the spectra of the matrices $A_\sigma$, $\sigma\in K$. The block $A_0$ is the adjacency matrix $A_B$ of the base graph. For $\sigma\ne0$, $A_\sigma$ is the signed adjacency matrix of $B$ with sign $\chi_\sigma(Y_e)$ on the edge $e$. Under $\mathsf U$ and for fixed $\sigma\ne0$, these signs are independent and uniform on $\{\pm1\}$, so $A_\sigma$ is a uniformly random signing of $B$.

To bound these matrices, we use the following three known results (\Cref{lemma:friedman,lemma:tangle-free,lemma:mop}). A graph is \emph{bicycle-free at radius $\ell$} if every ball of radius $\ell$ contains at most one cycle; this property is called $\ell$-tangle-free in~\cite{bordenave2015new}.

\begin{lemma}[Friedman's theorem~\cite{friedman2008proof}; the form of {\cite[Theorem~1]{bordenave2015new}}]
\label{lemma:friedman}
    Let $d\ge3$, and let $B$ be a uniformly random simple $d$-regular graph on $[N]$, with adjacency eigenvalues $d=\mu_1\ge\mu_2\ge\cdots\ge\mu_N$. For every fixed $\varepsilon>0$,
    \[
        \lim_{N\to\infty}\Pr\bigl(\max\{\mu_2,|\mu_N|\}\ge2\sqrt{d-1}+\varepsilon\bigr)=0,
    \]
    where the limit is along any sequence with $Nd$ even.
\end{lemma}

\begin{lemma}[{\cite[Lemma~9]{bordenave2015new}}]
\label{lemma:tangle-free}
    Let $d\ge3$ and $\ell\ge1$ be integers. The multigraph produced by the configuration model with $N$ vertices of degree $d$ is bicycle-free at radius $\ell$ with probability $1-O((d-1)^{4\ell}/N)$.
\end{lemma}

\begin{lemma}[{\cite[Theorem~3.1]{mohanty2020explicit}}]
\label{lemma:mop}
    Let $G$ be a $d$-regular graph on $N$ vertices, where $d\le\operatorname{polylog}N$, and assume that $G$ is bicycle-free at radius $\ell$ with $\ell\gg(\log\log N)^2$. Let $A_s$ be the signed adjacency matrix of $G$ with independent uniformly random signs on the edges. Then, with probability at least $1-N^{-100}$,
    \[
        \|A_s\|_{\mathrm{op}}\le2\sqrt{d-1}\,\left(1+O\left(\frac{(\log\log N)^4}{\ell^2}\right)\right).
    \]
\end{lemma}

\begin{lemma}[Expansion of a uniformly labeled simple lift]
\label{lem:lower-yes-simple}
    Let $d_s=d_0=512$ and fix $r\ge1$. Let $B$ be a uniformly random simple $d_0$-regular graph on $[N]$ with labels drawn from $\mathsf U$, and let $H$ be its lift. With probability $1-o(1)$ as $N\to\infty$, the following hold, where $\Phi_\textup{in}$ uses the degree bound $d=d_0$:
    \begin{enumerate}
        \item $\Phi_\textup{in}(H)\ge 465/1024$;
        \item $\Phi_\textup{in}(H-D')\ge 463/1024$ for every set $D'$ of vertices contained in a single fiber.
    \end{enumerate}
    In particular, both bounds exceed $9/20$.
\end{lemma}

\begin{proof}
    We first bound the eigenvalues of the blocks $A_\sigma$. By \Cref{lemma:friedman} with $\varepsilon=1$, with probability $1-o(1)$ we have $\max\{\mu_2,|\mu_N|\}<2\sqrt{d_0-1}+1=2\sqrt{511}+1<47$. Hence every eigenvalue of $A_B=A_0$ other than $\mu_1=d_0$ has absolute value less than $47$.

    For the signed blocks, let $\ell_N=\lfloor(\log_{d_0-1}N)/8\rfloor$, which is a positive integer for all sufficiently large $N$. By \Cref{lemma:tangle-free}, the configuration model produces a graph that is bicycle-free at radius $\ell_N$ with probability $1-O((d_0-1)^{4\ell_N}/N)=1-O(N^{-1/2})$. The configuration model produces a simple graph with probability tending to $e^{-(d_0^2-1)/4}>0$~\cite{bollobas1980probabilistic}, and conditional on this event its output is uniform. Hence $B$ is bicycle-free at radius $\ell_N$ with probability $1-O(N^{-1/2})$. Fix such a $B$. Since $d_0$ is a constant and $\ell_N=\Theta(\log N)\gg(\log\log N)^2$, \Cref{lemma:mop} applies to $B$. For each $\sigma\ne0$, the signs $\chi_\sigma(Y_e)$ are independent and uniform under $\mathsf U$, so except with probability at most $N^{-100}$ over the labels,
    \[
        \|A_\sigma\|_{\mathrm{op}} \le2\sqrt{d_0-1}\,\left(1+O\left(\frac{(\log\log N)^4}{\ell_N^2}\right)\right) \le2\sqrt{d_0-1}+1<47
    \]
    for all sufficiently large $N$. Here $\|A_\sigma\|_{\mathrm{op}}$ is the largest absolute value of an eigenvalue of the symmetric matrix $A_\sigma$. Since $q$ is fixed, a union bound over $\sigma$ shows that all these bounds hold simultaneously with probability $1-o(1)$. We assume this event in the rest of the proof.

    By the block decomposition, the eigenvalue $d_0$ of $A_H$ is simple, its eigenvectors are the constant functions in $W_0$, and every other eigenvalue has absolute value less than $47$. Hence $f^{\mathsf T}A_Hf\le47\|f\|_2^2$ for every $f\perp\mathbf{1}$. Let $S\subseteq V(H)$ with $0<|S|\le qN/2$, and let $f=\mathbf{1}_S-(|S|/qN)\mathbf{1}$. Then $f\perp\mathbf{1}$ and $\|f\|_2^2=|S|(1-|S|/qN)\ge|S|/2$. Since $H$ is $d_0$-regular,
    \[
        |E_H(S,V(H)\setminus S)| =\sum_{\{x,y\}\in E(H)}(f(x)-f(y))^2 =f^{\mathsf T}(d_0I-A_H)f \ge(d_0-47)\|f\|_2^2 \ge\frac{(d_0-47)|S|}{2}.
    \]
    Dividing by $d_0|S|$ gives $\Phi_\textup{in}(H)\ge(d_0-47)/(2d_0)=465/1024$.

    For the second item, let $D'$ be contained in one fiber, and let $S\subseteq V(H)\setminus D'$ with $0<|S|\le|V(H)\setminus D'|/2\le qN/2$. Every vertex of $S$ has at most one neighbor in $D'$, so
    \[
        |E_{H-D'}(S,V(H-D')\setminus S)| \ge|E_H(S,V(H)\setminus S)|-|S| \ge\Bigl(\frac{465}{1024}\,d_0-1\Bigr)|S| =\frac{463}{1024}\,d_0|S|.
    \]
\end{proof}

\subsection{Hidden color classes and the NO instances}
\label{subsec:lower-no}

Under the planted law, the potentials color the vertices of the lift by
\begin{equation}
\label{eq:lower-colors}
    C(v,a)=a+X_v\in K.
\end{equation}
The colors are not visible to the algorithm and are used only in the analysis.

\begin{lemma}[Color classes]
\label{lem:lower-colors}
    Under the planted law, the color classes partition $V(H)$ into $q$ sets of size $N$. Each lifted edge associated with a base edge $e$ has endpoints of the same color if and only if $Z_e=0$. Consequently, exactly $qB_{\rm bad}$ edges of $H$ join different color classes, where $B_{\rm bad}$ is the number of base edges with nonzero noise.
\end{lemma}

\begin{proof}
    For every base vertex $v$ and color $c\in K$, the unique copy of $v$ with color $c$ is $(v,c+X_v)$. So each color class contains exactly one vertex of each fiber and has $N$ vertices. For a base edge $e=uv$ and $a\in K$, the endpoints of the lifted edge $(u,a)(v,a+Y_e)$ satisfy
    \[
        C(v,a+Y_e)=a+Y_e+X_v=a+X_u+Z_e=C(u,a)+Z_e.
    \]
    Hence the $q$ lifted edges of $e$ join different classes exactly when $Z_e\ne0$. Summing over the base edges gives the count.
\end{proof}

Since the noises are independent of each other and of the base graph, $B_{\rm bad}\sim\operatorname{Bin}(d_sN/2,p)$ for every fixed simple base graph. A candidate core that is larger than the classes meets several of them, and its intersection with each class is one side of a cut of the core. The next lemma adds up these cuts.

\begin{lemma}[Averaging over small classes]
\label{lem:lower-class-cut}
    Let $G$ have maximum degree at most $d>0$, and suppose that its vertex set is partitioned into classes of size at most $s\ge1$. Let $F$ be the set of edges joining different classes. Every $U\subseteq V(G)$ of size $u>s$ satisfies
    \[
        \Phi_\textup{in}(G[U])\le\frac{2|F|}{d\,\min\{u,\,2(u-s)\}}.
    \]
    In particular, if $u\ge2s$, the bound is $2|F|/(du)$.
\end{lemma}

\begin{proof}
    Let $U_1,\ldots,U_k$ be the nonempty intersections of $U$ with the classes, and let $u_i=|U_i|$. Since $u>s$, we have $u_i<u$, so $(U_i,U\setminus U_i)$ is a cut of $G[U]$. Applying the definition of inner conductance to its smaller side gives
    \[
        \Phi_\textup{in}(G[U])\,d\min\{u_i,u-u_i\}\le|E_G(U_i,U\setminus U_i)|.
    \]
    We sum over $i$. On the right, every edge of $F$ inside $U$ is counted twice and no other edge is counted. On the left, if every $u_i\le u/2$, the minima add up to $u$. Otherwise, there is a unique index $j$ with $u_j>u/2$, and the minima add up to $2(u-u_j)\ge2(u-s)$. In both cases,
    \[
        \Phi_\textup{in}(G[U])\,d\min\{u,2(u-s)\} \le2|F\cap E(G[U])|\le2|F|.
        \qedhere
    \]
\end{proof}

    We apply \Cref{lem:lower-class-cut} with classes of size at most $N$. Since $q(N-1)$ is an integer smaller than $\lceil\alpha n\rceil$ by \eqref{eq:lower-padding}, we have
    \[
        q(N-1)<\alpha n\le qN.
    \]
    Let $a=\kappa q$, so that $a\ge2$ by the choice of $q$ in \eqref{eq:isotropic-explicit-coefficient}, and for $u>N$ let $D_N(u)=\min\{u,2(u-N)\}$. If $u\ge\kappa\alpha n$, then $u>a(N-1)\ge N$, and for $N\ge8$,
    \begin{equation}
    \label{eq:lower-rounding-denominator}
        D_N(u)>a(N-2),
        \qquad
        \frac{\kappa qN}{D_N(u)}<\frac{N}{N-2}\le\frac43.
    \end{equation}
    Indeed, $u>a(N-2)$, and $2(u-N)>2a(N-1)-2N=a(N-2)+(a-2)N\ge a(N-2)$. This estimate holds for all candidate cores at once and does not depend on $\mu$.

\subsection{Adaptive source indistinguishability}
\label{subsec:isotropic-lower}

We now show that an algorithm with few source vertex queries cannot tell $\mathsf U$ from $\mathsf P$. \Cref{subsubsec:lower-fixed-record} bounds the likelihood ratio of the labels on a fixed exposed graph, and \Cref{subsubsec:lower-adaptive-record} shows that an adaptive algorithm exposes, with high probability, a graph to which this bound applies.

\subsubsection{Likelihood ratios on a fixed exposed graph}
\label{subsubsec:lower-fixed-record}

Let $H$ be a finite multigraph. A \emph{spanning forest} of $H$ is an acyclic set $F\subseteq E(H)$ with the same connected components as $H$, and the edges in $E(H)\setminus F$ are the \emph{nonforest edges}. Each nonforest edge closes a cycle with $F$. We write $\operatorname{dist}_F(x,y)$ for the distance between $x$ and $y$ in $F$, which is $\infty$ for vertices in different trees. For $y\in K^{E(H)}$, the \emph{likelihood ratio} of the two laws is
\begin{equation}
\label{eq:isotropic-likelihood-def}
    D_H(y)
    =\frac{\Pr\nolimits^H_{\mathsf P}(y)}{\Pr\nolimits^H_{\mathsf U}(y)}
    =q^{|E(H)|}\Pr\nolimits^H_{\mathsf P}(y),
\end{equation}
with $\Pr\nolimits^H_{\mathsf U}$ and $\Pr\nolimits^H_{\mathsf P}$ as in \eqref{eq:lower-label-locality}. The next lemma shows that this ratio is close to $1$ when there are few nonforest edges and their endpoints are far apart in the forest.

\begin{lemma}[The likelihood ratio on a separated graph]
\label{lem:isotropic-closure}
    Let $H$ be a finite multigraph with a spanning forest $F$ and $R$ nonforest edges, and let $D_{\mathrm{sep}}\ge1$ be an integer. Assume that the $2R$ endpoints of the nonforest edges are distinct and pairwise at distance at least $D_{\mathrm{sep}}$ in $F$. If $0\le p\le1/4$, then every $y\in K^{E(H)}$ satisfies
    \begin{equation}
    \label{eq:isotropic-closure-bound}
        |D_H(y)-1| \le\bigl(1+(q-1)\rho^{D_{\mathrm{sep}}}\bigr)^R-1 \le\exp\bigl((q-1)R\rho^{D_{\mathrm{sep}}}\bigr)-1.
    \end{equation}
\end{lemma}

\begin{proof}
    For an assignment $\gamma\in K^{E(H)}$ of group elements to the edges, write
    \[
        \chi_\gamma(y)=\prod_{e\in E(H)}\chi_{\gamma_e}(y_e),
        \qquad
        \operatorname{supp}\gamma=\{e\in E(H):\gamma_e\ne0\}.
    \]
    The \emph{divergence} of $\gamma$ at a vertex $v$ is $\partial\gamma(v)=\sum_{e\ni v}\gamma_e$, computed in $K$, where a loop at $v$ is counted twice. We call $\gamma$ a \emph{circulation} if $\partial\gamma(v)=0$ for every $v$. For $K=\mathbb F_2$, the circulations are the indicator vectors of even subgraphs.

    We first expand the likelihood ratio in characters. By \eqref{eq:lower-fourier-inversion} and \eqref{eq:isotropic-bias}, the noise distribution is
    \[
        \Pr(Z=z)=\frac1q\sum_{\sigma\in K} \rho^{\mathbf{1}\{\sigma\ne0\}}\chi_\sigma(z).
    \]
    Given the potentials, the labels are independent, and $\Pr(Y_e=y_e\mid X)=\Pr(Z_e=y_e+X_u+X_v)$ for $e=uv$. Hence
    \[
         \Pr\nolimits^H_{\mathsf P}(y)
         =\E_X\left[
           \prod_{e=uv\in E(H)}
           \left(
            \frac1q\sum_{\gamma_e\in K}
            \rho^{\mathbf{1}\{\gamma_e\ne0\}}
            \chi_{\gamma_e}(y_e)
            \chi_{\gamma_e}(X_u)
            \chi_{\gamma_e}(X_v)
           \right)
         \right].
    \]
    Expanding the product, an assignment $\gamma$ contributes $q^{-|E(H)|}\chi_\gamma(y)\rho^{|\operatorname{supp}\gamma|} \prod_v\chi_{\partial\gamma(v)}(X_v)$. The potentials are independent and $\E[\chi_\tau(X_v)]=\mathbf{1}_{\{\tau=0\}}$, so only circulations survive the expectation, and
    \begin{equation}
    \label{eq:isotropic-likelihood}
        D_H(y)=\sum_{\gamma:\,\partial\gamma=0}\chi_\gamma(y)\rho^{|\operatorname{supp}\gamma|}.
    \end{equation}
    The zero circulation contributes $1$, so it remains to bound the contribution of the nonzero circulations.

    Next, a circulation is determined by its values on the nonforest edges. Indeed, if two circulations agree on the nonforest edges, their sum is a circulation supported on $F$. If this sum were nonzero, a leaf of its support would have exactly one incident nonzero value and hence nonzero divergence. Conversely, every choice of values on the nonforest edges extends to a circulation: for each nonforest edge $e=uv$ with value $g$, put $g$ on $e$ and on every edge of the $u$--$v$ path in $F$, which gives a circulation, and add these circulations over all nonforest edges. Hence, for $1\le k\le R$, exactly $\binom Rk(q-1)^k$ circulations are nonzero on exactly $k$ nonforest edges.

    We now bound the support of a nonzero circulation. Let $\gamma$ be a circulation that is nonzero on exactly $k\ge1$ nonforest edges, let $T_\gamma$ be the set of their $2k$ endpoints, and let $F_\gamma=\{f\in F:\gamma_f\ne0\}$. When $q>2$, the support of $\gamma$ need not be a union of cycles, since three nonzero elements of $K$ can sum to zero at a vertex. We therefore work with the components of $F_\gamma$. Since the endpoints of the nonforest edges are distinct, every $w\in T_\gamma$ lies on a unique nonzero nonforest edge, which we denote by $e(w)$. Let $C$ be a component of the forest $(V(H),F_\gamma)$ that contains a vertex of $T_\gamma$. In the sum of the divergences over $V(C)$, every edge of $F_\gamma$ at a vertex of $C$ lies in $C$ and is counted twice, so
    \[
        0=\sum_{w\in V(C)}\partial\gamma(w)=\sum_{w\in T_\gamma\cap V(C)}\gamma_{e(w)}.
    \]
    A single nonzero element is not $0$, so $t_C:=|T_\gamma\cap V(C)|\ge2$. Now walk around the tree $C$, starting at a vertex of $T_\gamma$ and traversing every edge once in each direction. This closed walk has length $2|E(C)|$. The first visits to the $t_C$ vertices of $T_\gamma\cap V(C)$ split it into $t_C$ segments, each joining two distinct vertices of $T_\gamma$. By assumption, each segment has length at least $D_{\mathrm{sep}}$, so $2|E(C)|\ge t_CD_{\mathrm{sep}}$. Summing over the components that meet $T_\gamma$ gives
    \[
        |\operatorname{supp}\gamma|\ge|F_\gamma|\ge\frac{D_{\mathrm{sep}}}{2}\sum_C t_C=kD_{\mathrm{sep}}.
    \]

    Finally, using \eqref{eq:isotropic-likelihood}, $|\chi_\gamma(y)|=1$, $0\le\rho\le1$, and the count of circulations, we obtain
    \[
        |D_H(y)-1| \le\sum_{k=1}^R\binom Rk(q-1)^k\rho^{kD_{\mathrm{sep}}} =\bigl(1+(q-1)\rho^{D_{\mathrm{sep}}}\bigr)^R-1.
    \]
    The second bound in \eqref{eq:isotropic-closure-bound} follows from $1+x\le e^x$.
\end{proof}

\subsubsection{Adaptive exposure}
\label{subsubsec:lower-adaptive-record}

A \emph{transcript} is the sequence of queries and answers seen by an algorithm.

\begin{lemma}[Adaptive source indistinguishability]
\label{lem:isotropic-source}
    Fix integers $d_s\ge3$ and $r\ge1$, let $q=2^r$, and let $b=1/(4\ln d_s)$. There are constants $s=s(d_s,q)>0$ and $N_0=N_0(d_s,q)$ such that the following holds for every $N\ge N_0$ with $d_sN$ even, every $p\in[0,1/4]$, and every $\xi\in[0,1/16]$ with
    \begin{equation}
    \label{eq:isotropic-finite-source-condition}
        b\lambda\ge2\xi,
    \end{equation}
where $\lambda$ is given by \eqref{eq:isotropic-noise}. Let the base graph be a uniformly random simple $d_s$-regular graph on $[N]$, labeled according to $\mathsf U$ or to $\mathsf P$ with noise rate $p$. Then every adaptive algorithm making at most $sN^{1/2+\xi}$ source vertex queries has transcript distributions under $\mathsf U$ and $\mathsf P$ at total variation distance less than $1/6$.
\end{lemma}

Condition \eqref{eq:isotropic-finite-source-condition} relates the query exponent to the noise: more noise makes long cycles less informative and allows more queries. For $p=0$, it forces $\xi=0$, and the lemma gives an $\Omega(\sqrt N)$ bound. The constants $s$ and $N_0$ do not depend on $p$ or $\xi$.

\begin{proof}
    We follow the exposure argument of~\cite[Section~4.3]{chiplunkar2018testing}. We first express the probability of a complete query record through the labels it exposes. We then show that, except with small probability, the exposed graph is a forest plus a few edges whose endpoints are far apart, and apply \Cref{lem:isotropic-closure} to such records.

    Let $D_{\mathrm{sep}}=\lfloor b\ln N\rfloor$ and
    \[
        T=\lfloor sN^{1/2+\xi}\rfloor,
        \qquad
        m=aN^{2\xi},
        \qquad
        M=4(d_s-1)^{D_{\mathrm{sep}}},
    \]
    where the constants $a,s\in(0,1]$ are chosen at the end of the proof. Inside this proof, $m$ denotes this cutoff rather than $\lceil\alpha n\rceil$. It need not be an integer, and if $m<1$, the condition $|R|\le m$ below means $R=\emptyset$. Since $b\ln(d_s-1)<1/4$, we have $M\le4N^{1/4}$. We take $N_0$ large enough that $D_{\mathrm{sep}}\ge1$ and $T\le N/8$ for all $N\ge N_0$; since $\xi\le1/16$, this does not depend on $p$ or $\xi$.

    By convexity of total variation distance, it suffices to consider a deterministic algorithm. We extend it arbitrarily to answers that cannot occur for a simple base graph, and stop it after $T$ source queries. We first work in the configuration model without conditioning on simplicity, and condition at the end.

    We reveal the matching as the algorithm proceeds. When the algorithm queries $v$, we expose the unmatched half-edges of $v$ one at a time, in increasing order, and for each of them reveal its partner and the label of the new edge occurrence. We call each such step a \emph{pairing}, and $v$ its \emph{querying vertex}. There are at most $d_sT$ pairings. The \emph{record} $\omega$ consists of the queries, the pairings in order, and the revealed labels. The transcript is a function of the record, so it suffices to compare the distributions $\nu_{\mathsf U}$ and $\nu_{\mathsf P}$ of the record. Let $H(\omega)$ be the exposed multigraph and $y(\omega)$ its labels.

    Fix a record $\omega$ that the algorithm can produce. Since the algorithm is deterministic, the record equals $\omega$ exactly when the matching contains the pairs of $\omega$ and the labels on $H(\omega)$ equal $y(\omega)$; the answers then determine all queries. By \eqref{eq:lower-label-locality},
    \[
         \Pr\nolimits_{\mathsf L}(\omega) =\pi(\omega)\Pr\nolimits^{H(\omega)}_{\mathsf L}(y(\omega)),
         \qquad \mathsf L\in\{\mathsf U,\mathsf P\},
    \]
    where $\pi(\omega)$ is the probability that a uniform perfect matching contains the pairs of $\omega$. Hence
    \begin{equation}
    \label{eq:isotropic-record-ratio}
        \frac{\Pr\nolimits_{\mathsf P}(\omega)}{\Pr\nolimits_{\mathsf U}(\omega)}=D_{H(\omega)}(y(\omega)).
    \end{equation}
    The same computation applies to partial records. Given the record so far, every completion of the matching has the same factor $\Pr\nolimits^{H}_{\mathsf L}(y)$, so the unmatched half-edges are uniformly matched given the past, under either law. In particular, the partner revealed in the next pairing is uniform among the other unmatched half-edges. Before each pairing, at most $2d_sT\le d_sN/4$ half-edges are matched. Hence, for every vertex set $W$ determined by the past,
    \begin{equation}
    \label{eq:isotropic-partner-prob}
        \Pr(\text{the partner lies in }W\mid\text{past}) \le\frac{d_s|W|}{\tfrac34d_sN-1} \le\frac{2|W|}{N}.
    \end{equation}

    Next, we keep track of the cycles created by the exposure. Quantities with index $t$ are taken just before the $t$-th pairing. Let $D_t$ be the set of \emph{discovered} vertices, namely the vertices queried so far, including the current querying vertex, and the partners revealed so far. Each query discovers at most $d_s+1$ vertices, so $|D_t|\le(d_s+1)T$.

    We maintain a spanning forest $F_t$ of the exposed graph. A newly queried vertex that was not discovered before forms a one-vertex tree. A pairing $vw$ is \emph{new} if $w\notin D_t$, a \emph{join} if $w\in D_t$ lies in a different tree than $v$, and a \emph{cycle pairing} otherwise. New and join edges are added to the forest, and the edges of cycle pairings are the nonforest edges. Let $R_t$ be the set of nonforest edges, $S_t$ the set of their endpoints, and $R$ the final set of nonforest edges. The set of \emph{guarded} vertices is
    \[
        B_t=\{x\in D_t:\operatorname{dist}_{F_t}(x,S_t)<D_{\mathrm{sep}}\}.
    \]
    Forest distances between discovered vertices never increase, and $D_t$ and $S_t$ only grow, so $B_t$ grows with $t$. A ball of radius $D_{\mathrm{sep}}-1$ in a forest of maximum degree $d_s$ has at most
    \[
        1+d_s\sum_{i=0}^{D_{\mathrm{sep}}-2}(d_s-1)^i\le M
    \]
    vertices, so
    \begin{equation}
    \label{eq:isotropic-guard-size}
        |B_t|\le|S_t|M\le2|R_t|M.
    \end{equation}

    We call a record \emph{bad} if one of the following events occurs, and \emph{good} otherwise:
    \begin{itemize}[leftmargin=3.2em]
        \item[$\mathcal E_1$:] a cycle pairing $vw$ has
         $\operatorname{dist}_{F_t}(v,w)<D_{\mathrm{sep}}$ (this includes loops);
        \item[$\mathcal E_2$:] a join or cycle pairing has its partner in $B_t$;
        \item[$\mathcal E_3$:] a cycle pairing has its querying vertex in $B_t$;
        \item[$\mathcal E_4$:] $|R|>m$.
    \end{itemize}
    On a good record, the endpoints of the nonforest edges are distinct and pairwise at distance at least $D_{\mathrm{sep}}$ in the forest. We check this by induction over the pairings. A new pairing adds a leaf and does not change the distances between earlier vertices. A join pairing $vw$ does not change distances inside the two merged trees. For an endpoint $x$ in the tree of $v$ and an endpoint $x'$ in the tree of $w$, the new distance is
    \[
        \operatorname{dist}_{F_t}(x,v)+1+\operatorname{dist}_{F_t}(w,x')\ge D_{\mathrm{sep}}+1,
    \]
    since $w\notin B_t$. A cycle pairing does not change the forest. Its endpoints are far from each other since $\mathcal E_1$ does not occur, and far from all earlier endpoints since $\mathcal E_2$ and $\mathcal E_3$ do not occur. Hence the exposed graph of a good record satisfies the assumptions of \Cref{lem:isotropic-closure}, with $R\le m$.

    We now bound the probability of a bad record. The following bounds hold under both laws, since they use only \eqref{eq:isotropic-partner-prob}. A cycle pairing needs a partner in $D_t$, which has probability at most $2(d_s+1)T/N$. Summing over the at most $d_sT$ pairings and applying Markov's inequality gives
    \begin{equation}
    \label{eq:isotropic-stopped-first-moment}
        \E|R|\le\frac{2d_s(d_s+1)T^2}{N},
        \qquad
        \Pr(\mathcal E_4)\le\frac{2d_s(d_s+1)T^2}{Nm}.
    \end{equation}
    For $\mathcal E_1$, the partner must be one of the at most $M$ discovered vertices at forest distance less than $D_{\mathrm{sep}}$ from the querying vertex, so $\Pr(\mathcal E_1)\le2d_sMT/N$.

    For $\mathcal E_2$ and $\mathcal E_3$, we consider only the pairings with $|R_t|\le m$; if $\mathcal E_4$ does not occur, these are all pairings. At such a pairing, $|B_t|\le2mM$ by \eqref{eq:isotropic-guard-size}, so the partner lies in $B_t$ with probability at most $4mM/N$. Hence $\Pr(\mathcal E_2\setminus\mathcal E_4)\le4d_smMT/N$.

    The event $\mathcal E_3$ needs a different argument, since the algorithm may choose to query a guarded vertex. Let $I_t$ be the indicator that the $t$-th pairing takes place, its querying vertex lies in $B_t$, and $|R_t|\le m$. It is determined by the past, and a cycle pairing needs a partner in $D_t$, so
    \[
     \Pr(\mathcal E_3\setminus\mathcal E_4)
     \le\sum_{t\le d_sT}
       \E\bigl[I_t\Pr(\text{the partner lies in }D_t\mid\text{past})\bigr]
     \le\frac{2(d_s+1)T}{N}\,
       \E\Bigl[\sum_tI_t\Bigr].
    \]
    Since $|R_t|$ is nondecreasing, the pairings with $|R_t|\le m$ form an initial segment; let $t^*$ be the last of them. Every querying vertex counted in $\sum_tI_t$ lies in $B_{t^*}$, and each vertex is the querying vertex of at most $d_s$ pairings, so $\sum_tI_t\le d_s|B_{t^*}|\le2d_smM$. The sum is $0$ if no pairing takes place. Therefore $\Pr(\mathcal E_3\setminus\mathcal E_4)\le4d_s(d_s+1)mMT/N$. Adding the four bounds gives, for $\mathsf L\in\{\mathsf U,\mathsf P\}$,
    \begin{equation}
    \label{eq:isotropic-stopped-graph-error}
     \Pr\nolimits_{\mathsf L}(\text{bad})
     \le\frac{2d_s(d_s+1)T^2}{Nm}
       +\frac{10d_s(d_s+1)(m+1)MT}{N}.
    \end{equation}

    With these bounds, we compare the two record distributions. Let $\varepsilon_F=\exp\bigl((q-1)m\rho^{D_{\mathrm{sep}}}\bigr)-1$. On a good record, \Cref{lem:isotropic-closure} gives $|D_{H(\omega)}(y(\omega))-1|\le\varepsilon_F$. By \eqref{eq:isotropic-record-ratio},
    \[
        \lVert\nu_{\mathsf U}-\nu_{\mathsf P}\rVert_{\mathrm{TV}}
        \le\frac12\sum_{\omega\text{ good}}\Pr\nolimits_{\mathsf U}(\omega)
        \bigl|D_{H(\omega)}(y(\omega))-1\bigr|
        +\frac{\Pr\nolimits_{\mathsf U}(\text{bad})+\Pr\nolimits_{\mathsf P}(\text{bad})}{2}
        \le\frac{\varepsilon_F}{2}
        +\frac{\Pr\nolimits_{\mathsf U}(\text{bad})+\Pr\nolimits_{\mathsf P}(\text{bad})}{2}.
    \]
    Now let $\mathrm{Simple}$ be the event that the matching yields a simple graph. By~\cite{bollobas1980probabilistic}, $\Pr(\mathrm{Simple})\to e^{-(d_s^2-1)/4}$, so $\Pr(\mathrm{Simple})\ge c_{d_s}:=\frac12e^{-(d_s^2-1)/4}$ for all large $N$. Since $\mathrm{Simple}$ depends only on the matching, the computation leading to \eqref{eq:isotropic-record-ratio} also applies to the laws conditioned on $\mathrm{Simple}$, with $\pi(\omega)$ replaced by the conditional probability that the matching contains the pairs of $\omega$. So the likelihood ratio of a record is still $D_{H(\omega)}(y(\omega))$. Moreover, $\Pr\nolimits_{\mathsf L}(\text{bad}\mid\mathrm{Simple})
    \le c_{d_s}^{-1}\Pr\nolimits_{\mathsf L}(\text{bad})$. Writing $\nu^{\mathrm s}_{\mathsf L}$ for the law of the record conditioned on $\mathrm{Simple}$, the same decomposition and \eqref{eq:isotropic-stopped-graph-error} give
    \begin{equation}
    \label{eq:isotropic-stopped-tv}
        \lVert\nu_{\mathsf U}^{\mathrm s}
           -\nu_{\mathsf P}^{\mathrm s}\rVert_{\mathrm{TV}}
        \le\frac{\varepsilon_F}{2}+\frac1{c_{d_s}}\left(\frac{2d_s(d_s+1)T^2}{Nm}+\frac{10d_s(d_s+1)(m+1)MT}{N}\right).
    \end{equation}
    
    It remains to choose the constants. Since $D_{\mathrm{sep}}\ge b\ln N-1$ and $\rho\ge1/2$, we have
    $\rho^{D_{\mathrm{sep}}}\le\rho^{-1}N^{-b\lambda}\le2N^{-b\lambda}$. Together with \eqref{eq:isotropic-finite-source-condition}, this gives
    \[
        m\rho^{D_{\mathrm{sep}}}\le2aN^{2\xi-b\lambda}\le2a, \qquad \varepsilon_F\le e^{2(q-1)a}-1.
    \]
    Moreover, $m+1\le2N^{2\xi}$ and $\xi\le1/16$ give
    \[
        \frac{T^2}{Nm}\le\frac{s^2}{a}, \qquad \frac{(m+1)MT}{N} \le8sN^{3\xi-1/4} \le8sN^{-1/16}.
    \]
    Substituting these bounds into \eqref{eq:isotropic-stopped-tv}, we get
    \[
        \lVert\nu_{\mathsf U}^{\mathrm s}
           -\nu_{\mathsf P}^{\mathrm s}\rVert_{\mathrm{TV}}
        \le\frac{e^{2(q-1)a}-1}{2}
        +\frac{2d_s(d_s+1)s^2}{a\,c_{d_s}}
        +\frac{80d_s(d_s+1)s}{c_{d_s}}N^{-1/16}.
    \]
    Let
    \[
        a=\frac{\ln(13/12)}{2(q-1)},\qquad
        s=\min\left\{1,\sqrt{\frac{a\,c_{d_s}}{48d_s(d_s+1)}}\right\}.
    \]
    Then each of the first two terms is at most $1/24$. Finally, choose $N_0=N_0(d_s,q)$ such that, in addition to the requirements above and $\Pr(\mathrm{Simple})\ge c_{d_s}$, the last term is at most $1/24$ for every $N\ge N_0$. Then the distance is at most $1/8<1/6$. None of these choices depends on $p$ or $\xi$. Since the transcript is a function of the record, and averaging over the random tape handles randomized algorithms, this proves the lemma.
    
    Replacing $1/24$ by $\tau/4$ in these choices gives distance less than any fixed $\tau\in(0,1)$, with $s$ and $N_0$ depending also on $\tau$.
\end{proof}

\subsection{Proof of \texorpdfstring{\Cref{thrm:isotropic-lower-bound-full}}{the lower bound}}
\label{subsubsec:lower-completion}

We now combine the parts. The noise rate is proportional to $\mu$ when $\mu\ge N^{-1/2}$ and is $0$ otherwise; this keeps all failure probabilities and the threshold $n_0$ independent of $\mu$.

\begin{proof}[Proof of \Cref{thrm:isotropic-lower-bound-full}]
    Write $c=c_{\kappa,\phi}$ and $b=1/(4\ln d_0)$. By the choice of $q$ in \eqref{eq:isotropic-explicit-coefficient}, $q\ge2/\kappa$. Since $q/(q-1)\le2$ and $\phi\le9/20$, we also have $0<c<1/16$. Let $n$ be large, let $0<\mu\le1$, define $m$ and $N$ by \eqref{eq:lower-padding}, and let $\psi=\mu\phi^2$. All lower bounds on $N$ below depend only on $\alpha$, $\kappa$ and $\phi$.

    We first choose the parameters of the source. Let
    \[
        (p_N,\xi_N)=
        \begin{cases}
        (\kappa\psi/2,\ c\mu), & \mu\ge N^{-1/2},\\
        (0,\ 0), & \mu<N^{-1/2},
        \end{cases}
    \]
    and let $\lambda_N$ be given by \eqref{eq:isotropic-noise} with $p=p_N$. Then $p_N\le\kappa\phi^2/2<1/4$ and $\xi_N\le c<1/16$. If $\mu\ge N^{-1/2}$, then \eqref{eq:isotropic-lambda-lower} gives
    \[
        b\lambda_N
        \ge\frac{1}{4\ln d_0}\cdot\frac{q}{q-1}\cdot
           \frac{\kappa\mu\phi^2}{2}
        =2c\mu
        =2\xi_N.
    \]
    If $\mu<N^{-1/2}$, both sides of \eqref{eq:isotropic-finite-source-condition} are $0$. So \Cref{lem:isotropic-source} applies with $d_s=d_0$, $p=p_N$, and $\xi=\xi_N$.

    Let $\mathcal D_{\rm Y}$ be the distribution of the padded lift of a uniformly random simple $d_0$-regular graph on $[N]$ with labels drawn from $\mathsf U$. Let $\mathcal D_{\rm N}$ be defined in the same way, with labels drawn from $\mathsf P$ with noise rate $p_N$. Since $d_0$ is even, these are defined for every $N$, and both are supported on simple graphs on $[n]$ of maximum degree at most $d_0$.

    By \Cref{lem:lower-yes-simple}, applied with $D'=D$, the core $H'$ of $G\sim\mathcal D_{\rm Y}$ satisfies $\Phi_\textup{in}(H')\ge463/1024>\phi$, except with probability $\varepsilon_{\rm Y}(N)$, where $\varepsilon_{\rm Y}(N)\to0$ does not depend on $\mu$. Since $|V(H')|=m\ge\alpha n$, we get $G\in\mathcal P_{\alpha,\phi}$ on this event.

    For the NO property, let $G\sim\mathcal D_{\rm N}$, and assume $N\ge8$. Partition $V(G)$ into the intersections of the color classes with $V(H')$ and the singletons of the isolated vertices. By \Cref{lem:lower-colors}, every class has at most $N$ vertices, and at most $qB_{\rm bad}$ edges join different classes, since deleting $D$ only removes edges. Let $U\subseteq V(G)$ with $u=|U|\ge\kappa\alpha n$. Then $u>N$, and \Cref{lem:lower-class-cut} gives
    \[
        \Phi_\textup{in}(G[U])\le\frac{2qB_{\rm bad}}{d_0D_N(u)}.
    \]
    If $\mu<N^{-1/2}$, then $B_{\rm bad}=0$, so every such $U$ has inner conductance $0$. Otherwise, $B_{\rm bad}\sim\operatorname{Bin}(d_0N/2,p_N)$ has mean $m_B=128\kappa\phi^2\mu N\ge128\kappa\phi^2\sqrt N$, and \Cref{lemma:Bernoulli-tails} gives
    \[
        \Pr\bigl(B_{\rm bad}\ge\tfrac32m_B\bigr)
        \le e^{-m_B/12}
        \le\exp\Bigl(-\frac{32}{3}\kappa\phi^2\sqrt N\Bigr)
        =:\varepsilon_{\rm N}(N).
    \]
    If $B_{\rm bad}<\frac32m_B$, then \eqref{eq:lower-rounding-denominator} shows that every such $U$ satisfies
    \[
        \Phi_\textup{in}(G[U])<\frac{3qp_NN}{2D_N(u)}=\frac34\psi\cdot\frac{\kappa qN}{D_N(u)}<\psi.
    \]
    In both cases, $G\notin\mathcal P_{\kappa\alpha,\psi}$ except with probability at most $\varepsilon_{\rm N}(N)$, which does not depend on $\mu$.

    Let $s=s(d_0,q)$ and $N_0=N_0(d_0,q)$ be the constants of \Cref{lem:isotropic-source}, and let
    \[
        A=\frac{s}{2}\left(\frac{\alpha}{q}\right)^{1/2+c}.
    \]
    We show that $An^{1/2+c\mu}\le sN^{1/2+\xi_N}$ for all large $N$, uniformly in $\mu$. Since $\alpha n\le m\le qN$, we have $n\le qN/\alpha$. If $\mu\ge N^{-1/2}$, then $\xi_N=c\mu$ and
    \[
        An^{1/2+c\mu}\le A\left(\frac q\alpha\right)^{1/2+c}N^{1/2+c\mu}=\frac{s}{2}N^{1/2+\xi_N}.
    \]
    If $\mu<N^{-1/2}$, then $\xi_N=0$, and $n^{c\mu}\le\exp\bigl(c\ln(qN/\alpha)/\sqrt N\bigr)\le2$ for all large $N$. Hence
    \[
        An^{1/2+c\mu}\le2A\left(\frac q\alpha\right)^{1/2}\sqrt N=s\left(\frac{\alpha}{q}\right)^{c}\sqrt N\le sN^{1/2+\xi_N}.
    \]

    Now suppose that $\mathcal A$ satisfies the correctness requirements of the theorem and makes at most $An^{1/2+c\mu}$ queries on every promised input and every random tape. Stop $\mathcal A$ after $\lfloor An^{1/2+c\mu}\rfloor$ queries on every input, with an arbitrary output if it attempts another query. This does not change its behavior on promised inputs. By \Cref{lem:lower-simple} and the query budget, the stopped algorithm can be simulated with at most $sN^{1/2+\xi_N}$ source vertex queries. For $N\ge N_0$, \Cref{lem:isotropic-source} shows that its acceptance probabilities under $\mathcal D_{\rm Y}$ and $\mathcal D_{\rm N}$ differ by less than $1/6$. On the other hand, correctness on promised inputs gives
    \[
        \Pr\nolimits_{\mathcal D_{\rm Y}}(\text{accept})
        \ge\frac23\bigl(1-\varepsilon_{\rm Y}(N)\bigr),
        \qquad
        \Pr\nolimits_{\mathcal D_{\rm N}}(\text{accept})
        \le\frac13+\frac23\varepsilon_{\rm N}(N),
    \]
    so the two probabilities differ by at least $\frac13-\frac23\bigl(\varepsilon_{\rm Y}(N)+\varepsilon_{\rm N}(N)\bigr)$.

    Let $N_1=N_1(\alpha,\kappa,\phi)\ge\max\{N_0,8\}$ be such that all estimates above hold and $\varepsilon_{\rm Y}(N)+\varepsilon_{\rm N}(N)<1/4$ for every $N\ge N_1$. Then the difference exceeds $1/6$, a contradiction. Finally, let $n_0=\lceil qN_1/\alpha\rceil$. For $n\ge n_0$, \eqref{eq:lower-padding} gives $N\ge\alpha n/q\ge N_1$. Hence the contradiction holds for every $n\ge n_0$ and every $0<\mu\le1$, and $A$ and $n_0$ depend only on $\alpha$, $\kappa$ and $\phi$.

    The coefficient $c_{\kappa,\phi}$ is not optimized; the same argument gives any coefficient smaller than $2c_{\kappa,\phi}$, with constants depending on the gap.
\end{proof}

\begin{remark}[Other degree bounds]
The choice $d_0=512$ serves two purposes: $d_0$ is even, so that $d_0N$ is even for every $N$ in \eqref{eq:lower-padding}, and the YES conductance exceeds $9/20$ (\Cref{lem:lower-yes-simple}). \Cref{lem:isotropic-source} holds for every fixed $d_s\ge3$ with $d_sN$ even. The NO argument also extends to these degrees: when conductance is normalized by $d=d_s$, the factor $d_s$ from the number $d_sN/2$ of base edges cancels the factor $1/d$ in \Cref{lem:lower-class-cut}. For any constant even $d\ge4$ and any $\phi$ with $0<\phi<\frac{d-2\sqrt{d-1}-2}{2d}$, the same proof with $d_s=d$, applying \Cref{lemma:friedman,lemma:mop} with a sufficiently small additive slack, gives the bound of \Cref{thrm:isotropic-lower-bound-full} with $\ln d_0$ replaced by $\ln d$ in $c_{\kappa,\phi}$ and with $A,n_0$ depending also on $d$; this range of $\phi$ is nonempty only for $d\ge8$. For $\alpha<1$, one may instead pad the full lift $H$, of order $qN<\alpha n+q+1\le n$ for all large $n$, with $n-qN$ isolated vertices. Then no vertices are deleted, \eqref{eq:lower-rounding-denominator} still holds since $q(N-1)<\alpha n$, and the condition relaxes to $\phi<\frac{d-2\sqrt{d-1}}{2d}$, which is nonempty for every even $d\ge4$. In particular, smaller degree bounds are possible at the cost of a smaller completeness threshold $\phi$, as in \Cref{thrm:exact-size-lower-bound-full}.
\end{remark}

\begin{remark}[Independent binary coordinates]
\label{rem:coordinate-hybrid}
    Independent binary noise on the $r$ coordinates also gives a bound of the form $n^{1/2+\Omega_{\kappa,\phi}(\mu)}$, with a smaller coefficient. Let the coordinates of $Z_e$ be independent Bernoulli variables with mean $\kappa\psi/(4r)$ when $\mu\ge N^{-1/2}$, and let $Z_e=0$ otherwise; then $\Pr(Z_e\ne0)\le\kappa\psi/4$, and the NO property follows as above. A hybrid argument over the $r$ coordinates reduces each step to the binary case $q=2$ of \Cref{lem:isotropic-source} with target distance $1/(12r)$: an algorithm can simulate the other coordinates itself, since a source vertex query reveals all edges at a vertex. This yields the coefficient $c^{\mathrm{coord}}_{\kappa,\phi}=\kappa\phi^2/(16r\ln d_0)$, smaller than $c_{\kappa,\phi}$ by the factor $rq/(q-1)=\Theta(\log(2/\kappa))$. The loss comes from the noise: at total rate $p$, independent coordinates give a single-coordinate parity bias about $1-2p/r$, whereas isotropic noise gives every nontrivial character bias $1-qp/(q-1)$.
\end{remark}

\section{Linear query complexity without size relaxation}
\label{sec:exact-size-obstruction}

This section proves the exact-size obstruction stated in \Cref{subsec:main-results}. A core on exactly $\lceil\alpha n\rceil$ vertices can be destroyed by isolating one hidden vertex. The two inputs differ in only constantly many edges, but every adaptive tester must make linearly many queries to detect the change. The argument also gives a lower bound on the expected number of queries on a fixed YES input.

\begin{theorem}[Linear lower bound without size relaxation]
\label{thrm:exact-size-lower-bound-full}
\label{thrm:exact-size-lower-bound}
    Fix $\alpha\in(0,1]$ and $\phi\in(0,1/2)$. Let $d\ge4$ be an even integer such that
    \begin{equation}
    \label{eq:exact-size-degree}
        \frac{d-2\sqrt{d-1}-1}{2d}>\phi.
    \end{equation}
    Such a choice $d=d(\phi)$ exists; for $\phi\le9/20$, one may take $d=512$. There is an integer $n_1=n_1(\alpha,\phi,d)$ with the following property. Let $n\ge n_1$, put $k=\lceil\alpha n\rceil$, and let $0<\mu_n\le1$. Suppose that an adaptive randomized algorithm $\mathcal A$, with unrestricted computation, accepts every simple graph in $\mathcal{P}_{\alpha,\phi}$ of order $n$ and maximum degree at most $d$ with probability at least $2/3$, and rejects every simple graph outside $\mathcal{P}_{\alpha,\mu_n\phi^2}$ of the same order and degree bound with probability at least $2/3$. Then there is a graph $G_{\rm Y}\in\mathcal{P}_{\alpha,\phi}$ of order $n$ and maximum degree at most $d$ on which the expected number of adjacency-list queries made by $\mathcal A$ is at least
    \[
        \frac{k}{3(d+1)}.
    \]
    In particular, its worst-case adjacency-list query complexity is at least this quantity. The lower bound holds even if uniform vertex samples are available at no cost.

    If $\mu_n\phi^2\le\phi$, then $dn$ adjacency-list queries suffice: read the entire adjacency table and decide membership in $\mathcal{P}_{\alpha,\phi}$ by exhaustive search.
\end{theorem}

For fixed $\alpha,\phi,d$, the lower bound is linear in $n$, so both the worst-case and the expected query complexities are $\Theta_{\alpha,\phi,d}(n)$. The bound does not depend on $\mu_n$.

\begin{proof}[Proof of \Cref{thrm:exact-size-lower-bound-full}]
    We first construct expanders of every sufficiently large order. The left side of \eqref{eq:exact-size-degree} tends to $1/2$ as $d\to\infty$, so a suitable even $d$ exists. For $d=512$, $2\sqrt{511}+1<46.3$ and $(512-46.3)/1024>9/20$. Fix $d$ as in the theorem.

    By \Cref{lemma:friedman} with $\varepsilon=1$, a uniformly random simple $d$-regular graph on $k$ vertices has second largest adjacency eigenvalue less than $2\sqrt{d-1}+1$ with probability $1-o(1)$ as $k\to\infty$; since $d$ is even, $dk$ is even for every $k$. Hence there is an integer $k_0=k_0(d)\ge d+1$ such that, for every $k\ge k_0$, there is a simple $d$-regular graph $H_k$ on $k$ vertices with
    \[
        \lambda_2(A_{H_k})\le2\sqrt{d-1}+1,
    \]
    where $\lambda_2$ denotes the second largest adjacency eigenvalue.

    For $S\subseteq V(H_k)$ with $0<|S|\le k/2$, put $f=\mathbf{1}_S-(|S|/k)\mathbf{1}$. Then $f\perp\mathbf{1}$, and
    \[
        \begin{aligned}
             |E_{H_k}(S,V(H_k)\setminus S)|
             &=f^{\mathsf T}(dI-A_{H_k})f\\
             &\ge\bigl(d-2\sqrt{d-1}-1\bigr)
                   |S|\left(1-\frac{|S|}{k}\right)\\
             &\ge\frac{d-2\sqrt{d-1}-1}{2}|S|
             >\phi d|S|.
        \end{aligned}
    \]
    Thus $\Phi_\textup{in}(H_k)>\phi$, with conductance normalized by the degree bound $d$.
 
    Next we define the YES graph and its local modifications. Choose $n_1$ so that $k=\lceil\alpha n\rceil\ge k_0$ and $\alpha n>1$ for all $n\ge n_1$, and fix such an $n$. Place $H_k$ on a fixed set $U\subseteq[n]$ of size $k$. Let $G_{\rm Y}$ consist of this copy of $H_k$ and the $n-k$ isolated vertices outside $U$, with fixed adjacency lists. Then $G_{\rm Y}\in\mathcal{P}_{\alpha,\phi}$, and $G_{\rm Y}$ does not depend on $\mathcal A$.
 
    For each $v\in U$, obtain $G_{\rm N}(v)$ by deleting the $d$ edges incident to $v$. Every connected component of $G_{\rm N}(v)$ has at most $k-1<\alpha n$ vertices. Hence every set $W$ with $|W|\ge\alpha n$ meets at least two components, and $G_{\rm N}(v)[W]$ is disconnected. A smallest component of this induced graph gives a nonempty cut side of size at most $|W|/2$ with no crossing edges, so $\Phi_\textup{in}(G_{\rm N}(v)[W])=0$. In particular,
    \begin{equation}
    \label{eq:exact-size-uniform-no}
        G_{\rm N}(v)\notin\mathcal{P}_{\alpha,\mu_n\phi^2}
        \qquad\text{for every }0<\mu_n\le1.
    \end{equation}
 
    We now couple the adjacency lists of $G_{\rm Y}$ and $G_{\rm N}(v)$. Give $G_{\rm N}(v)$ the same adjacency lists as $G_{\rm Y}$ outside $\{v\}\cup N_{H_k}(v)$. All entries at $v$ become null. At a neighbor of $v$, delete the entry for $v$ and order the remaining entries in any fixed valid way, followed by null entries. Thus a query $(x,i)$ can receive different answers on the two graphs only if $x\in\{v\}\cup N_{H_k}(v)$. For a fixed query vertex $x$, at most $d+1$ choices of $v\in U$ satisfy this condition.
 
    Fix the algorithm's random tape $\tau$, including the outcomes of any uniform vertex samples, and let $Q_\tau$ be the number of adjacency-list queries in its execution on $G_{\rm Y}$. If $\mathbb E_\tau[Q_\tau]=\infty$, the claimed lower bound is immediate. Otherwise $Q_\tau$ is finite for almost every $\tau$.
 
    Choose $v$ uniformly from $U$, independently of $\tau$, and couple the executions on $G_{\rm Y}$ and $G_{\rm N}(v)$ using the same random tape. Uniform vertex samples return the same vertices in both executions. The executions therefore remain identical until an adjacency-list query receives different answers. Before this first difference, every query also occurs in the fixed execution on $G_{\rm Y}$. A union bound over its $Q_\tau$ adjacency-list queries gives
    \begin{equation}
    \label{eq:exact-size-transcript}
        \Pr\nolimits_v[\text{the coupled transcripts differ}\mid\tau]\le \min\left\{1,\frac{(d+1)Q_\tau}{k}\right\}.
    \end{equation}
    If the transcripts agree, so do the outputs. Averaging over $\tau$ therefore yields
    \[
     \left|
     \Pr[\mathcal A\text{ accepts }G_{\rm Y}]
     -\mathbb E_v\!\left[
           \Pr[\mathcal A\text{ accepts }G_{\rm N}(v)]
         \right]
     \right|
     \le\frac{d+1}{k}\,\mathbb E_\tau[Q_\tau].
    \]
    All these graphs are simple, have order $n$, and have maximum degree at most $d$. By their YES and NO properties, the acceptance probabilities on the left differ by at least $1/3$. Consequently,
    \[
        \mathbb E_\tau[Q_\tau]\ge\frac{k}{3(d+1)}=\frac{\lceil\alpha n\rceil}{3(d+1)}.
    \]
    The same quantity also lower-bounds the worst-case query complexity.
 
    For the upper bound, reading the adjacency table takes $dn$ queries. With unrestricted computation, one can then test membership in $\mathcal{P}_{\alpha,\phi}$ by enumerating all candidate vertex sets and all their cuts. Since $\mu_n\le1$ and $\phi<1$, we have $\mu_n\phi^2\le\phi$, and hence $\mathcal{P}_{\alpha,\phi}\subseteq \mathcal{P}_{\alpha,\mu_n\phi^2}$. So this exact decision procedure satisfies both correctness requirements.
\end{proof}

The proof uses only that the core has maximum degree at most $d$ and inner conductance at least $\phi$: isolating one vertex then changes answers only at that vertex and its neighbors, and the same bound $k/(3(d+1))$ follows. Random regular graphs serve only to provide such a core of every sufficiently large order.
 
These hard instances are close to the YES property. Adding back the $d$ edges incident to $v$ transforms $G_{\rm N}(v)$ into $G_{\rm Y}\in\mathcal{P}_{\alpha,\phi}$. With edit distance normalized by $dn$, this gives
\[
    \textup{dist}_d(G_{\rm N}(v),\mathcal{P}_{\alpha,\phi})\le\frac1n.
\]
Thus this construction does not yield a lower bound for testers required to reject only graphs at a fixed positive distance from $\mathcal{P}_{\alpha,\phi}$. Indeed, under the distance promise the size threshold need not be relaxed: the tester of \Cref{thrm:upper-bound-distance-soundness-formal} rejects graphs that are $\varepsilon$-far from $\mathcal{P}_{\alpha,\mu_n^{\mathrm{dist}}\phi^2}$ using $O(\sqrt n\log n)$ queries for fixed $d,\alpha,\phi,\varepsilon$.
 
The argument also relies on retaining the size threshold $\alpha n$. For every fixed $\kappa\in(0,1)$, the remaining $k-1$ core vertices satisfy $k-1>\kappa\alpha n$ for all sufficiently large $n$. The component-size argument alone therefore no longer certifies the NO condition when the size threshold is relaxed to $\kappa\alpha n$. The noisy-label construction of \Cref{thrm:isotropic-lower-bound} establishes a lower bound even with this size relaxation.

\newpage
\appendix
\begin{center}
\LARGE\bfseries Appendix
\end{center}

\section{Proofs of the mixing and concentration bounds}
\label{sec:deferred-proofs}

This section proves two useful bounds, for completeness. \Cref{subsec:proof-of-mixing-time} proves the mixing bound of \Cref{lemma:mixing-time}, and \Cref{subsec:proof-of-Bernoulli-tails} proves the concentration bounds of \Cref{lemma:Bernoulli-tails}.

\subsection{Cheeger inequality and mixing}
\label{subsec:proof-of-mixing-time}

We prove \Cref{lemma:mixing-time} by the standard approach: the
conductance bound gives a spectral gap, and laziness turns the spectral gap into a bound on the
distance to $\pi$.

\begin{proof}[Proof of~\Cref{lemma:mixing-time}]

    Let $\Omega$ be the state space of $K$ and write $w_{xy}=\pi(x) K(x,y)$ and $w_{AB}=\sum_{x\in A, y\in B}{w_{xy}}$. Since $K$ is symmetric and stochastic, we get that $\pi$ is stationary and $w_{xy}=w_{yx}$. According to $\Phi(K)\ge h$, we get that for every $0<\pi(S)\le 1/2$, $\sum_{x\in S,y\in\Omega \setminus S}{w_{xy}}=w_{S,\Omega\setminus S}\ge h\pi(S)$.

    We first show that for every nonnegative function $g$ with $\pi(\{x:g(x)>0\})\le 1/2$, $\mathcal{E}_K(g)\ge \frac{h^2}{2}\lVert g\rVert_{2,\pi}^2$ holds. For $s\ge 0$, let $A_s=\{x:g(x)^2>s\}$. Note that $A_s\subseteq \{x:g(x)>0\}$ and $\pi(A_s)\le \pi(\{x:g(x)>0\})\le 1/2$. Integrating the conductance bound over $s$ gives
    \begin{align*}
        \sum_{x,y}{w_{xy}|g(x)^2-g(y)^2|}=2\int_0^\infty w_{A_s,\Omega\setminus A_s}ds\ge 2h\int_0^\infty \pi(A_s)ds=2h\lVert g\rVert_{2,\pi}^2.
    \end{align*}

    The first equality follows because the pair $(x,y)$ crosses the cut $A_s$ exactly when $s$ lies between $g(x)^2$ and $g(y)^2$.

    On the other hand, the Cauchy--Schwarz inequality gives
    \begin{align*}
        &\sum_{x,y}w_{xy}|g(x)^2-g(y)^2|\\
        &\le
        \left(\sum_{x,y}w_{xy}(g(x)-g(y))^2\right)^{1/2}
        \left(\sum_{x,y}w_{xy}(g(x)+g(y))^2\right)^{1/2}\\
        &\le\sqrt{2\mathcal{E}_K(g)}\sqrt{4\lVert g\rVert_{2,\pi}^2}=\sqrt{8\mathcal E_K(g)\|g\|_{2,\pi}^2}.
    \end{align*}
    
    For the last inequality, we used $(a+b)^2\le 2a^2+2b^2$, $\sum_y w_{xy}=\pi(x)$ and $\sum_x w_{xy}=\pi(y)$. Combining the two bounds proves the claim.

    Now let $f$ have $\pi$-mean zero. Choose a median $m$ of $f$ under $\pi$, so that
    \[
        \pi(\{x:f(x)>m\})\le \frac12,\qquad\pi(\{x:f(x)<m\})\le \frac12.
    \]
    Define
    \[
        g_+(x)=\max\{f(x)-m,0\},\qquad g_-(x)=\max\{m-f(x),0\}.
    \]
    Both functions satisfy the support condition above. Moreover, for every $x,y$,
    \[
        (f(x)-f(y))^2\ge (g_+(x)-g_+(y))^2+(g_-(x)-g_-(y))^2.
    \]
    Therefore,
    \begin{align*}
        \mathcal E_K(f)
        &\ge \mathcal E_K(g_+)+\mathcal E_K(g_-)\\
        &\ge \frac{h^2}{2}
           \bigl(\|g_+\|_{2,\pi}^2+\|g_-\|_{2,\pi}^2\bigr)\\
        &=\frac{h^2}{2}\|f-m\|_{2,\pi}^2\\
        &=\frac{h^2}{2}\bigl(\|f\|_{2,\pi}^2+m^2\bigr)\\
        &\ge \frac{h^2}{2}\|f\|_{2,\pi}^2.
    \end{align*}

    This proves the first assertion.

    Now suppose that $K$ is lazy, meaning that $K(x,x)\ge 1/2$ for every $x$. Define the inner product $\langle u,v\rangle_\pi=\sum_y\pi(y)u(y)v(y)$. Expanding the Dirichlet form gives
    \[
        \mathcal E_K(f)=\langle f,(I-K)f\rangle_\pi.
    \]
    Since $K$ is symmetric, it has an orthonormal eigenbasis under this inner product. Laziness makes all its eigenvalues nonnegative: indeed, $2K-I$ is a symmetric stochastic matrix, whose eigenvalues lie in $[-1,1]$. The first assertion thus implies that every eigenvalue on the mean-zero subspace lies in $[0,1-h^2/2]$. Consequently, for every mean-zero function $u$,
    \[
        \|K^t u\|_{2,\pi} \le (1-h^2/2)^t\|u\|_{2,\pi} \le e^{-h^2t/2}\|u\|_{2,\pi}.
    \]

    Fix a starting state $x$ and define
    
    \[
        r_t(y)=\frac{K^t(x,y)}{\pi(y)}-1 =|\Omega|K^t(x,y)-1.
    \]
    By symmetry, $r_t=K^t r_0$. Also, $r_0$ has $\pi$-mean zero and
    \[
        \|r_0\|_{2,\pi}^2 =\frac{1}{|\Omega|}\bigl((|\Omega|-1)^2+(|\Omega|-1)\bigr)=|\Omega|-1.
    \]
    Applying Cauchy--Schwarz once more, we obtain
    \begin{align*}
        \|e_xK^t-\pi\|_{\mathrm{TV}}
        &=\frac12\sum_y\pi(y)|r_t(y)|\\
        &\le \frac12\|r_t\|_{2,\pi}\\
        &\le \frac{1}{2}\sqrt{|\Omega|-1}\,e^{-h^2t/2}\\
        &\le \frac{1}{2}\sqrt{|\Omega|}\,e^{-h^2t/2},
    \end{align*}
    which ends the proof.
\end{proof}

\subsection{Bernoulli concentration bounds}
\label{subsec:proof-of-Bernoulli-tails}

We prove the Bernoulli concentration bounds (\Cref{lemma:Bernoulli-tails}) used in \Cref{lemma:score-estimation}. 

\begin{proof}[Proof of~\Cref{lemma:Bernoulli-tails}]
    Write $S=\sum_{i=1}^k X_i$ and $m=kp$. If $m=0$, both bounds are immediate, so assume $m>0$. Independence and $1+x\le e^x$ give, for every $t\in\mathbb R$,
    \[
        \mathbb E[e^{tS}]
        =\prod_{i=1}^k\mathbb E[e^{tX_i}]
        =(1-p+pe^t)^k
        \le \exp\bigl(m(e^t-1)\bigr).
    \]

    For the upper tail, apply Markov's inequality with $t=\ln(1+a)>0$ to obtain
    \[
        \Pr[S\ge(1+a)m] \le \exp\bigl(m(e^t-1)-t(1+a)m\bigr) =\exp\bigl(-m((1+a)\ln(1+a)-a)\bigr).
    \]
    For $0\le a\le1$, the inequality $\ln(1+x)\ge x-x^2/2$ for $x\ge0$ implies
    \[
        (1+a)\ln(1+a)-a =\int_0^a\ln(1+x)\,dx \ge\frac{a^2}{2}-\frac{a^3}{6} \ge\frac{a^2}{3}.
    \]
    This proves the upper-tail bound.

    For the lower tail, suppose first that $0<a<1$. Applying Markov's inequality to $e^{tS}$ with $t=\ln(1-a)<0$ yields
    \[
        \Pr[S\le(1-a)m] \le \exp\bigl(m(e^t-1)-t(1-a)m\bigr) =\exp\bigl(-m(a+(1-a)\ln(1-a))\bigr).
    \]
    Since $-\ln(1-x)\ge x$ for $0\le x<1$,
    \[
        a+(1-a)\ln(1-a) =\int_0^a-\ln(1-x)\,dx \ge\frac{a^2}{2}.
    \]
    Finally, at $a=1$,
    \[
        \Pr[S\le0]=(1-p)^k\le e^{-kp}\le e^{-kp/2}.
    \]
    Thus the lower-tail bound also holds at the endpoint.
\end{proof}

\section{Graphs close to expanders}
\label{sec:graphs-close-to-expanders}

This appendix quantifies the relation between having a large expanding core and being close in edit distance to an expander, as mentioned in \Cref{subsec:related-work}. Subdivision in \Cref{sec:distance} enlarges a core with few edits; pruning gives the complementary implication, showing that few deleted edges can destroy expansion on only a controlled number of vertices.

We use the standard pruning argument (see, for example, \cite{saranurak2019expander}). Repeatedly remove the smaller side of a cut whose surviving boundary is too sparse. Expansion in the original graph forces the total removed set to have a large boundary, unless sufficiently many of its boundary edges were deleted. The first-crossing argument below prevents a single removal from discarding a large part of the graph before this charge can be made.

\begin{lemma}[Pruning after edge deletions]
\label{lem:pruning}
    Let $H$ be a graph on $m\ge2$ vertices with maximum degree at most $d$ and $\Phi_\textup{in}(H)\ge\phi>0$. Delete $r\le\phi dm/32$ edges from $H$ to obtain $H^-$. Then there is a set $U'\subseteq V(H)$ with
    \[
        |U'|\ge m-\frac{2r}{\phi d}
        \qquad\text{and}\qquad
        \Phi_\textup{in}(H^-[U'])\ge\phi/4.
    \]
\end{lemma}

\begin{proof}
    Start with $U'=V(H)$. While $\Phi_\textup{in}(H^-[U'])<\phi/4$, choose a nonempty set $S\subseteq U'$ with $|S|\le|U'|/2$ and fewer than $\phi d|S|/4$ edges of $H^-$ between $S$ and $U'\setminus S$, and remove $S$ from $U'$. Let $D=V(H)\setminus U'$ be the set of removed vertices and $s=|D|$. Every edge of $H^-$ between $D$ and $U'$ was counted in the cut when its endpoint in $D$ was removed, and at most $r$ edges of $H$ are missing from $H^-$. Hence
    \begin{equation}
    \label{eq:prune-boundary}
        |E_H(D,V(H)\setminus D)|\le r+\phi ds/4.
    \end{equation}

    The removed set never reaches $m/4$ vertices. Otherwise, consider the first step at which it does, and let $s_0<m/4$ be its previous size. Because the removed side has at most half the remaining vertices, the new size satisfies $m/4\le s<5m/8$. Both sides of the original cut then contain at least $m/4$ vertices, so expansion gives a boundary of at least $\phi dm/4$. But \eqref{eq:prune-boundary} bounds it by $\phi dm/32+5\phi dm/32=3\phi dm/16$, a contradiction.

    We may therefore apply original expansion to the removed set itself: $\phi ds\le r+\phi ds/4$. Rearranging gives $s\le4r/(3\phi d)\le2r/(\phi d)$. Each pruning step removes a vertex, but the preceding argument forbids removing a quarter of the graph. The process must stop with a nonempty set, and its stopping condition gives conductance at least $\phi/4$.
\end{proof}

\begin{corollary}[A core survives small edit noise]
\label{coro:close-witness}
    If $\textup{dist}_d(G,\mathcal P_{\alpha,\phi})\le\zeta$, $\zeta\le\phi\alpha/32$ and $\alpha n\ge 2$, then $G\in\mathcal P_{\alpha-2\zeta/\phi,\,\phi/4}$.
\end{corollary}

\begin{proof}
    Let $G'\in\mathcal P_{\alpha,\phi}$ with $\textup{dist}_d(G,G')\le\zeta$, and let $U$ be a core in $G'$ with $m=|U|\ge\alpha n$. At most $\zeta dn\le\zeta dm/\alpha\le\phi dm/32$ edges of $G'[U]$ are missing in $G$. Apply \Cref{lem:pruning} to $H=G'[U]$ with these deletions. The resulting set $U'$ has $|U'|\ge m-2\zeta n/\phi\ge(\alpha-2\zeta/\phi)n$, and $\Phi_\textup{in}(G[U'])\ge\Phi_\textup{in}(H^-[U'])\ge\phi/4$, because $H^-[U']$ is a subgraph of $G[U']$: every cut of $G[U']$ has at least as many edges as the same cut of $H^-[U']$, and the denominator $d|S|$ is the same.
\end{proof}

For $\alpha=1$, \Cref{coro:close-witness} says that a graph $\zeta$-close to a $\phi$-expander contains a $\phi/4$-expanding core on at least $(1-2\zeta/\phi)n$ vertices, provided $\zeta\le\phi/32$. In the other direction, apply \Cref{coro:size-slack-implies-small-edit-distance} with $\alpha=1$ and $\kappa=1-\eta$. For $d\ge3$, $0<\eta\le1/4$, and $n\ge4/\eta$, every graph in $\mathcal P_{1-\eta,\theta}$ is $(5\eta/2)$-close to $\mathcal P_{1,\theta/(2d+2)}$. Thus the two notions agree up to constant-factor losses in the parameters. For $\eta=0$, the latter implication is immediate without edits.

\paragraph*{AI methodology statement.}
We used ChatGPT (OpenAI, GPT-6 Astra) to assist with exploring and checking proofs, including drafting parts of the lower-bound analysis and independently rederiving the circulation and adaptive-exposure arguments in \Cref{lem:isotropic-closure,lem:isotropic-source}, and with writing code for numerical sanity checks. We used ChatGPT and Claude (Anthropic, Claude Opus 5.5) to assist with restructuring and rewriting parts of the exposition, including the lower-bound section. The authors verified the correctness and originality of all content, including references, and take full responsibility for the paper.

\clearpage
\phantomsection
\addcontentsline{toc}{section}{References}
\bibliographystyle{alphaurl}
\bibliography{references}

\end{document}